\documentclass[a4paper,11pt]{article}
\pdfoutput=1 
\usepackage{jheppub} 
\usepackage{amsmath,amsfonts,amssymb,mathrsfs,graphicx,color,longtable,bm,wasysym}
\usepackage{hyperref}
\usepackage{graphicx}
\usepackage{array}
\usepackage{color}
\usepackage[usenames,dvipsnames,table]{xcolor}
\usepackage{tikz}
\usepackage[utf8]{inputenc}
\usepackage{epsfig,relsize,xspace}
\usepackage[T1]{fontenc}
\usepackage{bbold}
\usepackage{float}
\usepackage{amsmath}
\usepackage{empheq}
\usepackage{booktabs}
\usepackage{color}
\usepackage[utf8]{inputenc}
\usepackage{xspace}
\usepackage{scalerel}
\usepackage[most]{tcolorbox}
\usepackage{multirow}
\usepackage{scalefnt}
\usepackage{bold-extra}
\usepackage[shortlabels]{enumitem}
\usepackage[tikz]{bclogo}
\usepackage{subcaption}
\usepackage{tikz-feynman}
\usepackage{multirow}
\usepackage{cancel}
\usepackage{verbatim}
\usetikzlibrary{arrows,shapes}
\usepackage{afterpage}
\usepackage{graphicx,grffile}
\usepackage{slashed}
\usepackage{soul}
\usepackage{braket}
\usepackage{xifthen}
\usepackage{afterpage}
\usepackage{multirow}

\allowdisplaybreaks
 
\definecolor{nicered}{rgb}{0.7,0.1,0.1}
\definecolor{nicegreen}{rgb}{0.1,0.5,0.1}
\definecolor{niceblue}{rgb}{0.0,0.1,0.7}
\hypersetup{colorlinks,citecolor=niceblue,linkcolor=niceblue,urlcolor=niceblue}

\usepackage[normalem]{ulem}
 
\def \bm#1{\mbox{\boldmath$#1$\unboldmath}}
\def \beq{\begin{equation}}
\def \eeq{\end{equation}}
\def \bea{\begin{eqnarray}}
\def \eea{\end{eqnarray}}
\def \LOplusPS {LO$+$PS}
\def \NLOplusPS {NLO$+$PS}
\def \NNLOplusPS {NNLO$+$PS}
\def \NLOplusPSbm {NLO$\bm{+}$PS}
\def \NNLOplusPSbm {NNLO$\bm{+}$PS}

\def \MiNNLOPS {MiNNLO$_{\rm PS}$}

\newcommand{\PYTHIA}[1]{{\tt Pythia~{#1}}}
\newcommand{\MiNLO}{MiNLO$^{\prime}$}

\newcommand{\Spab}[3]{ \langle #1 | #2 | #3 ]}

\newcommand{\Spaa}[3][]{
 \ifthenelse{\isempty{#1}}
 {\langle #2 #3 \rangle}
 {\langle #2 | #1 | #3 \rangle}
}

\newcommand{\Spbb}[3][]{
 \ifthenelse{\isempty{#1}}
 {[ #2 #3 ]}
 {[ #2 | #1 | #3 ]}
}

\usepackage{amsmath}
\usepackage{graphicx}

\newcommand{\ghzz}[1]{g_{hZZ}^{(#1)}}
\newcommand{\ghaz}[1]{g_{h\gamma Z}^{(#1)}}

\def \ggF {ggF}
\def \qqF {${\rm q \bar q F}$}

\begin{document}
\pagecolor{white}

\def\arraystretch{1.25}

\preprint{MPP-2023-261}

\title{SMEFT at \NNLOplusPSbm: $\bm{Vh}$ production}

\author[1]{Rhorry Gauld,}
\author[1]{Ulrich Haisch}
\author[1,2]{and Luc Schnell}

\affiliation[1]{Max Planck Institute for Physics, \\ F{\"o}hringer Ring 6, 80805 M{\"u}nchen, Germany}
\affiliation[2]{Physik-Department, Technische Universit\"at M\"unchen, \\ James-Franck-Strasse 1, 85748 Garching, Germany}

\emailAdd{rgauld@mpp.mpg.de}
\emailAdd{haisch@mpp.mpg.de}
\emailAdd{schnell@mpp.mpg.de}

\abstract{In the context of the Standard Model effective field theory~(SMEFT) the next-to-next-to-leading~(NNLO) QCD corrections to the Higgsstrahlungs~($Vh$) processes in hadronic collisions are calculated and matched to a parton shower~(PS). \NNLOplusPS~precision is achieved for the complete sets of SMEFT operators that describe the interactions between the Higgs and two vector bosons and the couplings of the Higgs, a $W$ or a $Z$ boson, and light fermions. A~{\tt POWHEG-BOX} implementation of the computed NNLO~SMEFT corrections is provided that allows for a realistic exclusive description of $Vh$ production at the level of hadronic events. This feature makes it an essential tool for future Higgs characterisation studies by the ATLAS and CMS collaborations. Utilising our new Monte Carlo code the numerical impact of \NNLOplusPS~corrections on the kinematic distributions in $pp \to Zh \to \ell^+ \ell^- h$ production is explored, employing well-motivated SMEFT benchmark~scenarios.}

\maketitle

\section{Introduction}
\label{sec:introduction}

The dominant decay of the Standard Model~(SM) Higgs boson is into a pair of bottom quarks, with an expected branching ratio close to 60\% for a Higgs mass of $125 \, {\rm GeV}$. Large~QCD~backgrounds from multi-jet production, however, make a search in the dominant gluon-gluon fusion~(\ggF) Higgs production channel very challenging at the LHC. The~most sensitive production mode for detecting $h \to b \bar b$ decays is the associated production of a Higgs boson and a massive gauge boson~($Vh$), where the leptonic decay of the vector boson enables a clean selection, leading to a significant background reduction. The $h \to b \bar b$ decay mode has been observed by both ATLAS and CMS in LHC~Run~II~\cite{ATLAS:2018kot,CMS:2018nsn}, and these measurements constrain the $h \to b \bar b$ signal strength in the Higgsstrahlungs processes $\big ( \mu^{b \bar b}_{Vh} \big )$ to be SM-like within about~25\%. With LHC~Run~III ongoing and the high-luminosity upgrade~(HL-LHC) on the horizon, the precision of the $\mu^{b \bar b}_{Vh}$ measurements is expected to improve significantly, resulting in an ultimate projected HL-LHC accuracy of $15 \%$ ($5\%$) in the case of the $Wh$~($Zh)$ production channel~\cite{ATLAS:2018jlh,CMS:2018qgz}.

Besides providing a probe of the dominant decay mode of the Higgs boson, precision $Vh$ measurements also play an important role in the Higgs characterisation programme which is commonly performed in the framework of the SM effective field theory~(SMEFT)~\cite{Buchmuller:1985jz,Grzadkowski:2010es,Brivio:2017vri}. In~fact, radiative corrections in the SMEFT to both $Vh$ production~\cite{Mimasu:2015nqa,Degrande:2016dqg,Alioli:2018ljm,Bishara:2020vix,Bishara:2020pfx,Bishara:2022vsc} and the $h \to f \bar f$ decays~\cite{Gauld:2015lmb,Gauld:2016kuu,Cullen:2019nnr,Cullen:2020zof,Haisch:2022nwz} have been calculated. The~existing studies for $Vh$ production have mostly focused on the subset of higher-dimensional interactions that modify the couplings of the Higgs to two vector bosons achieving next-to-leading order~(NLO)~\cite{Maltoni:2013sma,Mimasu:2015nqa,Degrande:2016dqg,Greljo:2017spw,Alioli:2018ljm} and next-to-next-to-leading~order~(NNLO)~\cite{Bizon:2021rww} in QCD, respectively, while in the case of $h \to b \bar b$ both NLO~QCD and NLO~electroweak~(EW) corrections to the total decay width have been calculated for the full set of relevant dimension-six SMEFT~operators~\cite{Gauld:2015lmb,Gauld:2016kuu,Cullen:2019nnr}. In the publications~\cite{Bishara:2020vix,Bishara:2020pfx,Bishara:2022vsc} special attention has finally been paid to the class of SMEFT operators that lead to interactions between a Higgs, $W$ or $Z$ boson, and light quarks. 

The goal of the present work is to generalise and to extend the recent SMEFT~calculation~\cite{Haisch:2022nwz} which has achieved NNLO plus parton shower~(\NNLOplusPS)~accuracy for the dimension-six operators that contribute to the subprocesses $pp \to Zh$ and $h \to b \bar b$ directly in~QCD. This class of operators includes effective Yukawa- and chromomagnetic dipole-type interactions of the bottom quark that modify the $h \to b \bar b$ decay but do not play a role in $pp \to Zh$ production. Purely EW effective interactions that alter the couplings of the Higgs to gauge bosons are instead not included in the \NNLOplusPS~Monte~Carlo~(MC) generator presented in~\cite{Haisch:2022nwz}. Since these types of SMEFT contributions can lead to phenomenologically relevant effects in the Higgsstrahlungs processes~\cite{Maltoni:2013sma,Mimasu:2015nqa,Degrande:2016dqg,Greljo:2017spw,Alioli:2018ljm,Bizon:2021rww}, we include these type of interactions in the current article, extending the NLO~SMEFT calculations~\cite{Mimasu:2015nqa,Degrande:2016dqg,Alioli:2018ljm} to the NNLO level. Likewise, we improve the precision of the calculations of SMEFT corrections to $pp \to Vh$ production that are associated to couplings between a Higgs, $W$ or $Z$ boson, and light quarks~\cite{Bishara:2020vix,Bishara:2020pfx,Bishara:2022vsc} to NNLO in QCD. The obtained fixed-order SMEFT predictions are implemented into the {\tt POWHEG-BOX}~\cite{Alioli:2010xd} and consistently matched to a parton shower~(PS) using the \MiNNLOPS~method~\cite{Monni:2019whf,Monni:2020nks}. In this way, NNLO~QCD accuracy is retained for both production and decays, while the matching to the PS ensures a realistic exclusive description of the $pp \to Zh \to \ell^+ \ell^- h$ and the $pp \to Wh \to \ell \nu h$ process at the level of hadronic events. These features make our new \NNLOplusPS~generator~\cite{GitLabPowheg} a precision tool for future LHC Higgs characterisation studies in the SMEFT framework. 

This paper is structured as follows: in~Section~\ref{sec:operators} we specify the dimension-six SMEFT operators that are relevant in the context of our work. A comprehensive description of the basic ingredients~of the SMEFT calculation of $pp \to Vh$ production and their implementation in the context of our \NNLOplusPS~event generator is presented in~Section~\ref{sec:calculation}. We~motivate simple SMEFT benchmark scenarios in Section~\ref{sec:anatomy} by discussing the leading constraints on the Wilson coefficients of the relevant operators. The impact of the SMEFT corrections on kinematic distributions in $pp \to Zh \to \ell^+ \ell^- h$ production at the LHC is studied in~Section~\ref{sec:numerics} employing the simple benchmark scenarios for the Wilson coefficients identified earlier. In~Section~\ref{sec:higgsdecay} we outline how the \NNLOplusPS~calculations of $pp \to Vh$ and $h \to b \bar b$ can be combined, while Section~\ref{sec:conclusions} contains our conclusions and an outlook. The~analytic formulae~for the parameters and couplings that have been implemented into our MC code~\cite{GitLabPowheg} are relegated to~Appendix~\ref{app:parameters}. Details on our implementation of the $gg \to Zh$ process can be found in~Appendix~\ref{app:gginitiated}. In~Appendix~\ref{app:NLOPSvsNNLOPS} we finally discuss the impact of NNLO~QCD effects by comparing results for $pp \to Zh \to \ell^+ \ell^- h$ production obtained at \NLOplusPS~and~\NNLOplusPS, respectively. 

\section{SMEFT operators}
\label{sec:operators}

\begin{figure}[t!]
\begin{center}
\includegraphics[width=0.9\textwidth]{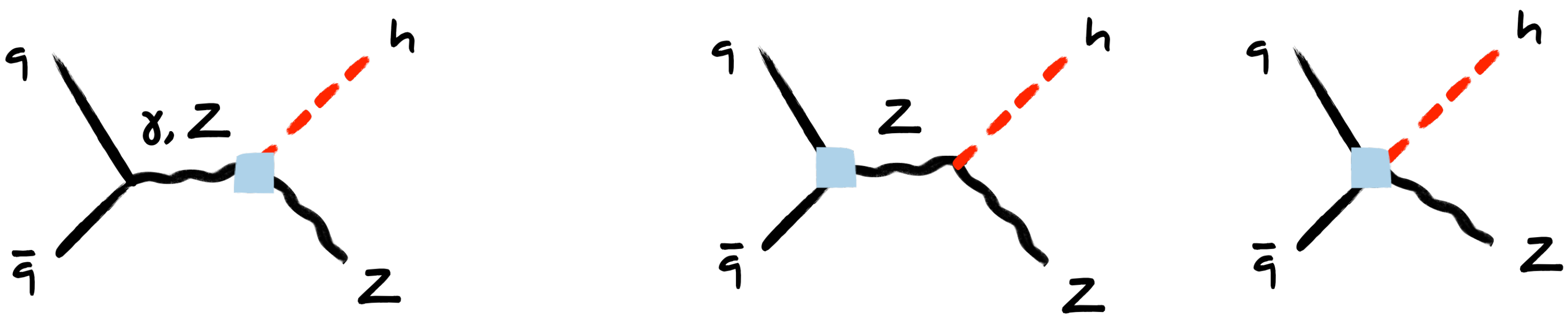}
\end{center}
\vspace{-2mm} 
\caption{\label{fig:tree} Tree-level SMEFT contributions to $q \bar q \to Zh$ production. The diagram on the left involves an insertion of one of the operators defined in~(\ref{eq:operators1}), while the two graphs on the right stem from an insertion of one of the operators given in~(\ref{eq:operators2}). The operator insertions are indicated by the blue squares.}
\end{figure}

Throughout this work we neglect all light fermion masses in both the SM and SMEFT corrections to the $pp \to Zh$ and $pp \to Wh$ processes. The~full~set of dimension-six SMEFT operators has been presented in the so-called Warsaw basis in the article~\cite{Grzadkowski:2010es}. This basis contains the following three independent operators 
\beq \label{eq:operators1}
Q_ {H\hspace{-0.3mm}B} = H^\dagger H \hspace{0.5mm} B_{\mu \nu} B^{\mu \nu} \,, \quad 
Q_{HW} = H^\dagger H \hspace{0.5mm} W_{\mu \nu}^a W^{a, \mu \nu} \,, \quad 
Q_{HW\!B} = H^\dagger \sigma^a H \hspace{0.5mm} W_{\mu \nu}^a B^{\mu \nu} \,,
\eeq
that modify the couplings between the Higgs and two vector bosons at tree level. The~SM Higgs doublet is denoted by $H$, while $B_{\mu \nu}$ and $W_{\mu \nu}^a$ are the $U(1)_Y$ and $SU(2)_L$ gauge field strength tensors and $\sigma^a$ are the Pauli matrices. In the case of the operators that result in couplings between the Higgs, a $W$ or a $Z$ boson, and light quarks, we consider the following five effective interactions
\beq \label{eq:operators2}
\begin{split}
& Q_{Hq}^{(1)} = (H^\dagger i \overset{\leftrightarrow}{D}_\mu H)(\bar q \gamma^\mu q) \,, \qquad 
Q_{Hq}^{(3)} = (H^\dagger i \overset{\leftrightarrow}{D} \ \!\!\!^{\; a}_\mu H) (\bar q \gamma^\mu \sigma^a q) \,, \\[1mm]
& \hspace{1mm} Q_{Hd} = (H^\dagger i \overset{\leftrightarrow}{D}_\mu H) (\bar d \gamma^\mu d) \,, \qquad 
Q_{Hu} = (H^\dagger i \overset{\leftrightarrow}{D}_\mu H) (\bar u \gamma^\mu u) \,, \\[2.75mm]
& \hspace{2.5cm} Q_{Hud} = (\widetilde H^\dagger i D_\mu H) (\bar u \gamma^\mu d) \,, 
\end{split}
\eeq 
where $H^\dagger i \overset{\leftrightarrow}{D}_\mu H = i H^\dagger \big (D_\mu - \overset{\leftarrow}{D}_\mu \big ) H$ and $H^\dagger i \overset{\leftrightarrow}{D} \ \!\!\!^{\; a}_\mu H = i H^\dagger \big (\sigma^a D_\mu - \overset{\leftarrow}{D}_\mu \sigma^a \big ) H$ with $D_\mu$ the usual covariant derivative and the shorthand notation $\widetilde H_i = \epsilon_{ij} (H_j)^\ast$ with $\epsilon_{ij}$ totally antisymmetric and $\epsilon_{12}=1$ has been used. The symbol~$q$ denotes left-handed quark doublets, while $u$ and~$d$ are the right-handed quark~singlets of up and down type, respectively. Illustrative diagrams that contribute to $Zh$ production and involve an insertion of one of the operators in~(\ref{eq:operators1}) or~(\ref{eq:operators2}) are displayed in~Figure~\ref{fig:tree}. Notice that $Q_{Hud}$ only contributes to~$pp \to Wh$ production and the dimension-six SMEFT Lagrangian includes the sum of the operator $Q_{Hud}$ and its hermitian conjugate.

Besides the two sets of operators~(\ref{eq:operators1}) and~(\ref{eq:operators2}) that alter the $pp \to Vh$ production process, we also consider effective interactions that modify the $Z \to \ell^+ \ell^-$ and $W \to \ell \nu$ decays at tree level. In the Warsaw basis there are three such operators, namely 
\bea \label{eq:operators3}
Q_{H\ell}^{(1)} = (H^\dagger i \overset{\leftrightarrow}{D}_\mu H) (\bar \ell \gamma^\mu \ell) \,, \quad 
Q_{H\ell}^{(3)} = (H^\dagger i \overset{\leftrightarrow}{D} \ \!\!\!^{\; a}_\mu H) (\bar \ell \gamma^\mu \sigma^a \ell) \,, \quad 
Q_{He} = (H^\dagger i \overset{\leftrightarrow}{D}_\mu H) (\bar e \gamma^\mu e) \,. \hspace{6mm} 
\eea
Here $\ell$ and $e$ denote a left-handed lepton doublet and right-handed lepton singlet field, respectively. Notice that in writing~(\ref{eq:operators2}) and~(\ref{eq:operators3}) we have assumed that the full SMEFT Lagrangian respects an approximate $U(3)^5$ flavour symmetry which allows us to drop all flavour indices and that the operator $Q_{Hud}$ is forbidden if the $U(3)^5$ flavour symmetry is~exact. 

The final type of SMEFT corrections that change the Higgsstrahlungs processes indirectly is provided by the Wilson coefficients of the operators that shift the Higgs kinetic term and/or the EW SM input parameters. In order to fully describe these shifts the following three additional operators are needed at tree level:
\beq \label{eq:operators4}
Q_{H\Box} = (H^\dagger H) \Box (H^\dagger H) \,, \quad 
Q_{H\hspace{-0.25mm}D} = (H^\dagger D_\mu H)^\ast(H^\dagger D^\mu H) \,, \quad 
Q_{\ell\ell} = (\bar \ell \gamma_\mu \ell) (\bar \ell \gamma^\mu \ell) \,. 
\eeq

\section{Calculation in a nutshell}
\label{sec:calculation}

In this section, we sketch the different ingredients of our \NNLOplusPS~SMEFT calculation of $pp \to V h$ production. We begin by recalling the basic steps of the NNLO~QCD computation within the SM and then detail the general method that we employ to calculate the relevant squared matrix elements in the SMEFT and their implementation into the {\tt POWHEG-BOX}. It~is then explained how the fixed-order NNLO SMEFT calculations of the $pp \to Vh$ processes are consistently matched to a PS using the \MiNNLOPS~method. 

\subsection{SM calculation}
\label{sec:SMfixedorder}

A core input of the NNLO QCD calculation are the squared matrix elements up to $\mathcal{O}(\alpha_s^2)$ in the SMEFT. To better explain how the calculation of these objects is performed, we first revisit the structure of the NNLO computation in the~SM, which we have repeated, and also implemented into the {\tt POWHEG-BOX}. Before doing so, we note that we will generically refer to the process $pp \to V h$ (and its corresponding subprocesses) in both the text and corresponding figures in what follows, but it should be understood that $V=W,Z$ refers to a final-state lepton pair, and that the calculation does include spin-correlation effects in the gauge-boson decays.

\begin{figure}[t!]
\begin{center}
\includegraphics[width=0.975\textwidth]{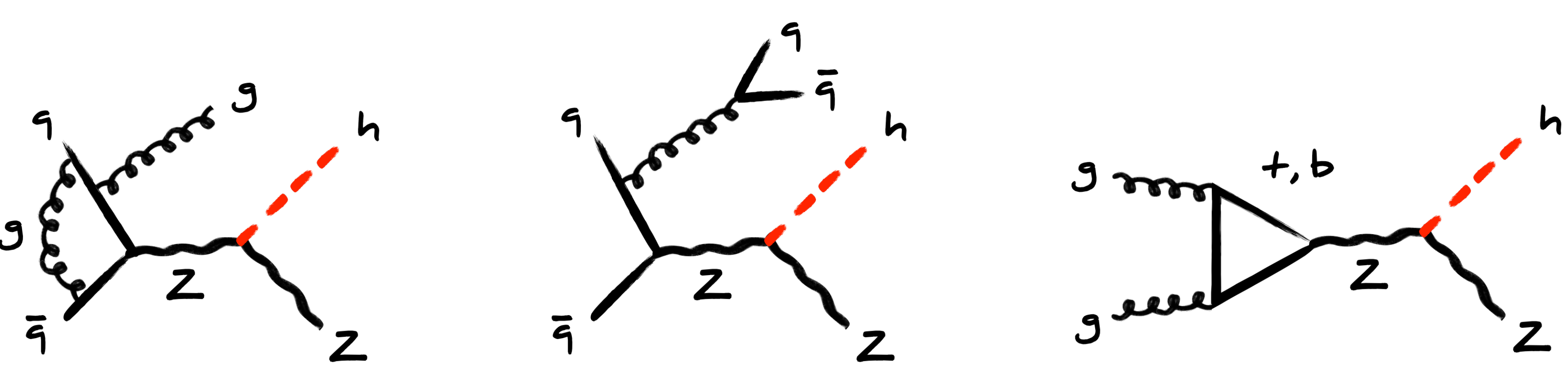}
\end{center}
\vspace{-2mm} 
\caption{\label{fig:corrections} Examples of higher-order QCD corrections to $p p \to Zh$ production within the SM. The diagram on the left features additional virtual and real gluon lines (\texttt{B}-type), the diagram in the middle involves a second quark line (\texttt{C}- and \texttt{D}-type) and the diagram on the right is a \ggF~contribution (\texttt{A}-type). Consult the main text for further details.}
\end{figure}

In the NNLO calculation of $pp \to V h$, the contributing partonic channels can be classified according to the number of external quark lines ($\texttt{A}=0,\texttt{B}=1,\texttt{C}=\texttt{D}=2$), external gluons, and also by the number of loops at the squared amplitude level --- see~also~\cite{Majer:2020kdg}. Starting with the \texttt{B}-type corrections (i.e.~those with a single external quark line), the required squared matrix elements are called \texttt{B0g0V, B1g0V, B0g1V, B1g1V, B2g0V, B0g2V}, where the number before \texttt{g} refers to the number of additional external gluons relative to the leading-order~(LO) contribution for that type, and the number after the {\tt g} refers to the number of loops at the squared level. For example, the left most diagram in Figure~\ref{fig:corrections} contains one additional gluon (relative to the Born-level contribution in the quark-antiquark fusion or \qqF~channel) and is a one-loop graph, and would therefore contribute to \texttt{B1g1V} through its interference with the corresponding tree-level amplitude. In the SM case, the analytic expressions for the corresponding spinor-helicity amplitudes can be found in~\cite{Kramer:1986sg,Hamberg:1990np,Gehrmann:2011ab,Majer:2020kdg}. To obtain the desired squared matrix elements, the spinor-helicity amplitudes can be squared and then summed over all contributing helicities numerically --- an explicit example of this procedure is given below. The \texttt{C}-type corrections feature two external quarks lines, or in other words a double real emission contribution with two final-state quarks, and arise from the interference of diagrams such as that represented in the centre of Figure~\ref{fig:corrections}. The \texttt{D}-type corrections account for the additional structures that can appear when same-flavour quarks are considered. These squared matrix elements are called \texttt{C0g0V} and \texttt{D0g0V} and within the SM the analytic expressions for the corresponding spinor-helicity amplitudes are provided in the work~\cite{Majer:2020kdg}. Finally, the \ggF~contributions shown on the right in~Figure~\ref{fig:corrections} constitute the third type of correction which are considered (\texttt{A}-type). They are referred to as \texttt{A0g2V} and the corresponding~SM spinor-helicity amplitudes are given in~\cite{Campbell:2016jau}. Notice that due to charge conservation the third type of corrections only contributes to the $pp \to Zh$ but not the $pp \to Wh$ process. We add that the corrections called $V_{I,II}$ and $R_{I,II}$ that are related to top-quark loops and involve one external quark line~\cite{Brein:2011vx} are neglected in our SM calculation. Since in total the numerical effect of these contributions amounts to only around~$1\%$~\cite{Brein:2011vx,Zanoli:2021iyp,Haisch:2022nwz}, ignoring the $V_{I,II}$ and $R_{I,II}$ terms seems justified at present.

The corresponding calculation including the impact of SMEFT operators (which will be discussed in the following subsection) can also be performed using spinor-helicity techniques. That calculation requires new helicity amplitudes which can (in part) be obtained from knowledge of the~SM amplitudes. For clarity of explanation, it will be useful to first consider an explicit example in the~SM. To do that we consider the case of~\texttt{B1g0Z} which involves a single external quark line and one external gluon at tree level. A corresponding SM Feynman diagram is displayed on the left-hand side in~Figure~\ref{fig:real}. Note that we consider the leptons (quarks) to be outgoing (incoming). The~corresponding spinor-helicity amplitude with left-handed fermion chiralities and a physical gluon with a negative helicity reads
\beq \label{eq:B1g0ZHelLLminus} 
{\cal A}_{\texttt{B1g0Z}} \left( 1_q^-, 2_g^-, 3^+_{\bar q} ; 4_\ell^-, 5_{\bar \ell}^+ \right) = \frac{\braket{3 4}}{\braket{1 2} \braket{2 3}} \Big ( \braket{1 3} \left [ 5 1 \right ] + \braket{2 3} \left [ 5 2 \right ] \Big )\,,
\eeq
where $\braket{i j}$ and $[ i j ]$ denote the usual spinor products --- see for example~\cite{Dixon:1996wi} for a review of the spinor-helicity formalism. Notice that the semicolon in the expression on the left-hand side of~(\ref{eq:B1g0ZHelLLminus}) separates the particles with incoming and outgoing convention, respectively. The amplitudes for the remaining helicity combinations can be obtained via the following parity and charge conjugation relations
\beq \label{eq:B1g0ZHel} 
\begin{split}
{\cal A}_{\texttt{B1g0Z}}\left( 1_q^-, 2_g^+, 3^+_{\bar q} ; 4_\ell^-, 5_{\bar \ell}^+ \right) & = -{\cal A}_{\texttt{B1g0Z}}\left( 3_q^-, 2_g^-, 1^+_{\bar q} ; 5_\ell^-, 4_{\bar \ell}^+ \right)^* \,, \\[2mm]
{\cal A}_{\texttt{B1g0Z}}\left( 1_q^-, 2_g^{h_g}, 3^+_{\bar q} ; 4_\ell^+, 5_{\bar \ell}^- \right) & = {\cal A}_{\texttt{B1g0Z}}\left( 1_q^-, 2_g^{h_g}, 3^+_{\bar q} ; 5_\ell^-, 4_{\bar \ell}^+ \right) \,, \\[2mm]
{\cal A}_{\texttt{B1g0Z}}\left( 1_q^+, 2_g^{h_g}, 3^-_{\bar q} ; 4_\ell^-, 5_{\bar \ell}^+ \right) & = -{\cal A}_{\texttt{B1g0Z}}\left( 3_q^-, 2_g^{h_g}, 1^+_{\bar q} ; 4_\ell^-, 5_{\bar \ell}^+ \right) \,, \\[2mm]
{\cal A}_{\texttt{B1g0Z}}\left( 1_q^+, 2_g^{h_g}, 3^-_{\bar q} ; 4_\ell^+, 5_{\bar \ell}^- \right) & = -{\cal A}_{\texttt{B1g0Z}}\left( 3_q^-, 2_g^{h_g}, 1^+_{\bar q} ; 5_\ell^-, 4_{\bar \ell}^+ \right) \,.
\end{split}
\eeq
The resulting spin-averaged matrix element \texttt{B1g0Z} then takes the form 
\bea \label{eq:B1g0Z} 
\texttt{B1g0Z} = \frac{8 \pi \hspace{0.25mm} \alpha_s \hspace{0.25mm} C_F}{C_A} \sum_{h_q, h_g, h_\ell = \pm} \left| \frac{g_{Zq}^{h_q} \hspace{0.5mm} g_{Z\ell}^{h_\ell} \hspace{0.5mm} g_{hZZ}}{D_Z (s_{123}) \hspace{0.5mm} D_Z (s_{45})} \, {\cal A}_{\texttt{B1g0Z}}\left( 1_q^{h_q}, 2_g^{h_g}, 3^{-h_q}_{\bar q} ; 4_\ell^{h_\ell}, 5_{\bar \ell}^{-h_\ell} \right) \right|^2 , \hspace{6mm}
\eea
where 
\beq \label{eq:Mandelstams}
s_{ij} = \left( p_i + p_j \right)^2 \,, \qquad s_{ijk} = s_{ij} + s_{jk} + s_{ki} \,, 
\eeq
are the usual Mandelstam invariants with $p_i$ the four-momentum of particle $i$. We~have furthermore introduced
\beq \label{eq:DZ}
D_Z(s) = s - m_Z^2 + i \hspace{0.25mm} m_Z \hspace{0.125mm} \Gamma_Z \,, 
\eeq
with $\Gamma_Z$ denoting the total decay width of the $Z$ boson. In~(\ref{eq:B1g0Z}) the variable $\alpha_s$ denotes the strong coupling constant while $C_F = 4/3$ and $C_A = 3$ are the relevant colour factors. The symbols $g_{Zf}^{h_f}$ and $g_{hZZ}$ represent the $Z f \bar f$ and $hZZ$ coupling strengths, respectively. The explicit expressions for these quantities are given in~Appendix~\ref{app:parameters}. 

To compute the SMEFT contributions that involve modified couplings between the Higgs and two vector bosons, it is important to notice that by using the spinor identity 
\beq \label{eq:spinor_helicity_1}
\Spaa[]{i}{j} \Spbb[]{k}{l} = \frac{1}{2}\Spab{j}{\gamma^\mu}{k} \Spab{i}{\gamma_\mu}{l} \,,
\eeq
the result~(\ref{eq:B1g0ZHelLLminus}) can be rewritten as 
\beq \label{eq:B1g0ZHelLLminusuncontract}
{\cal A}_{\texttt{B1g0Z}}\left( 1_q^-, 2_g^-, 3^+_{\bar q} ; 4_\ell^-, 5_{\bar \ell}^+ \right) = \Spab{4}{\gamma_\mu}{5} \, {\cal A}_{qgq}^\mu (1_q^-, 2_g^-, 3_{\bar q}^+) \,.
\eeq
Here the spinor-helicity amplitude corresponding to the $q \bar q g$ subprocess with the indicated~helicities is given by 
\beq \label{eq:qgq} 
{\cal A}_{qgq}^\mu (1_q^-, 2_g^-, 3_{\bar q}^+) = \frac{\Spaa[]{1}{3} \Spab{3}{\gamma^\mu}{1} + \Spaa{2}{3} \Spab{3}{\gamma^ \mu}{2}}{2 \Spaa[]{1}{2} \Spaa[]{2}{3}} \,.
\eeq

\subsection{SMEFT calculation}
\label{sec:SMEFTfixedorder}

The technically most involved part of the SMEFT calculation results from insertions of the three operators introduced in~(\ref{eq:operators1}) since $Q_ {H\hspace{-0.3mm}B}$, $Q_{HW}$ and $Q_{HW\!B}$ generate modified~$hVV$ vertices with helicity structures different from the one present in the SM,~i.e.~the spinor chain~$\Spab{4}{\gamma_\mu}{5}$ in~(\ref{eq:B1g0ZHelLLminusuncontract}). These modifications can be included at the level of~(\ref{eq:B1g0ZHelLLminus}) by means of generalised currents that describe the splitting of the initial vector boson $V_1$ into the outgoing vector boson $V_2$ and the Higgs boson $h$~\cite{Mimasu:2015nqa}. If the initial-state quarks and final-state leptons are left-handed the relevant generalised neutral currents are given by 
\bea \label{eq:new_structures}
\begin{split}
{\cal A}_{hZZ}^\mu (p_{123}, 4^-_{\ell}, 5^+_{\bar \ell}) & = \frac{g_{Zq}^- \hspace{0.5mm} g_{Z \ell}^-}{D_Z(s_{123}) \hspace{0.5mm} D_Z(s_{45})} \, \Bigg \{ \Spab{4}{\gamma^\mu}{5} \left(g_{hZZ} + \delta \ghzz{2} \left(s_{123} + s_{45} \right) + \delta \ghzz{3} \right) \\[2mm]
& \hspace{-1cm} - \delta \ghzz{2} \, p_{123}^\mu \Spab{4}{\slashed{p}_{123}}{5} -\frac{\delta \ghzz{1}}{2} \left(\Spaa[\gamma^\mu \slashed{p}_{123}]{4}{4} \Spbb{4}{5} + \Spaa{4}{5} \Spbb[\slashed{p}_{123} \gamma^\mu]{5}{5} \right) \Bigg \}\,, \\[4mm]
{\cal A}_{h\gamma Z}^\mu (p_{123}, 4^-_{\ell}, 5^+_{\bar \ell}) & = \frac{g_{\gamma q}^- \hspace{0.5mm} g_{Z \ell}^-}{s_{123} \hspace{0.5mm} D(s_{45})} \, \Bigg \{ -\frac{\delta \ghaz{1}}{2} \bigg( \Spab{4}{\gamma^\mu}{5} \left( \Spab{4}{\slashed{p}_{123}}{4} + \Spab{5}{\slashed{p}_{123}}{5} \right) \\[2mm]
& \hspace{-1cm} - 2 \left( p_4^\mu + p_5^\mu \right) \Spab{4}{\slashed{p}_{123}}{5} \bigg) + \delta \ghaz{2} \, \left( \Spab{4}{\gamma^\mu}{5} \, s_{123} - p_{123}^\mu \, \Spab{4}{\slashed{p}_{123}}{5} \right) \Bigg \}\,, 
\end{split}
\eea
where the structures ${\cal A}_{hZZ}^\mu$ and ${\cal A}_{h\gamma Z}^\mu$ encode the modified $hZZ$ and $h\gamma Z$ vertices, respectively, and $p_{123}$ denotes the four-momentum of the incoming vector boson. The symbols $g_{\gamma q}^{h_q}$ are the $\gamma q \bar{q}$ coupling strengths while $\delta \ghzz{1}$, $\delta \ghzz{2}$, $\delta \ghzz{3}$, $\delta \ghaz{1}$ and $\delta \ghaz{2}$ are anomalous couplings that describe the interactions between the Higgs boson and the relevant vector bosons as indicated by the subscript. The explicit expressions for all the couplings appearing in~(\ref{eq:new_structures}) can be found in~Appendix~\ref{app:parameters}. We stress that although the anomalous couplings $\delta \ghzz{2}$ and $\delta \ghaz{2}$ do not receive corrections from the Wilson coefficients $C_ {H\hspace{-0.3mm}B}$, $C_{HW}$ and $C_{HW\!B}$ our {\tt POWHEG-BOX} implementation contains the full generalised neutral currents~(\ref{eq:new_structures}). The presented MC code can therefore be used to extend the Higgsstrahlungs computations in the anomalous-coupling framework~\cite{Maltoni:2013sma,Greljo:2017spw,Bizon:2021rww} to the \NNLOplusPS~level. 

\begin{figure}[t!]
\begin{center}
\includegraphics[width=0.765\textwidth]{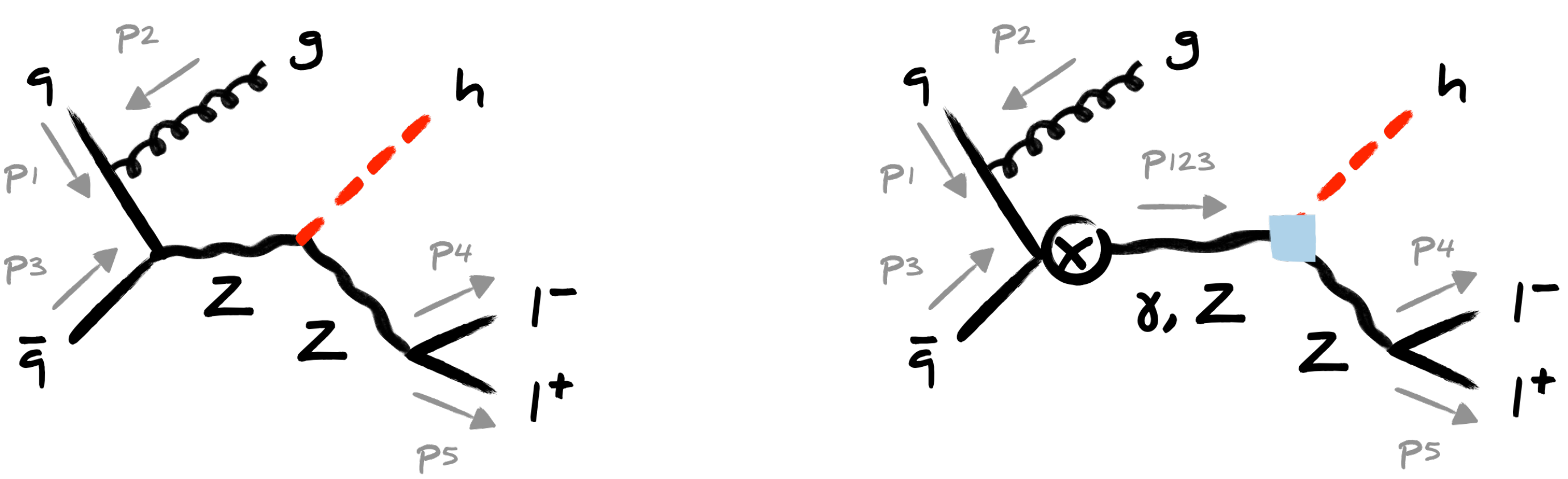}
\end{center}
\vspace{-2mm} 
\caption{\label{fig:real} Example graphs that contribute to the \texttt{B1g0Z} matrix elements. The~diagram on the left shows a SM contribution. On the right we instead depict a SMEFT correction that receives contributions from the generalised $hZZ$ and $h\gamma Z$ currents introduced in~(\ref{eq:new_structures}). The four-momentum flow is indicated by the grey arrows and labels. See the main text for additional explanations.}
\end{figure}

By looking at~(\ref{eq:B1g0ZHelLLminusuncontract}) and~(\ref{eq:new_structures}) it is now readily seen that in order to obtain the spin-averaged squared matrix element~\texttt{B1g0Z} that contains the contributions from the SM as well as the Wilson coefficients $C_ {H\hspace{-0.3mm}B}$, $C_{HW}$ and $C_{HW\!B}$ one just has to replace the coupling and propagator dressed helicity amplitude appearing in the modulus of~(\ref{eq:B1g0Z}) by the following spinor contraction 
\beq \label{eq:contraction} 
{\cal A}_{qgq, \mu} \left(1_q^{h_q}, 2_g^{h_g}, 3_{\bar q}^{-h_q} \right ) \left [ {\cal A}_{hZZ}^\mu (p_{123}, 4_\ell^{h_\ell}, 5_{\bar \ell}^{-h_\ell} ) + {\cal A}_{h\gamma Z}^\mu (p_{123}, 4_\ell^{h_\ell}, 5_{\bar \ell}^{-h_\ell} ) \right ] \,.
\eeq
A schematic depiction of~(\ref{eq:contraction}) is given on the right in~Figure~\ref{fig:real}. Notice that all helicity configurations of ${\cal A}_{qgq}^\mu$ can be obtained from~(\ref{eq:B1g0ZHelLLminusuncontract}) and~(\ref{eq:qgq}) using the relations~(\ref{eq:B1g0ZHel}) while in the case of ${\cal A}_{hZZ}^\mu$ and ${\cal A}_{h\gamma Z}^\mu$ one just has to perform the replacements $g_{Vf}^- \to g_{Vf}^{h_f}$ for $f = q, \ell$ and $V = Z,\gamma$.

Insertions of the operators~(\ref{eq:operators2}) and~(\ref{eq:operators3}) lead to the Feynman diagrams shown on the right-hand side in~Figure~\ref{fig:tree} at tree level. In order to capture this contribution in the case of the squared matrix element $\texttt{B1g0Z}$, one simply has to add the following term 
\beq \label{eq:23contribution} 
\left ( \frac{\delta g_{hZq}^{(1) \hspace{0.25mm} h_q} \hspace{0.5mm} g_{Z\ell}^{h_\ell}}{D_Z(s_{45})} +\frac{g_{Zq}^{h_q} \hspace{0.75mm} \delta g_{hZ \ell}^{(1) \hspace{0.25mm} h_\ell}}{D_Z(s_{123})} \right ) \, {\cal A}_{\texttt{B1g0Z}}\left( 1_q^{h_q}, 2_g^{h_g}, 3^{-h_q}_{\bar q} ; 4_\ell^{h_\ell}, 5_{\bar \ell}^{-h_\ell} \right) \,,
\eeq
to the corresponding dressed SM amplitude appearing within the modulus of~(\ref{eq:B1g0Z}). The analytic expressions for the couplings $\delta g_{hZf}^{(1) \hspace{0.25mm} h_f}$ are given in~Appendix~\ref{app:parameters}. In~(\ref{eq:23contribution}) the first term in the brackets describes the contribution from $Q_{Hq}^{(1)}$, $Q_{Hq}^{(3)}$, $Q_{Hd}$ and $Q_{Hu}$, while the second term is induced by $Q_{H\ell}^{(1)}$, $Q_{H\ell}^{(3)}$ and $Q_{He}$. Notice that compared to the corresponding SM contribution in~(\ref{eq:B1g0Z}) the SMEFT correction proportional to $\delta g_{hZq}^{(1) \hspace{0.25mm} h_q}$ in~(\ref{eq:23contribution}) is missing the $Z$-boson propagator depending on $s_{123}$. This feature explains the high-energy growth~\cite{Bishara:2020vix,Bishara:2020pfx,Bishara:2022vsc} of the SMEFT $pp \to Vh$ amplitudes involving the Wilson coefficients $C_{Hq}^{(1)}$, $C_{Hq}^{(3)}$, $C_{Hd}$ and $C_{Hu}$. 

The last type of SMEFT corrections to the squared matrix element $\texttt{B1g0Z}$ is associated to the tree-level shifts of the SM parameters and couplings. EW input scheme corrections from $C_{H\hspace{-0.25mm}D}$, $C_{HW\!B}$, $C_{H\ell}^{(3)}$ and $C_{\ell\ell}$ lead to the shifts $\delta g_1$, $\delta g_2$ and $\delta v$ of the $U(1)_Y$ and $SU(2)_L$ gauge coupling and the Higgs vacuum expectation value (VEV), respectively, that in turn induce the shifts $\delta g_{hZZ}^{(0)}$ and $\delta g_{Zf}^{(0) \hspace{0.25mm} h_f}$ in the respective couplings of the $Z$ boson. The expressions for these shifts are listed in~Appendix~\ref{app:parameters}. In~practice, the input scheme corrections can be accounted for by applying the replacements $g_{hZZ} \to g_{hZZ} + \delta g_{hZZ}^{(0)}$ and $g_{Zf}^{h_f} \to g_{Zf}^{h_f} + \delta g_{Zf}^{(0) \hspace{0.25mm} h_f}$ to~(\ref{eq:B1g0Z}). Similarly, the Wilson coefficients $C_{Hq}^{(1)}$, $C_{Hq}^{(3)}$, $C_{Hd}$, $C_{Hu}$, $C_{H\ell}^{(1)}$, $C_{H\ell}^{(3)}$ and $C_{He}$ lead to direct shifts in the $Zf \bar f$ couplings that we include through the shifts $g^{h_f}_{Z f} \to g^{h_f}_{Z f} + \delta g^{(1) \hspace{0.25mm} h_f}_{Z f}$ in~(\ref{eq:B1g0Z}). The~expressions for the latter shifts are again given in~Appendix~\ref{app:parameters}. 

While we have used the spinor-helicity amplitudes ${\cal A}_{\texttt{B1g0Z}}$ in this section as examples to illustrate the general approach that we have employed in our SMEFT calculation of $pp \to V h$ at NNLO in QCD, it is important to realise that the computation of all other spinor-helicity amplitudes and squared matrix elements proceeds in an analogous manner. In the case of the SMEFT corrections arising from the choice of EW input scheme as well as those associated to insertions of the operators~(\ref{eq:operators2}) and~(\ref{eq:operators3}) this is clear in view of the factorisation properties of these contributions~$\big($cf.~(\ref{eq:23contribution})$\big)$. Likewise, since the spinor-helicity structure of the partonic part of a given SMEFT amplitude remains the same as in the SM, it can simply be extracted from the SM expressions and contracted with the part of the SMEFT amplitude that does change. It follows that an explicit calculation of the full partonic structures for the higher-order corrections to $pp \to Vh$ in the SMEFT can always be avoided, because the relevant amplitudes can be obtained from those in the SM by applying relations \`a la~(\ref{eq:B1g0ZHelLLminusuncontract}) and~(\ref{eq:contraction}) which involves only spinor algebra. Still the calculation of all spinor contractions needed to achieve NNLO accuracy for the $pp \to Vh$ processes in the SMEFT is a non-trivial task and the final expressions for the spinor-helicity amplitudes turn out to be too lengthy to be reported here. All algebraic manipulations of spinor products needed in the context of this work have been performed with the aid of the {\tt Mathematica} package~{\tt S@M}~\cite{Maitre:2007jq}.

\subsection{Squared matrix element library}
\label{sec:MEs}

We provide all squared matrix elements discussed in the previous sections in a self-contained \texttt{Fortran} library~\cite{GitLabAmplitudes}. Our library includes the spinor-helicity amplitudes for the dimension-four SM and dimension-six SMEFT contributions as well as the definitions for the couplings and the propagators depending on the EW input scheme, which are combined and evaluated numerically in the squared matrix elements up to the desired SMEFT power counting order. Since several of these parts may be relevant in a broader context, we briefly outline the structure of the library. The file \texttt{Bridge} contains the general routines that are required to evaluate the squared matrix elements for a phase-space point, which is represented internally by an object of type \texttt{Event\_t}. These routines allow to set up the numerical expressions for the spinor-helicity brackets, pass the input parameters to the \texttt{Event\_t} object and calculate the dependent parameters for a chosen EW input scheme. The file \texttt{squaredamps} contains the squared matrix elements. The squared matrix element \texttt{B1g0Z} discussed above, for example, has the form \texttt{B1g0Z(i1,i2,i3,i4,i5,K,f1,f2)}, where the integers $\texttt{i1},...,\texttt{i5} \in \{1,...,5 \}$ allow to specify the crossing of the external legs, \texttt{K} is the \texttt{Event\_t} object of the event, and \texttt{f1} and \texttt{f2} indicate the flavours of the quark and lepton lines present in the relevant topology, respectively. Our implementation employs the Monte Carlo Particle Numbering Scheme conventions of the~PDG~\cite{ParticleDataGroup:2022pth}. The dimension-four, -six and -eight contributions to the squared matrix elements are calculated individually and their inclusion can be controlled via the flags \texttt{SM}, \texttt{Linear} and \texttt{Quadratic} of the \texttt{Event\_t} object, respectively. Notice that the dimension-six or linear (dimension-eight or quadratic) SMEFT contributions arise from the interference of the SMEFT and SM amplitudes (self-interference of the SMEFT amplitudes). The spinor-helicity amplitudes, the loop coefficients and the functions implementing the parity and charge conjugation relations are collected in the \texttt{amplib} file. We point out that the required spinor-helicity amplitudes for $pp \to Z \to \ell^+ \ell^-$ and $pp \to W \to \ell \nu$ production at NNLO QCD in the SMEFT are just special cases of the amplitudes in our library. They can be obtained by setting up the \texttt{Event\_t} object in the exact same way (for example~giving it the four physical momenta of the external fermions in the case of \texttt{B0g0Z}), only that in this case the momenta will be linearly dependent because of momentum conservation. 

Two further comments seem to be in order. First, besides including the squared matrix elements described above, we also provide the corresponding colour- and spin-correlated squared matrix elements that are required to build the infrared (IR) subtraction terms in the~\NNLOplusPS~implementation of $pp \to Vh$ production. In the case of \texttt{B1g0V} for instance the colour- and spin-correlated squared matrix elements are called \texttt{B1g0V\_colour} and \texttt{B1g0V\_spin}, respectively. The definition of these squared matrix elements follows the \texttt{POWHEG} conventions specified in~(2.6) and (2.8) of the publication~\cite{Alioli:2010xd}. While the elements of \texttt{B1g0V\_colour} are simply equal to \texttt{B1g0V} times colour factors, calculating \texttt{B1g0V\_spin} requires a bit more care. In our notation, it takes the form 
\bea \label{eq:B1g0Zspin}
\begin{split}
\texttt{B1g0Z\_spin}^{\mu \nu} & = \frac{8 \pi \hspace{0.25mm} \alpha_s \hspace{0.25mm} C_F}{C_A} \sum_{h_q, h_\ell = \pm} \left| \frac{g_{Zq}^{h_q} \hspace{0.5mm} g_{Z\ell}^{h_\ell} \hspace{0.5mm} g_{hZZ}}{D_Z (s_{123}) \hspace{0.5mm} D_Z (s_{45})} \right |^2 ~ \sum_{h_{g_1}, h_{g_2} = \pm} \epsilon^{\mu \, \ast}_{h_{g_1}} \epsilon^\nu_{h_{g_2}} \, \\[2mm]
& \phantom{xx} \times {\cal A}_{\texttt{B1g0Z}}\left( 1_q^{h_q}, 2_g^{h_{g_1}}, 3^{-h_q}_{\bar q} ; 4_\ell^{h_\ell}, 5_{\bar \ell}^{-h_\ell} \right) {\cal A}^\dagger_{\texttt{B1g0Z}}\left( 1_q^{h_q}, 2_g^{h_{g_2}}, 3^{-h_q}_{\bar q} ; 4_\ell^{h_\ell}, 5_{\bar \ell}^{-h_\ell} \right)\,,
\end{split}
\eea
where the $\epsilon_\pm^\mu$ are polarisation vectors normalised as 
\beq \label{eq:polnorm}
\sum_{\mu, \nu} g_{\mu \nu} \, \epsilon_{h_{g_1}}^{\mu \, \ast} \epsilon_{h_{g_2}}^\nu = - \delta_{h_{g_1} h_{g_2}} \,,
\eeq
with $g_{\mu \nu} = \text{diag} \left (1, -1, -1, -1 \right)$. The polarisation vectors $\epsilon_\pm^\mu$ are implemented in \texttt{POWHEG} as
\beq \label{eq:polPWG}
\epsilon^\mu_{[\texttt{PWG}]\pm} = \mp \frac{1}{\sqrt{2}} \left( \epsilon^\mu_1 \pm i \epsilon^\mu_2\right) \,,
\eeq
where the form of the four-vectors $\epsilon_1^\mu$ and $\epsilon_2^\mu$ can be found in (A.12) of \cite{Hagiwara:1988pp}. In order to obtain~$\mathcal{A}_\texttt{B1g0Z}$ as given in~(\ref{eq:B1g0ZHelLLminus}), we have however employed 
\beq \label{eq:polours}
\epsilon^\mu_{[\texttt{GHS}] +} = \frac{\left \langle 3 | \gamma^\mu | 2 \right ]}{\sqrt{2} \left \langle 3 2 \right \rangle}\,, \qquad \epsilon^\mu_{[\texttt{GHS}] -} = \frac{\left \langle 2 | \gamma^\mu | 1 \right ]}{\sqrt{2} \left [ 2 1 \right]} \,,
\eeq
and therefore we have to take into account a complex phase 
\beq
e^{ i \varphi} \equiv \sum_{\mu, \nu} g_{\mu \nu} \, \epsilon^{\mu \ast}_{[\texttt{PWG}] +} \epsilon_{[\texttt{GHS}] -}^\nu \,.
\eeq
See the discussion around (3.22) of \cite{Hasegawa:2009tx}. We compute the phase $\varphi$ numerically and by taking it into account in our definition of the polarisation vectors allows us to compare our results for the squared matrix elements with previous implementations that are based on~(\ref{eq:polPWG}). Our implementation is further validated by the fact that the IR poles are properly cancelled in the \texttt{POWHEG} procedure. 
 
Second, for the squared matrix element \texttt{B1g1Z}, we follow the discussion presented in the article~\cite{Gehrmann:2011ab}. There, the spinor-helicity amplitude is split into three primitive amplitudes~$\mathcal{A}_\Omega$~with $\Omega = \alpha, \beta, \gamma$ and their corresponding loop coefficients $\Omega$:
\beq \label{eq:TGdecomp}
\mathcal{A}_\texttt{B1g1Z} = \alpha \mathcal{A}_\alpha + \beta \mathcal{A}_\beta + \gamma \mathcal{A}_\gamma \,.
\eeq
The primitive amplitudes $\mathcal{A}_\Omega$ are given in (2.22) of the work~\cite{Gehrmann:2011ab} for the case of $q\bar{q} \to V \gamma$ production, where momentum conservation between the initial-state quarks and the final-state vector bosons is assumed. Despite sharing the same chirality structure as the $q\bar{q} \to V \gamma$ process, the $q\bar{q} \to Vhj$ process additionally involves an Higgs boson in the final state (which carries away some momentum). 
Accordingly, the expressions for $\mathcal{A}_\Omega$ from~\cite{Gehrmann:2011ab} have to be modified to suit our purposes. Going back to the most general partonic current in~(2.13) of~\cite{Gehrmann:2011ab}, we have derived the primitive amplitudes $\mathcal{A}_\Omega$ using~(2.17)~to~(2.20) without imposing momentum conservation. The resulting amplitudes in our conventions agree with~(A.45)~to~(A.47) of~\cite{Majer:2020kdg} and contain additional terms proportional to the structure
\beq \label{eq:Anc}
\mathcal{A}_\text{nc} = \langle 13 \rangle [21] \frac{\langle14 \rangle [51] + \langle 24 \rangle [52] + \langle 34 \rangle [53]}{2s_{123} \langle12\rangle} \,. 
\eeq
Using (\ref{eq:spinor_helicity_1}) it is clear that $\mathcal{A}_\text{nc}$ vanishes for $p_1 + p_2 + p_3 = p_4 + p_5$, i.e.~when the momentum of the Higgs boson vanishes. The loop coefficients $\Omega$ can be further divided into an IR divergent and an IR finite part as follows 
\beq \label{eq:Omegadecomp}
\Omega = I^{(1)}(\epsilon) \,\Omega^{(0)} + \Omega^{(1),\, \text{finite}} \,,
\eeq
with $I^{(1)}(\epsilon)$ the usual IR singularity operator as given for example in~(C.9) of~\cite{Gehrmann:2011ab}. We~include the $\mathcal{O}(1/\epsilon^2)$, $\mathcal{O}(1/\epsilon)$ and $\mathcal{O}(1)$ pieces of $I^{(1)}(\epsilon) \,\Omega^{(0)} $ in the array entries $1$, $2$ and $3$ of the loop coefficients $\Omega$, respectively. Note that $\alpha^{(0)} = \beta^{(0)} = 1$ and $\gamma^{(0)} = 0$. The finite parts $\Omega^{(1), \, \text{finite}}$ are instead decomposed as
\beq \label{eq:Omega1decomp}
\Omega^{(1), \, \text{finite}} = C_A \, \Omega^{(1), \, \text{finite}}_1 + \frac{1}{C_A} \, \Omega^{(1), \, \text{finite}}_2 + \beta_0 \, \Omega^{(1), \, \text{finite}}_3 \,,
\eeq
and include the leading colour, the subleading colour and the $\beta_0 = (11 \hspace{0.25mm} C_A - 4 \, T_F \hspace{0.125mm} N_f)/6$ pieces individually. Here $T_F = 1/2$ and $N_f = 4$ denotes the number of active quark flavours. Analytic expressions for the coefficients $\Omega^{(1), \, \text{finite}}_i$ with $i = 1,2,3$ were provided as {\tt FORM} output in the arXiv submission of \cite{Gehrmann:2011ab} for the three kinematical regions (i.e.~$s_{13} > 0$, $s_{12} > 0$ and $s_{23} > 0$) relevant at hadron colliders. They are expressed as one- and two-dimensional harmonic polylogarithms (HPLs) and we translate the appearing HPLs to logarithms and dilogarithms using the relevant formulae in~\cite{Gehrmann:2000zt, Gehrmann:2001jv}, computing the latter numerically with the help of~\texttt{LoopTools}~\cite{Hahn:1998yk}. The finite contributions $\Omega^{(1), \, \text{finite}}$ are added to the array entry~3 of the loop coefficients $\Omega$ in our code.

\subsection{\NNLOplusPSbm~implementation}
\label{sec:NNLOPS}

In the following, we briefly describe how we consistently match the fixed-order NNLO SMEFT calculations of $pp \to Zh \to \ell^+ \ell^- h$ and $pp \to Wh \to \ell \nu h$ production to a~PS. We~first recall that within the SM computations of the $pp \to Vh$ processes have recently reached \NNLOplusPS~accuracy~\cite{Astill:2018ivh,Alioli:2019qzz,Bizon:2019tfo,Zanoli:2021iyp}. In fact, here we follow the approach presented in~\cite{Zanoli:2021iyp} which employs the \MiNNLOPS~method to match fixed-order QCD and PS~effects. The~interested reader is referred to the latter work and the articles~\cite{Monni:2019whf,Monni:2020nks} for additional technical details not covered below. 

In the case of Higgsstrahlung from \qqF, the starting point of our \NNLOplusPS~implementation is the computation of the $q \bar q \to Vh$ channel in association with one light QCD parton at NLO according to the {\tt POWHEG} method~\cite{Nason:2004rx,Frixione:2007vw}, inclusive in the radiation of a second light QCD parton. The computation of the relevant matrix elements has been outlined in~Section~\ref{sec:SMfixedorder} and relies on the SM spinor-helicity amplitudes calculated in~\cite{Kramer:1986sg,Hamberg:1990np,Gehrmann:2011ab,Majer:2020kdg}. In a second step, an appropriate Sudakov form factor and higher-order corrections are applied such that the calculation remains finite in the unresolved limit of the light partons and NNLO accurate for inclusive $q \bar q \to Vh$ production. In the third step, the kinematics of the second radiated parton (accounted for inclusively in the first step) is generated following the {\tt POWHEG} method to preserve the NLO accuracy of the~$Vh$ plus jet cross section, including subsequent radiation through \PYTHIA{8.2}~\cite{Sjostrand:2014zea}. We~stress that since all emissions are ordered in transverse momentum~($p_T$) and the used Sudakov form factor matches the leading logarithms~generated by \PYTHIA{8.2}, the \MiNNLOPS~approach maintains the leading-logarithmic accuracy of the PS. The \ggF~one-loop contributions to Higgsstrahlung, i.e.~the $gg \to Zh$ process, can instead be computed independently at \LOplusPS~and simply added to the \qqF~results. Following the general discussion in~Section~\ref{sec:SMEFTfixedorder} the relevant SMEFT spinor-helicity amplitudes can be obtained from those in the SM. We take the SM amplitudes from the {\tt MCFM} implementation~\cite{Boughezal:2016wmq} of the results obtained in~\cite{Campbell:2016jau}. Further details on our implementation of the $gg \to Zh$ process are given in~Appendix~\ref{app:gginitiated}.

Our~\MiNNLOPS~generator~\cite{GitLabPowheg} for $q \bar q \to Vh$ production has been implemented into the {\tt POWHEG-BOX}. While it uses the infrastructure of the \NNLOplusPS~SM Higgsstrahlungs generator~\cite{Zanoli:2021iyp} the parts of the code that calculate the squared matrix elements are entirely new and independent. To validate our implementation of the SM computation we have performed extensive numerical checks against~\cite{Zanoli:2021iyp}. The~individual spinor-helicity amplitudes were furthermore compared to a private implementation of the results presented in~\cite{Majer:2020kdg} (which entered the calculation~\cite{Gauld:2019yng}), and results for the squared matrix elements were numerically validated with {\tt OpenLoops~2}~\cite{Buccioni:2019sur}. We exploit the Frixione-Kunszt-Signer~subtraction~\cite{Frixione:1995ms,Frixione:1997np} to deal with soft and collinear singularities of the real contributions and to cancel the IR~poles of the virtual corrections. In fact, the full {\tt POWHEG-BOX} machinery is used that automatically builds the soft and collinear counterterms and remnants, and also verifies the behaviour in the soft and collinear limits of the real squared matrix elements against their soft and collinear approximations. These checks provide non-trivial tests of our SMEFT calculation of Higgsstrahlung in the \qqF~channel. In~the case of the $g g \to Zh$ channel we have instead written a simple \LOplusPS~generator~\cite{GitLabPowheg} using the {\tt POWHEG-BOX} framework. Our implementation of the corresponding SM spinor-helicity amplitudes has been validated numerically against {\tt OpenLoops~2} at the level of the squared matrix elements. 

Besides the direct SMEFT contributions described in~Section~\ref{sec:SMEFTfixedorder} the $pp \to Vh$ processes also receive corrections from the propagators of the~$Z$ and $W$ boson, because SMEFT operators generically modify the masses and the total decay widths of all unstable particles. Our {\tt POWHEG-BOX} implementation contains the complete tree-level shifts of the relevant masses and total decay widths that are induced by the Wilson coefficients~of the operators~(\ref{eq:operators1}) to~(\ref{eq:operators4}) for the $\alpha$, the $\alpha_\mu$ as well as the LEP~scheme --- cf.~Appendix~\ref{app:parameters} for a comprehensive discussion of the different EW input schemes in the SMEFT. For instance, in the case of the total decay width of the $Z$ and $W$ boson in the~LEP scheme we find the relevant shifts 
\beq \label{eq:deltawidthZW}
\begin{split}
\delta \Gamma_Z & \simeq -\frac{v^2}{\Lambda^2} \, \Big [ \, 1.99 \hspace{0.25mm} C_{HW\!B} - 1.14 \hspace{0.25mm} C_{Hq}^{(1)} - 3.89 \hspace{0.25mm} C_{Hq}^{(3)} + 0.46 \hspace{0.25mm} C_{Hd}
- 0.62 \hspace{0.25mm} C_{Hu} \\[1mm] 
& \hspace{1.4cm} + 0.46 \hspace{0.25mm} C_{H\ell}^{(1)} + 4.98 \hspace{0.25mm} C_{H\ell}^{(3)} + 0.46 \hspace{0.25mm} C_{He} + 1.63 \hspace{0.25mm} C_{H\hspace{-0.25mm}D} - 3.25 \hspace{0.25mm} C_{\ell\ell} \, \Big ] \, {\rm GeV} \,, \\[2mm]
\delta \Gamma_W & \simeq -\frac{v^2}{\Lambda^2} \, \Big [ \, 4.77 \hspace{0.25mm} C_{HW\!B} - 2.67 \hspace{0.25mm} C_{Hq}^{(3)} + 5.31 \hspace{0.25mm} C_{H\ell}^{(3)} + 2.16 \hspace{0.25mm} C_{H\hspace{-0.25mm}D} - 3.32 \hspace{0.25mm} C_{\ell\ell} \, \Big ] \, {\rm GeV} \,.
\end{split}
\eeq
These results agree with those presented in the literature~$\big($cf.~for instance~\cite{Dawson:2019clf}$\big)$. We stress that we do not perform an expansion in the SMEFT corrections to the propagators of the~$Z$ and $W$ boson. In consequence, the dependence on the Wilson coefficients of our numerical results is generally non-linear. The resulting non-linear effects are however always very~small.

\section{Anatomy of SMEFT effects}
\label{sec:anatomy}

In this section, we motivate simple SMEFT benchmark scenarios by discussing the leading experimental constraints on the Wilson coefficients of the dimension-six operators introduced in~(\ref{eq:operators1}) to~(\ref{eq:operators4}). In this discussion the choice of an EW input scheme is an important ingredient. While in our {\tt POWHEG-BOX} implementation~\cite{GitLabPowheg}  the user can choose between the $\alpha$, $\alpha_\mu$, and LEP schemes (see Appendix~\ref{app:parameters} for a brief discussion of EW input schemes in the SMEFT) the following discussion is based on the LEP scheme which uses as inputs $\{\alpha, G_F, m_Z\}$. Here $\alpha$ is the fine-structure constant, $G_F$ is the Fermi constant as extracted from muon decay and $m_Z$ is the mass of the $Z$ boson in the on-shell scheme. 

In the LEP scheme, the weak mixing angle $\theta_w$, the $U(1)_Y$ and $SU(2)_L$ couplings $g_1$ and $g_2$ and the VEV of the Higgs boson $v$ can all be written in terms of the EW input parameters $\{\alpha, G_F, m_Z\}$. One finds the following relations 
\beq \label{eq:sin2thetaw}
s_w^2 = \frac{1}{2} \left [ 1 - \sqrt{1 - \frac{2 \sqrt{2} \pi \alpha}{G_F \hspace{0.125mm} m_Z^2}} \right ] \simeq 0.23 \,, 
\eeq
and 
\beq \label{eq:g1g2v}
g_1 = \frac{\sqrt{4 \pi \alpha}}{c_w} \simeq 0.36 \,, \qquad 
g_2 = \frac{\sqrt{4 \pi \alpha}}{s_w} \simeq 0.65 \,, \qquad 
v = \frac{1}{\sqrt[4]{2} \hspace{0.25mm} \sqrt{G_F}} \simeq 246.22 \, {\rm GeV} \,,
\eeq
where the given numerical values correspond to $\alpha = 1/127.951$, $G_F = 1.1663788 \cdot 10^{-5} \, {\rm GeV}^{-2}$ and $m_Z = 91.1876 \, {\rm GeV}$~\cite{ParticleDataGroup:2022pth}. Notice that in~(\ref{eq:sin2thetaw}) and~(\ref{eq:g1g2v}) we have used the abbreviations~$s_w$ and~$c_w$ to denote the sine and cosine of the weak mixing angle, respectively. 

\subsection[$W$-boson mass]{$\bm{W}$-boson mass}
\label{sec:wbosonmass}

The on-shell mass of the $W$ boson is a predicted quantity in the LEP scheme as well. In terms of the results~(\ref{eq:sin2thetaw}) and~(\ref{eq:g1g2v}) one has 
\beq \label{eq:MW}
m_W = c_w \hspace{0.25mm} m_Z \simeq 79.83 \, {\rm GeV} \,. 
\eeq
The corresponding relative tree-level shift of $m_W$ induced in the $U(3)^5$ symmetric SMEFT is given by 
\beq \label{eq:dMWoMWtheo} 
\begin{split}
\frac{\delta m_W}{m_W} & = -\frac{c_w \hspace{0.25mm} s_w}{2 \left ( c_w^2 - s_w^2 \right ) } \frac{v^2}{\Lambda^2} \left [ 2 \hspace{0.25mm} C_{HW\!B} + \frac{s_w}{c_w} \left ( 2 \hspace{0.25mm} C_{H\ell}^{(3)} - C_{\ell\ell} \right ) + \frac{c_w}{2 \hspace{0.25mm} s_w} \, C_{H\hspace{-0.25mm}D} \right ] \\[2mm] 
& \simeq - 0.048 \hspace{0.5mm} C_{HW\!B} - 0.027 \hspace{0.25mm} C_{H\ell}^{(3)} - 0.022 \hspace{0.25mm} C_{H\hspace{-0.25mm}D} + 0.013 \hspace{0.25mm} C_{\ell\ell} \,,
\end{split}
\eeq
where we have employed~(\ref{eq:sin2thetaw}),~(\ref{eq:g1g2v}) and assumed $\Lambda = 1 \, {\rm TeV}$ for the energy scale that suppresses the dimension-six operators to obtain the numerical results presented in the second line. Employing the latest world average of the measured $W$-boson mass obtained by the PDG~\cite{ParticleDataGroup:2022pth} and the state-of-the-art SM prediction~\cite{Chen:2020xot}, we obtain the following 95\% confidence level~(CL) limit 
\beq \label{eq:dMWoMWexp}
\frac{\delta m_W}{m_W} \in [ -0.9, 5.6 ] \cdot 10^{-4} \,,
\eeq
on the allowed relative shift of $m_W$. This result in general sets stringent constraints on the Wilson coefficients entering~(\ref{eq:dMWoMWtheo}). For instance, in the case of the operator $Q_{HW\!B}$ we find the bound 
\beq \label{eq:CHWBbound} 
\frac{C_{HW\!B}}{\Lambda^2} \in [-1.2, 0.2] \cdot 10^{-2} \, {\rm TeV}^{-2} \,,
\eeq
if all other Wilson coefficients in~(\ref{eq:dMWoMWtheo}) are set to zero. Similar though weaker limits also apply in the case of $C_{H\hspace{-0.25mm}D}$, $C_{H\ell}^{(3)}$ and $C_{\ell\ell}$ if each of the Wilson coefficients is treated as the only non-zero contribution. 

\subsection[$Z$-boson couplings]{$\bm{Z}$-boson couplings}
\label{sec:Zbosoncouplings}

The linear combinations of Wilson coefficients that modify the couplings of the $W$ and $Z$ boson to fermions at tree level are strongly constrained by the EW precision measurements performed at SLC and LEP. In the case of the $Z$ boson the shifts of the left- and right-handed couplings to a fermion $\psi = \{e, \nu, d, u\}$ take the following form 
\beq \label{eq:dgLdgR}
\begin{split}
\delta g_L^\psi & = \frac{g_2}{c_w} \, \frac{v^2}{\Lambda^2} \left [ g_{T^3_\psi} \hspace{0.25mm} T^3_\psi - g_{Q_\psi} \hspace{0.25mm} Q_\psi - \frac{1}{2} \left ( C_{H\psi_L}^{(1)} - 2 \hspace{0.25mm} T_\psi^3 \hspace{0.25mm} C_{H\psi_L}^{(3)} \right ) \right ] \,, \\[2mm]
\delta g_R^\psi & = \frac{g_2}{c_w} \, \frac{v^2}{\Lambda^2} \left [ -g_{Q_\psi} \hspace{0.25mm} Q_\psi - \frac{1}{2} \left ( 1 - \delta_{\psi\nu} \right ) C_{H\psi_R} \right ] \,,
\end{split}
\eeq
with 
\beq \label{eq:gT3gQ}
g_{T^3_\psi} = -C_{H\ell}^{(3)} - \frac{C_{H\hspace{-0.25mm}D}}{4} + \frac{C_{\ell\ell}}{2}\,, \qquad 
g_{Q_\psi} = \frac{c_w \hspace{0.25mm} s_w}{c_w^2 - s_w^2} \left [ C_{HW\!B} - \frac{s_w}{c_w} \, g_{T^3_\psi} \right ] \,.
\eeq
Here the coupling $g_2$ is defined via~(\ref{eq:g1g2v}), $\psi_L = \{\ell, q\}$ and $\psi_R = \{e, d, u\}$, while $T^3_\psi = \pm 1/2$ denotes the weak isospin eigenvalue, $Q_\psi$ is the electric charge of the fermion $\psi$ in units of~$e = \sqrt{4 \pi \alpha}$. Notice that the expressions given in~(\ref{eq:gT3gQ}) correspond to the LEP~scheme and that we have assumed that the Wilson coefficients of the operators that involve fermionic fields are flavour universal. Below we will assume that such a flavour universality holds for all SM fermions apart from the charm, bottom and top quark. 

In the case of the electron the formulae~(\ref{eq:dgLdgR}) and~(\ref{eq:gT3gQ}) lead to the following left- and right-handed coupling shifts
\beq \label{eq:DeltagLRe} 
\begin{split}
\delta g_L^e & \simeq 0.036 \hspace{0.5mm} C_{HW\!B} - 0.022 \hspace{0.25mm} C_{H\ell}^{(1)} + 0.020 \hspace{0.25mm} C_{H\ell}^{(3)} + 0.011 \hspace{0.25mm} C_{H\hspace{-0.25mm}D} - 0.021 \hspace{0.25mm} C_{\ell\ell} \,, \\[2mm]
\delta g_R^e & \simeq 0.036 \hspace{0.5mm} C_{HW\!B} + 0.020 \hspace{0.25mm} C_{H\ell}^{(3)} - 0.022 \hspace{0.25mm} C_{H e} + 0.005 \hspace{0.25mm} C_{H\hspace{-0.25mm}D} - 0.010 \hspace{0.25mm} C_{\ell\ell} \,, 
\end{split}
\eeq
when~(\ref{eq:sin2thetaw}),~(\ref{eq:g1g2v}) and $\Lambda = 1 \, {\rm TeV}$ are used as numerical inputs. The EW precision measurements performed by SLC and LEP~\cite{ALEPH:2005ab,ParticleDataGroup:2022pth} imply that at 95\%~CL one has 
\beq \label{eq:DeltagLReLEPSLD} 
\delta g_L^e \in [-7.1, 2.0] \cdot 10^{-4} \,, \qquad 
\delta g_R^e \in [-7.0, 1.6] \cdot 10^{-4} \,.
\eeq
Ignoring cancellations these limits again put severe constraints on the Wilson coefficients that appear in~(\ref{eq:DeltagLRe}). Numerically, we obtain for instance the bounds 
\beq \label{eq:CHl3CHebounds}
\frac{C_{H\ell}^{(3)}}{\Lambda^2} \in [-3.6, 1.0] \cdot 10^{-2} \, {\rm TeV}^{-2} \,, \qquad
\frac{C_{He}}{\Lambda^2} \in [-0.7, 3.1] \cdot 10^{-2} \, {\rm TeV}^{-2} \,, 
\eeq
from the constraint on $\delta g_L^e$ and $\delta g_R^e$, respectively, if we allow only the considered Wilson coefficient to take a~non-vanishing value. We add that the Wilson coefficients $C_{H\hspace{-0.25mm}D}$, $C_{H\ell}^{(1)}$ and $C_{\ell\ell}$ can also be constrained by the SLC and LEP measurements of the $Z$-boson coupling to neutrinos. The obtained bounds, however, turn out to be weaker than those that derive from~(\ref{eq:DeltagLRe}) and~(\ref{eq:DeltagLReLEPSLD}). 

Employing again~(\ref{eq:sin2thetaw}),~(\ref{eq:g1g2v}),~(\ref{eq:dgLdgR}) and~(\ref{eq:gT3gQ}) the left- and right-handed coupling shifts in the case of the down and up quark take the following form 
\beq \label{eq:DeltagLRdu} 
\begin{split}
\delta g_L^d & \simeq 0.012 \hspace{0.5mm} C_{HW\!B} - 0.022 \left ( C_{Hq}^{(1)} + C_{Hq}^{(3)} \right ) + 0.029 \hspace{0.25mm} C_{H\ell}^{(3)} + 0.007 \hspace{0.25mm} C_{H\hspace{-0.25mm}D} - 0.015 \hspace{0.25mm} C_{\ell\ell} \,, \\[2mm]
\delta g_R^d & \simeq 0.012 \hspace{0.5mm} C_{HW\!B} - 0.022 \hspace{0.25mm} C_{Hd} + 0.007 \hspace{0.25mm} C_{H\ell}^{(3)} + 0.002 \hspace{0.25mm} C_{H\hspace{-0.25mm}D} - 0.003 \hspace{0.25mm} C_{\ell\ell} \,, \\[2mm]
\delta g_L^u & \simeq -0.024 \hspace{0.5mm} C_{HW\!B} - 0.022 \left ( C_{Hq}^{(1)} - C_{Hq}^{(3)} \right ) - 0.036 \hspace{0.25mm} C_{H\ell}^{(3)} - 0.009 \hspace{0.25mm} C_{H\hspace{-0.25mm}D} + 0.018 \hspace{0.25mm} C_{\ell\ell} \,, \\[2mm]
\delta g_R^u & \simeq -0.024 \hspace{0.5mm} C_{HW\!B} - 0.022 \hspace{0.25mm} C_{Hu} - 0.013 \hspace{0.25mm} C_{H\ell}^{(3)} - 0.003 \hspace{0.25mm} C_{H\hspace{-0.25mm}D} + 0.007 \hspace{0.25mm} C_{\ell\ell} \,, 
\end{split}
\eeq
for a common operator suppression scale of $\Lambda = 1 \, {\rm TeV}$. The measurements by SLC and LEP performed at the $Z$-pole lead to the following limits 
\beq \label{eq:DeltagLRduLEPSLD} 
\begin{split}
& \delta g_L^d \in [-6.2, 2.0] \cdot 10^{-2} \,, \qquad \delta g_R^d \in [-4.8, 3.4] \cdot 10^{-2} \,, \\[2mm]
& \hspace{1mm} \delta g_L^u \in [0.2, 6.8] \cdot 10^{-2} \,, \qquad \delta g_R^u \in [-1.3, 5.3] \cdot 10^{-2} \,,
\end{split}
\eeq
at 95\%~CL. These bounds translate into 
\beq \label{eq:CHq1CHq3CHdCHubounds}
\begin{split}
& \frac{C_{Hq}^{(1)}}{\Lambda^2} \in [-0.9, 2.8] \; {\rm TeV}^{-2} \,, \qquad \frac{C_{Hq}^{(3)}}{\Lambda^2} \in [-0.9, 2.8] \; {\rm TeV}^{-2} \,, \\[2mm]
& \frac{C_{Hd}}{\Lambda^2} \in [-1.5, 2.1] \; {\rm TeV}^{-2} \,, \qquad \frac{C_{Hu}}{\Lambda^2} \in [-2.4, 0.6] \; {\rm TeV}^{-2} \,, 
\end{split}
\eeq
if each Wilson coefficient is treated independently and the limit on $\delta g_L^d$ as given in~(\ref{eq:DeltagLRduLEPSLD}) is used to constrain $C_{Hq}^{(1)}$ and $C_{Hq}^{(3)}$. We emphasise that notably more stringent limits on the Wilson coefficients in~(\ref{eq:CHq1CHq3CHdCHubounds}) would be obtained if one were to perform a complete EW fit including heavy flavour measurements and assume a SMEFT Lagrangian with an approximate $U(3)^5$ symmetry. This is a simple consequence of the fact that in such a case one has $\delta g_{L,R}^b = \delta g_{L,R}^d$ ($\delta g_{L,R}^c = \delta g_{L,R}^u$) and that the bottom (charm) couplings were significantly better determined than the down (up) couplings at SLC and LEP --- see for example~Figure~F.3~in~\cite{ALEPH:2005ab}. The limits~(\ref{eq:CHq1CHq3CHdCHubounds}) are relaxed because they assume flavour universality only for the down and strange quark as far as quarks are concerned. 

\subsection{Unitarity of the quark-mixing matrix}
\label{sec:CAA}

As pointed out in~\cite{Blennow:2022yfm,Cirigliano:2022qdm} the consistency of $\beta$-decay measurements with the unitarity of the Cabibbo-Kobayashi-Maskawa~(CKM) quark-mixing matrix imposes stringent constraints on certain dimension-six SMEFT Wilson coefficients. In fact, deviations of first-row CKM unitarity can be parameterised by $\Delta_{\rm CKM} \equiv |V_{ud}|^2 + |V_{us}|^2 - 1$ with $V_{ij}$ the relevant elements of the quark-mixing matrix, and this quantity can be expressed under the assumption of a~$U(3)^5$ flavour symmetry as 
\beq \label{eq:CKMuni}
\Delta_{\rm CKM} = \frac{2 \hspace{0.25mm} v^2}{\Lambda^2} \left ( C_{Hq}^{(3)} - C_{H\ell}^{(3)} + C_{\ell \ell} - C_{\ell q}^{(3)} \right ) \,. 
\eeq
The first three Wilson coefficients have already been introduced in~(\ref{eq:operators2}) to~(\ref{eq:operators4}), while~$C_{\ell q}^{(3)}$ corresponds to the Wilson coefficient of the four-fermion operator $Q_{\ell q}^{(3)} = (\bar \ell \gamma_\mu \sigma^a \ell) (\bar q\gamma^\mu \sigma^a q)$. A recent combined global fit to measurements of super-allowed nuclear $\beta$ transitions and kaon decays led to~\cite{Cirigliano:2022yyo}
\beq \label{eq:CKMuniexp}
\Delta_{\rm CKM} = \left (-1.48 \pm 0.53 \right ) \cdot 10^{-3} \,, 
\eeq
which deviates from zero at the $3\sigma$ level. The result~(\ref{eq:CKMuniexp}), together with other tensions in the SM interpretation of semileptonic decays and Cabibbo universality, is collectively called the Cabibbo angle anomaly (CAA)~\cite{Grossman:2019bzp}. Ignoring cancellations in~(\ref{eq:CKMuni}), the constraint~(\ref{eq:CKMuniexp}) puts severe bounds on the relevant Wilson coefficients. At $95\%$~CL we find 
\beq \label{eq:boundCKM}
\frac{C_{Hq}^{(3)}}{\Lambda^2} \in [-2.0, -0.4] \; {\rm TeV}^{-2} \,, 
\eeq
and up to a possible overall sign the same limit applies also to $C_{H\ell}^{(3)}$, $C_{\ell \ell}$ and $C_{\ell q}^{(3)}$. We~add that the studies~\cite{Cirigliano:2022yyo,Cirigliano:2023nol} indicate that right-handed charged-current interactions could provide a viable explanation for the CAA. In the SMEFT such effects can be induced at the dimension-six level only by the operator~$Q_{Hud}$ that has been introduced in~(\ref{eq:operators2}). This~operator contributes to $pp \to Wh$ production but since its Wilson coefficient vanishes in the $U(3)^5$ symmetric limit, and it is strongly Yukawa suppressed in minimal-flavour violating scenarios~\cite{DAmbrosio:2002vsn}, we will employ $C_{Hud}=0$ in all of our benchmark scenarios for the Wilson~coefficients. Our squared matrix element library, however, includes the $Q_{Hud}$ contribution to $pp \to Wh$ production up to NNLO in QCD, meaning that it can be used to extend the \NLOplusPS~analyses~\cite{Alioli:2018ljm,Cirigliano:2023nol} to the \NNLOplusPS~level. 

\subsection{Higgs-boson observables}
\label{sec:Higgsphysics}

The Wilson coefficients of the operators~(\ref{eq:operators1}) modify the Higgs signal strengths in final states with two vector bosons. In~the LEP scheme and including only the tree-level SMEFT corrections due to $C_ {H\hspace{-0.3mm}B}$, $C_{HW}$ and $C_{HW\!B}$ the coupling modifiers relevant for $h \to WW$ and~$h \to ZZ$ can be written in the~$\kappa$~framework as follows~\cite{Falkowski:2019hvp}
\beq \label{eq:treekappas}
\begin{split}
\delta \kappa_{WW} & \simeq -\frac{v^2}{\Lambda^2} \, \Big [ 0.76 \hspace{0.25mm} C_{HW} + 3.18 \hspace{0.25mm} C_{HW\!B} \Big ] \,, \\[2mm]
\delta \kappa_{ZZ} & \simeq \frac{v^2}{\Lambda^2} \, \Big [ 2.71 \hspace{0.25mm} C_ {H\hspace{-0.3mm}B} - 3.18 \hspace{0.25mm} C_{HW} + 1.23 \hspace{0.25mm} C_{HW\!B} \Big ] \,.
\end{split}
\eeq
In the tree-level approximation the corresponding expressions for the $h \to \gamma\gamma$ and $h \to \gamma Z$ decays do not depend on the choice of the EW input scheme. One finds 
\beq \label{eq:loopkappas}
\begin{split}
\delta \kappa_{\gamma \gamma} & \simeq \frac{1}{g_{h\gamma\gamma}} \frac{v^2}{\Lambda^2} \, \Big [ c_w^2 \hspace{0.25mm} C_ {H\hspace{-0.3mm}B} + s_w^2 \hspace{0.25mm} C_{HW} - c_w \hspace{0.25mm} s_w \hspace{0.25mm} C_{HW\!B} \Big ] \,, \\[2mm]
\delta \kappa_{\gamma Z} & \simeq - \frac{1}{g_{h\gamma Z}} \frac{v^2}{\Lambda^2} \, \Big [ 2 \hspace{0.25mm} c_w \hspace{0.25mm} s_w \left ( C_ {H\hspace{-0.3mm}B} - C_{HW} \right ) + \left ( c_w^2 - s_w^2 \right ) C_{HW\!B} \Big ] \,,
\end{split}
\eeq
where $g_{h\gamma\gamma} \simeq -2.02 \cdot 10^{-3}$ and $g_{h\gamma Z} \simeq -7.23 \cdot 10^{-3}$ parameterise the loop-induced $h\gamma\gamma$ and $h \gamma Z$ couplings within the SM. Explicit analytic formulae for $g_{h\gamma\gamma}$ and $g_{h\gamma Z}$ can be found for instance in~\cite{Brivio:2020onw}. 

Assuming that the \ggF~Higgs production cross section is SM-like and taking the SM predictions for the total decay width of the Higgs and its branching ratios from~\cite{LHCHiggsWG4}, we obtain for $\Lambda = 1\, {\rm TeV}$ the following linearised expressions for the relevant signal strenghts
\beq \label{eq:mustheo}
\begin{split}
& \mu_{\rm \ggF}^{WW} \simeq 1 + 0.074 \hspace{0.25mm} C_ {H\hspace{-0.3mm}B} - 0.0086 \hspace{0.25mm} C_{HW} \,, \qquad 
\mu_{\rm \ggF}^{ZZ} \simeq 1 + 0.40 \hspace{0.25mm} C_ {H\hspace{-0.3mm}B} - 0.30 \hspace{0.25mm} C_{HW} \,, \\[2mm]
& \hspace{4mm} \mu_{\rm \ggF}^{\gamma \gamma} \simeq 1 - 46.0 \hspace{0.25mm} C_ {H\hspace{-0.3mm}B} - 14.0 \hspace{0.25mm} C_{HW} \,, \qquad 
\mu_{\rm \ggF}^{\gamma Z} \simeq 1 + 14.3 \hspace{0.25mm} C_ {H\hspace{-0.3mm}B} - 14.1 \hspace{0.25mm} C_{HW} \,.
\end{split}
\eeq
Notice that in view of~(\ref{eq:CHWBbound}) we have neglected the contributions of the Wilson coefficient~$C_{HW\!B}$ in~(\ref{eq:mustheo}). The corresponding measured Higgs signal strenghts are
\beq \label{eq:musexp}
\mu_{\rm \ggF}^{WW} = 1.18 \pm 0.13 \,, \quad
\mu_{\rm \ggF}^{ZZ} = 0.96 \pm 0.08 \,, \quad 
\mu_{\rm \ggF}^{\gamma \gamma} = 1.05 \pm 0.09 \,, \quad 
\mu_{\rm \ggF}^{\gamma Z} = 2.2 \pm 0.7 \,.
\eeq
Here the first three results represent unofficial weighted averages of the ATLAS~\cite{ATLAS:2020qdt} and CMS~\cite{CMS:2020gsy} measurements, while the fourth result stems from an official combination performed by the ATLAS and CMS collaborations in~\cite{CMS:2023mku}. 

By comparing~(\ref{eq:mustheo}) and~(\ref{eq:musexp}) it is evident that under the assumption of $C_{HW\!B} \simeq 0$ the constraints on the Wilson coefficients $C_ {H\hspace{-0.3mm}B}$ and $C_{HW}$ from $h \to \gamma \gamma$ and $h \to \gamma Z$ are in general more stringent than those that arise from $h \to WW$ and $h \to ZZ$. In fact, since the signal strength $\mu_{\rm \ggF}^{\gamma \gamma}$ is at present significantly better measured than $\mu_{\rm \ggF}^{\gamma Z}$, all combinations of Wilson coefficients that obey 
\beq \label{eq:flat}
C_ {H\hspace{-0.3mm}B} \simeq -\frac{s_w^2}{c_w^2} \, C_{HW} \simeq -0.30 \hspace{0.5mm} C_{HW} \,, 
\eeq
are most weakly constrained by~(\ref{eq:musexp}). Notice that for Wilson coefficients $C_ {H\hspace{-0.3mm}B}$ and $C_{WB}$ satisfying~(\ref{eq:flat}) and $|C_{HW}|$ sufficiently small the signal strengths $\mu_{\rm \ggF}^{WW}$, $\mu_{\rm \ggF}^{ZZ}$ and $\mu_{\rm \ggF}^{\gamma \gamma}$ are all predicted to be SM-like, while $\mu_{\rm \ggF}^{\gamma Z}$ can at the same time be notably different from one. 

\subsection{Discussion}
\label{sec:discussion}

We are now in a position to identify simple benchmark scenarios for the full set of Wilson coefficients of the operators defined in~(\ref{eq:operators1}) to~(\ref{eq:operators4}). As a common operator suppression scale we hereafter take $\Lambda = 1 \, {\rm TeV}$. In view of the stringent constraints arising from~(\ref{eq:dMWoMWtheo}),~(\ref{eq:dMWoMWexp}),~(\ref{eq:DeltagLRe}) and~(\ref{eq:DeltagLReLEPSLD}) we always employ
\beq \label{eq:benchall}
C_{H\ell}^{(1)} = C_{H\ell}^{(3)} = C_{He} = C_{H\Box} = C_{H\hspace{-0.25mm}D} = C_{\ell\ell} = 0 \,,
\eeq
when studying the numerical impact of the SMEFT dimension-six operators~(\ref{eq:operators1}) to~(\ref{eq:operators4}) on $Zh$ production at the LHC in~Section~\ref{sec:numerics}. 

In the case of the dimension-six operators that modify the couplings between the Higgs and two vector bosons at tree level, we choose the following values for the Wilson coefficients
\beq \label{eq:bench1}
C_ {H\hspace{-0.3mm}B} = 0.015 \,, \qquad C_{HW} = -0.05 \,, \qquad C_{HW\!B} = 0 \,, 
\eeq
where the choice of $C_{HW\!B}$ is motivated by~(\ref{eq:CHWBbound}). The values~(\ref{eq:bench1}) lead to the following Higgs signal strengths 
\beq \label{eq:musbench1}
\mu_{\rm \ggF}^{WW} \simeq 1.00 \,, \qquad
\mu_{\rm \ggF}^{ZZ} \simeq 1.02 \,, \qquad 
\mu_{\rm \ggF}^{\gamma \gamma} \simeq 1.01 \,, \qquad 
\mu_{\rm \ggF}^{\gamma Z} \simeq 2.13 \,,
\eeq
if one assumes that the \ggF~Higgs production cross section in the SMEFT is equal to that of the SM and sets to zero the Wilson coefficients of all the operators appearing in~(\ref{eq:operators2}) to~(\ref{eq:operators4}). Notice that the predictions~(\ref{eq:musbench1}) are all SM-like apart from $\mu_{\rm \ggF}^{\gamma Z}$ which is close to the central value of the measured signal strength given in~(\ref{eq:musexp}). When considering the benchmark scenario~(\ref{eq:bench1}) we will, besides employing~(\ref{eq:benchall}), also set all the Wilson coefficients of the operators~(\ref{eq:operators2}) to zero. 

In the case of the operators that result in couplings between the Higgs, a $W$ or a $Z$ boson, and light quarks, we consider the four benchmark scenarios 
\begin{align} 
& C_{Hq}^{(3)} = 0.05 \,, \qquad C_{Hq}^{(1)} = C_{Hd} = C_{Hu} = C_{Hud} = 0 \,, \label{eq:bench3} \\[2mm]
& \! C_{Hd} = -0.1 \,, \qquad C_{Hq}^{(1)} = C_{Hq}^{(3)} = C_{Hu} = C_{Hud} = 0 \,, \label{eq:bench4} \\[2mm]
& \, C_{Hu} = 0.1 \,, \qquad C_{Hq}^{(1)} = C_{Hq}^{(3)} = C_{Hd} = C_{Hud} = 0 \,, \label{eq:bench5} \\[2mm]
& C_{Hq}^{(1)} = 0.05 \,, \qquad C_{Hq}^{(3)} = C_{Hd} = C_{Hu} = C_{Hud} = 0 \,, \label{eq:bench2}
\end{align} 
where one of the five relevant Wilson coefficients is non-zero while the remaining ones vanish identically. As in the case of~(\ref{eq:bench1}) we set all Wilson coefficients not present in~(\ref{eq:bench3}) to~(\ref{eq:bench2}) to zero when considering the class of operators introduced in~(\ref{eq:operators2}).

\section{Phenomenological analysis}
\label{sec:numerics}

In the following, we present \NNLOplusPS~accurate results for $pp \to Zh \to \ell^+ \ell^- h$ production with a stable Higgs boson including the SMEFT effects discussed in~Section~\ref{sec:calculation}. All~SM input parameters are taken from the most recent PDG~review~\cite{ParticleDataGroup:2022pth}. In~particular, we use $\alpha = 1/127.951$, $G_F = 1.1663788 \cdot 10^{-5} \, {\rm GeV}^{-2}$, $m_Z = 91.1876 \, {\rm GeV}$, $\Gamma_Z^{\rm SM} = 2.4952 \, {\rm GeV}$, $m_h = 125.09 \, {\rm GeV}$ and $\Gamma_h^{\rm SM} = 4.1 \, {\rm MeV}$. The~values of the weak mixing angle, the $U(1)_Y$ and $SU(2)_L$ gauge coupling and the Higgs VEV are calculated from~(\ref{eq:sin2thetaw}) and~(\ref{eq:g1g2v}), respectively, meaning that we work in the LEP scheme. The {\tt NNPDF31\_nnlo\_as\_01180} parton distribution functions~(PDFs)~\cite{NNPDF:2017mvq} are employed in our MC~simulations and events are showered with~\PYTHIA{8.2}~\cite{Sjostrand:2014zea} utilising the Monash tune~\cite{Skands:2014pea}. To target the $Z \to \ell^+ \ell^-$ decay, we select events with two charged leptons~(electrons or muons). The leptons are defined at the dressed level, meaning the lepton four-momentum is combined with the four-momenta of nearby prompt photons arising in the shower using a dressing-cone size of $\Delta R_{\ell \gamma} < 0.1$. The leptons are required to have $p_{T,\ell} > 15 \, {\rm GeV}$ and a pseudorapidity of $|\eta_\ell| < 2.5$. The invariant mass of the dilepton pair is restricted to $m_{\ell^+ \ell^-} \in [75, 105] \, {\rm GeV}$. These restrictions are close to those imposed by the existing ATLAS and CMS studies of~\cite{ATLAS:2018kot,CMS:2018nsn,ATLAS:2019yhn,ATLAS:2020fcp,CMS:2022exo} and we will simply refer to them as ``fiducial~cuts'' in what follows. 

\afterpage{
\begin{figure}[h]
\begin{center}
\includegraphics[height=0.575\textwidth]{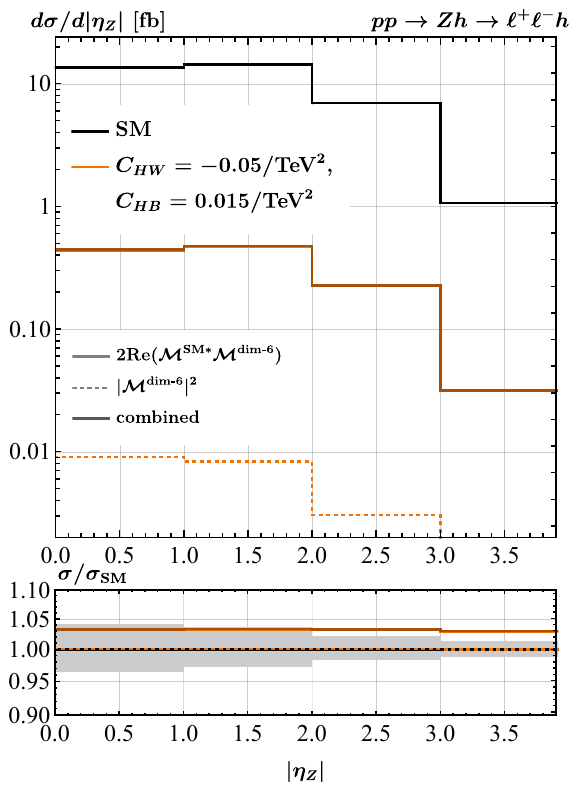} \qquad \quad 
\includegraphics[height=0.575\textwidth]{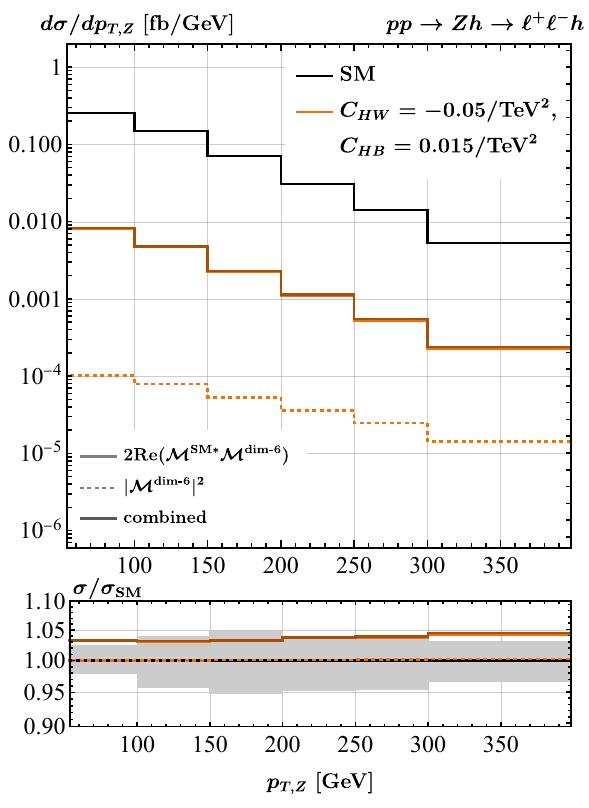}

\vspace{4mm}

\includegraphics[height=0.575\textwidth]{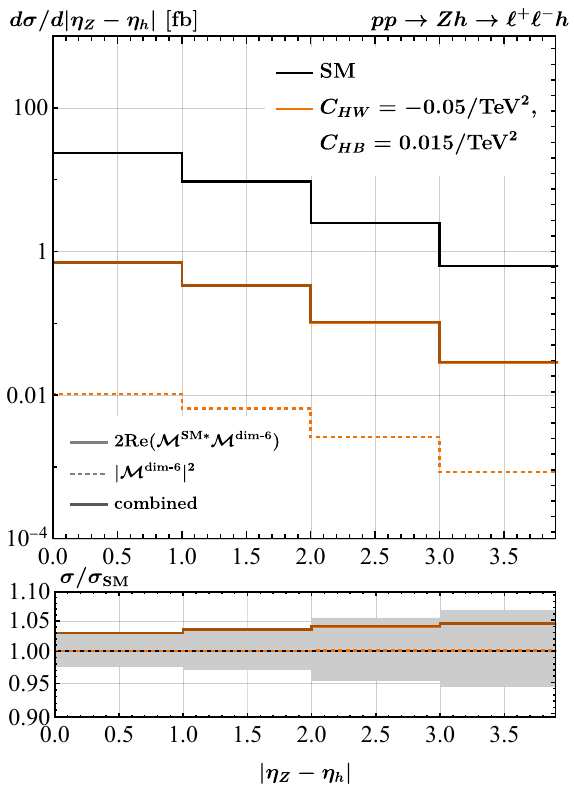} \qquad \quad 
\includegraphics[height=0.575\textwidth]{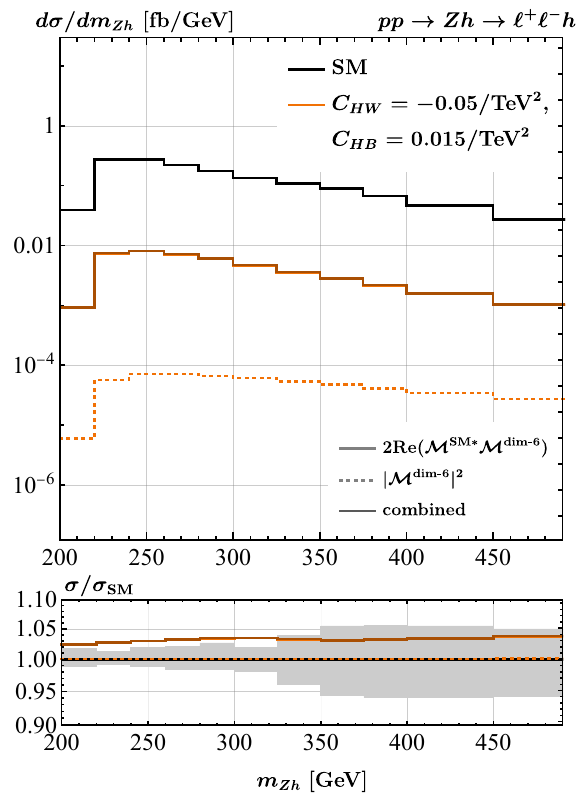}
\end{center}
\vspace{-2mm} 
\caption{\label{fig:bench1} \NNLOplusPS~predictions for $pp \to Zh \to \ell^+ \ell^- h$ production in the SMEFT benchmark scenario~(\ref{eq:bench1}) assuming a common operator suppression scale of $\Lambda = 1\, {\rm TeV}$. The~four panels show the fiducial cross section differential in $|\eta_Z|$~(upper~left), $p_{T,Z}$~(upper~right), $|\eta_{Z}-\eta_{h}|$~(lower~left) and $m_{Zh}$~(lower~right) for proton-proton ($pp$) collisions at~$13 \, {\rm TeV}$. The~SM predictions are indicated by the solid black lines while the solid (dotted) orange curves represent the SMEFT contribution linear (quadratic) in the Wilson coefficients. The solid dark orange lines correspond to the sums of the linear and quadratic SMEFT contributions. The lower panels depict the ratios between the BSM and the SM distributions with the grey band representing the SM scale uncertainties. See main text for further details.}
\end{figure}
\clearpage
}

The four panels in~Figure~\ref{fig:bench1} display our \NNLOplusPS~predictions for $pp \to Zh \to \ell^+ \ell^- h$ in the SMEFT benchmark scenario~(\ref{eq:bench1}) assuming $\Lambda = 1\, {\rm TeV}$ and LHC collisions with a centre-of-mass energy of $\sqrt{s} = 13 \, {\rm TeV}$. The fiducial cross sections differential in the $Z$-boson pseudorapidity~$|\eta_Z|$~(upper~left) and transverse momentum $p_{T,Z}$~(upper~right) as well as the~pseudorapidity difference $|\eta_{Z}-\eta_{h}|$~(lower~left) and the invariant mass $m_{Zh}$~(lower~right) of the $Zh$ system are shown. The central renormalisation scale~$\mu_R$ and factorisation scale~$\mu_F$ are set according to the \MiNNLOPS~procedure~\cite{Monni:2019whf,Monni:2020nks} and the grey bands in the lower panels represent the corresponding perturbative uncertainties in the SM. These uncertainties have been obtained from seven-point scale variations enforcing the constraint $1/2 \leq \mu_R/\mu_F \leq 2$. Within the SM they do not exceed $5\%$ for what concerns the considered distributions, and relative scale uncertainties of very similar size are also found in the case of the SMEFT spectra. The~SM predictions are indicated by the solid black lines while the solid~(dotted) orange curves represent the SMEFT contributions that are linear~(quadratic) in the Wilson coefficients. Notice that the linear~(quadratic) SMEFT contributions arise from the interference of the SMEFT and SM amplitudes (self-interference of the SMEFT amplitudes). The solid dark orange lines finally correspond to the sums of the linear and quadratic SMEFT contributions. We observe that in the case of~the~$|\eta_Z|$ distribution the SMEFT effects related to $C_{H\hspace{-0.3mm}B}$ and $C_{HW}$ to first approximation simply shift the spectra by a constant amount. In the cases of the $p_{T,Z}$, the~$|\eta_{Z}-\eta_{h}|$ and the $m_{Zh}$ spectra the relative sizes of the SMEFT corrections instead grow with increasing~$p_{T,Z}$, $|\eta_{Z}-\eta_{h}|$ and~$m_{Zh}$, respectively. It is also evident from all panels that the quadratic SMEFT contributions are negligibly small compared to the linear terms. Numerically, we find that the benchmark scenario~(\ref{eq:bench1}) leads to relative corrections of around $+2\%$ to $+5\%$ in the studied distributions --- the fiducial cross section is enhanced by $+3.5\%$ compared to the~SM. Notice that while~the predicted deviations are sometimes larger than the corresponding~SM~scale uncertainties, they are typically smaller than the ultimate projected HL-LHC accuracy in $Zh$ production channel that amounts to $5\%$~\cite{ATLAS:2018jlh,CMS:2018qgz}. From this numerical exercise one can conclude that future constraints on the Wilson coefficients $C_{H\hspace{-0.3mm}B}$, $C_{HW}$ and $C_{HW\!B}$ from $Vh$ production are unlikely to be as stringent as the limits that future determinations of the Higgs signal strengths in $h \to \gamma \gamma$ and $h \to \gamma Z$ will allow to set. This will in particular be the case if both Higgs signal strengths turn out to be SM-like~$\big($see~(\ref {eq:mustheo})~and~(\ref {eq:musexp})$\big)$. 

\begin{figure}[!t]
\begin{center}
\includegraphics[height=0.575\textwidth]{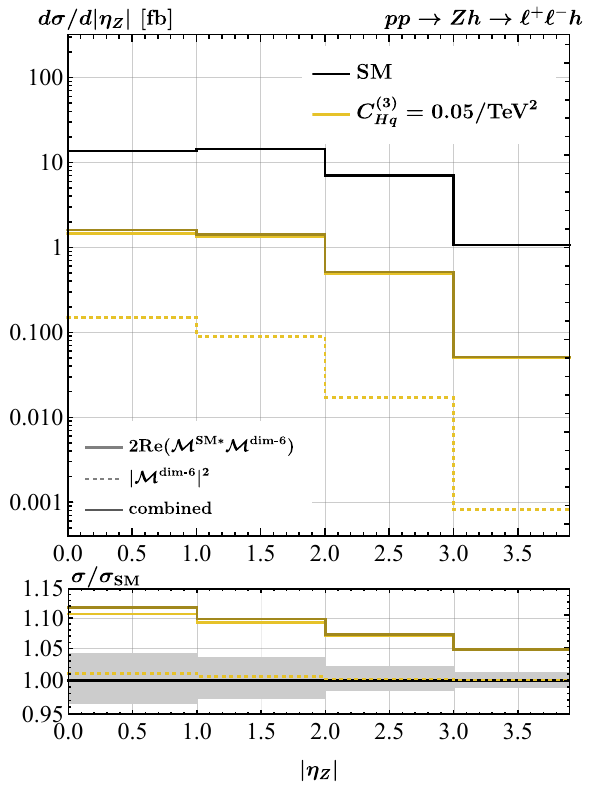} \qquad \quad 
\includegraphics[height=0.575\textwidth]{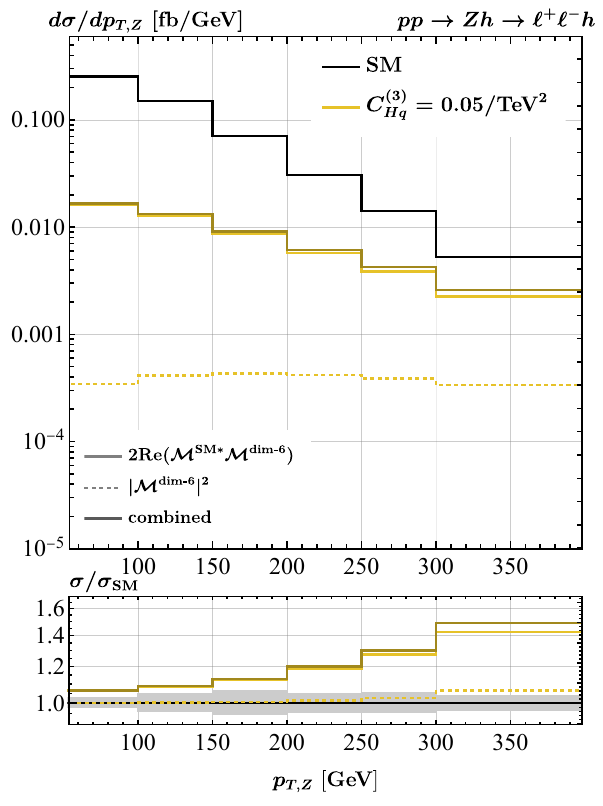}

\vspace{4mm}

\includegraphics[height=0.575\textwidth]{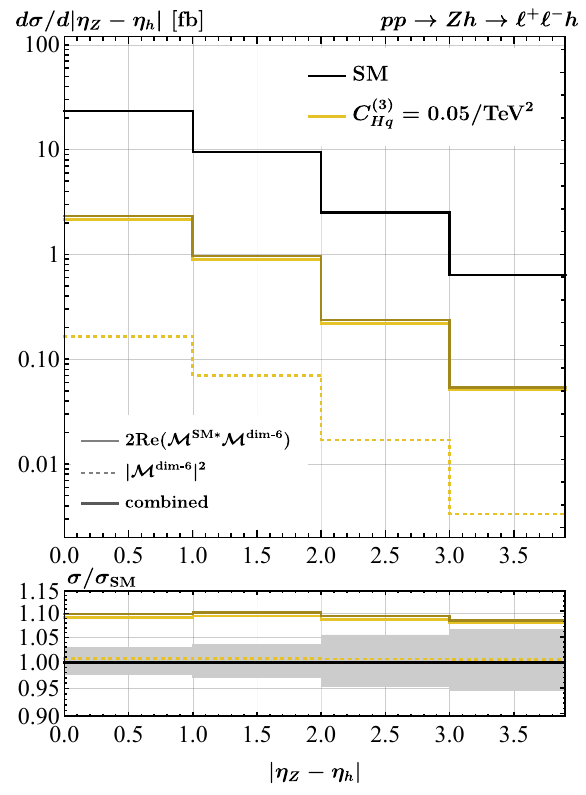} \qquad \quad 
\includegraphics[height=0.575\textwidth]{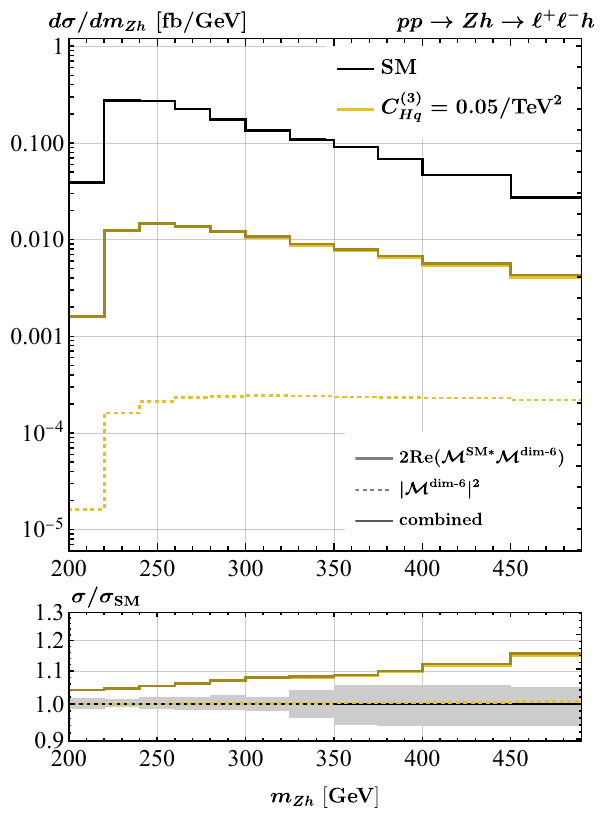}
\end{center}
\vspace{-2mm} 
\caption{\label{fig:bench3} As Figure~\ref{fig:bench1} but for benchmark scenario~(\ref{eq:bench3}) with $\Lambda=1\, {\rm TeV}$. The yellow lines correspond to the BSM results.}
\end{figure}

\begin{figure}[!t]
\begin{center}
\includegraphics[height=0.575\textwidth]{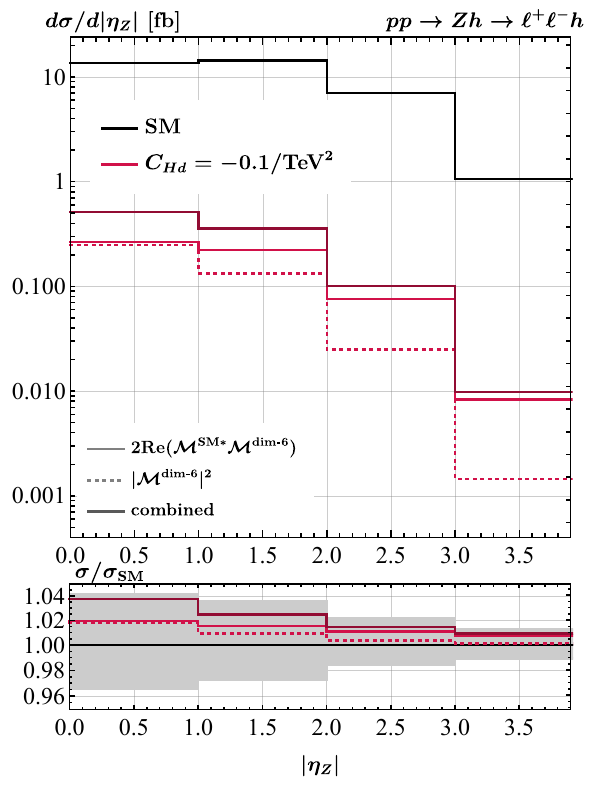} \qquad \quad 
\includegraphics[height=0.575\textwidth]{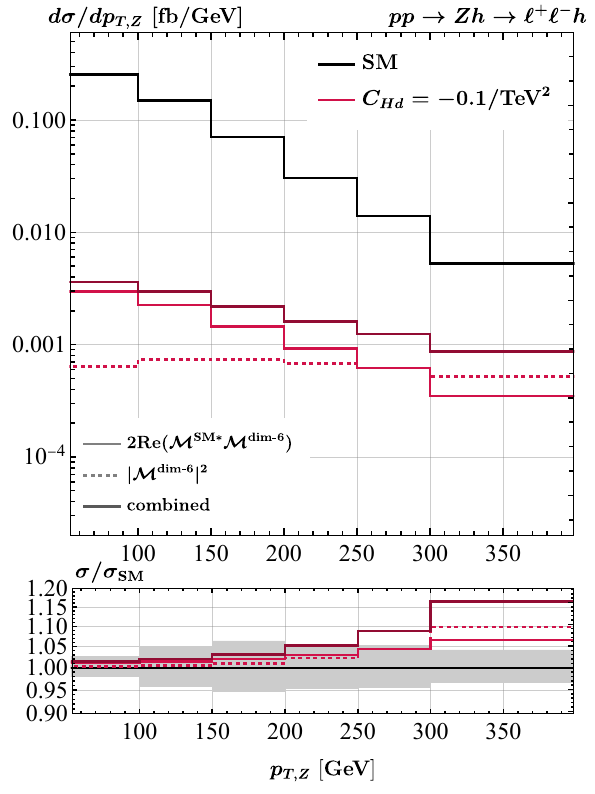}

\vspace{4mm}

\includegraphics[height=0.575\textwidth]{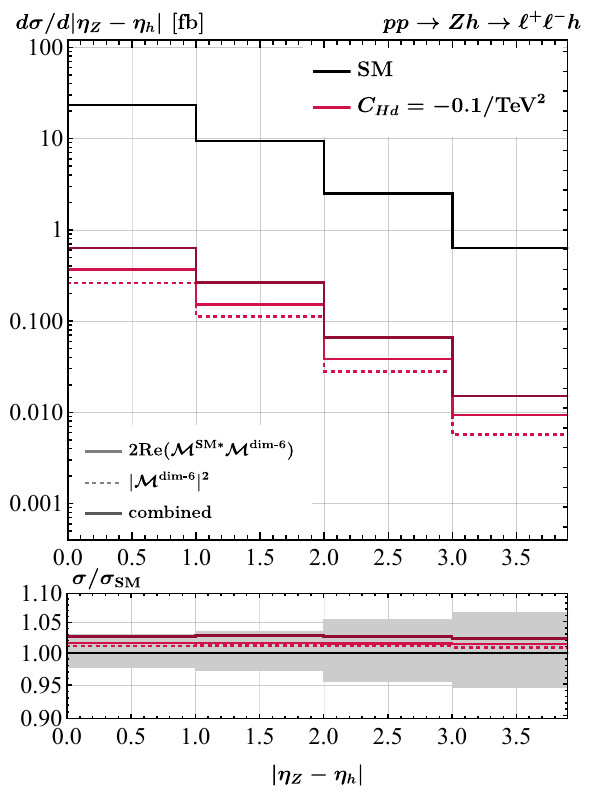} \qquad \quad 
\includegraphics[height=0.575\textwidth]{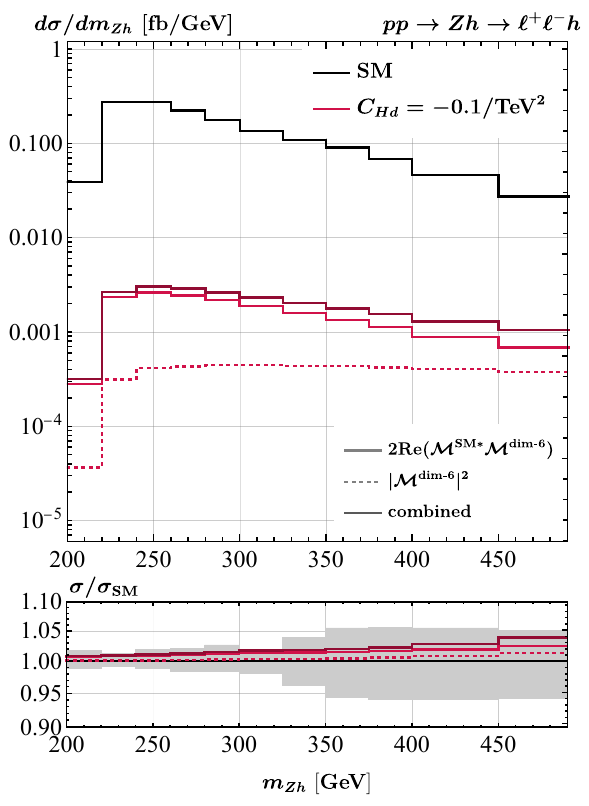}
\end{center}
\vspace{-2mm} 
\caption{\label{fig:bench4} As Figure~\ref{fig:bench1} but for benchmark scenario~(\ref{eq:bench4}) assuming $\Lambda=1\, {\rm TeV}$. The red curves represent the BSM results.}
\end{figure}

\begin{figure}[!t]
\begin{center}
\includegraphics[height=0.575\textwidth]{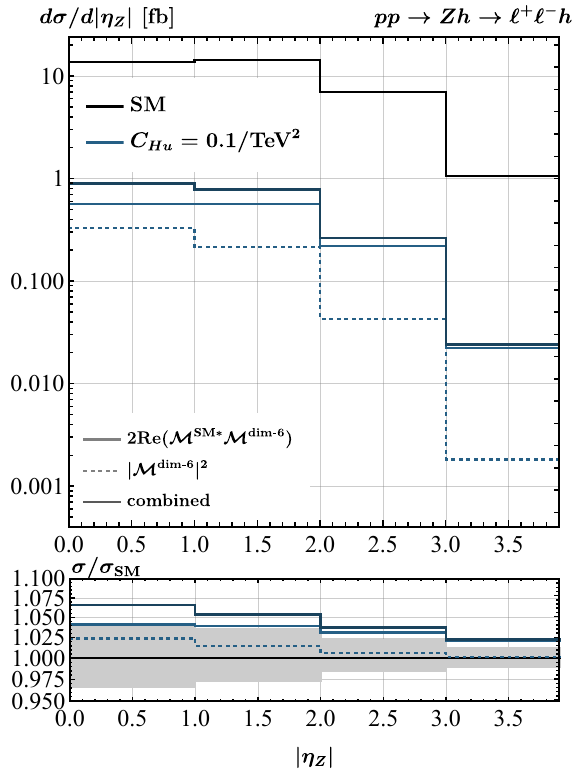} \qquad \quad 
\includegraphics[height=0.575\textwidth]{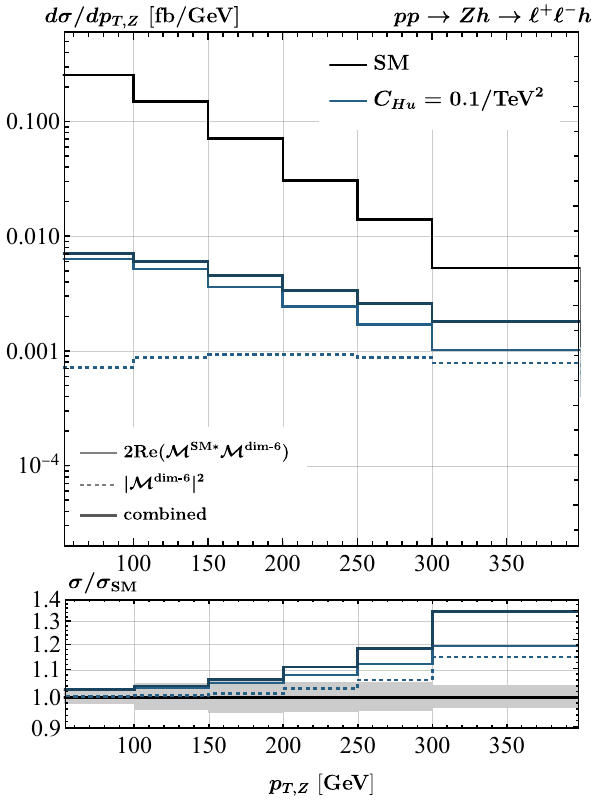}

\vspace{4mm}

\includegraphics[height=0.575\textwidth]{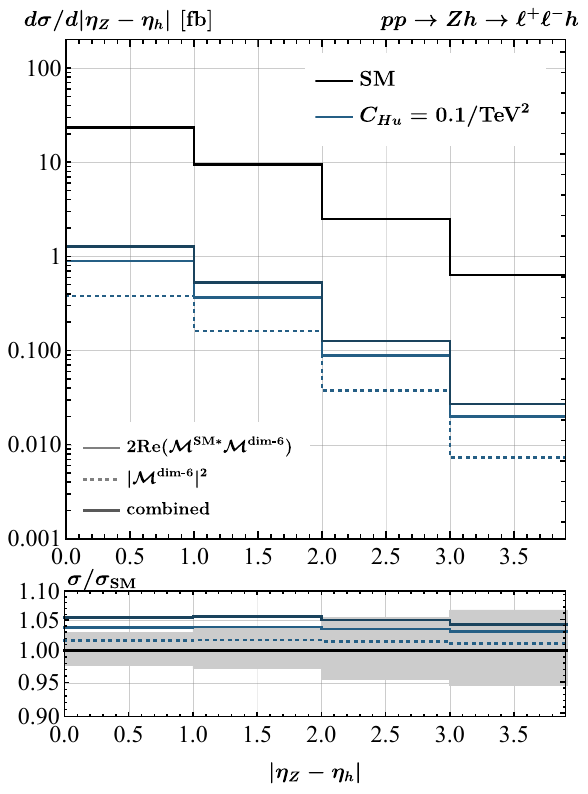} \qquad \quad 
\includegraphics[height=0.575\textwidth]{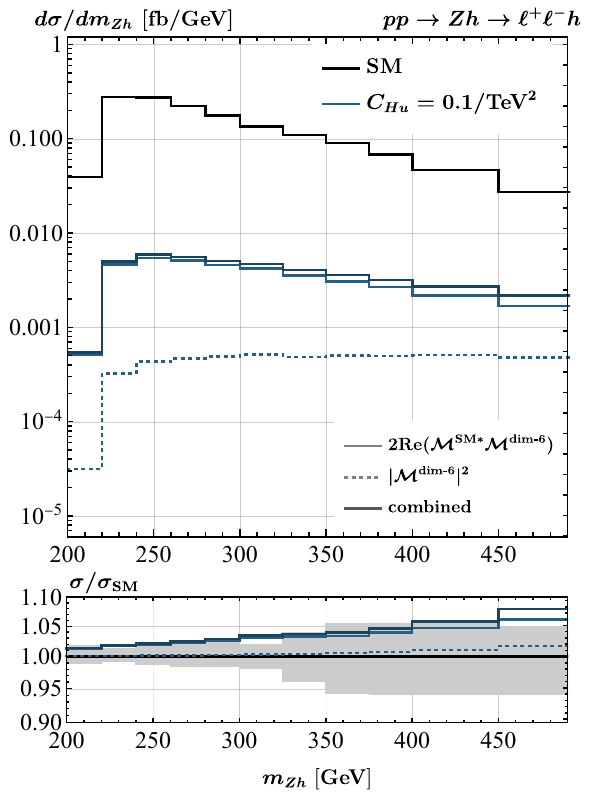}
\end{center}
\vspace{-2mm} 
\caption{\label{fig:bench5} As Figure~\ref{fig:bench1} but for benchmark scenario~(\ref{eq:bench5}) with $\Lambda=1\, {\rm TeV}$. The blue lines are the BSM results.}
\end{figure}

\begin{figure}[!t]
\begin{center}
\includegraphics[height=0.575\textwidth]{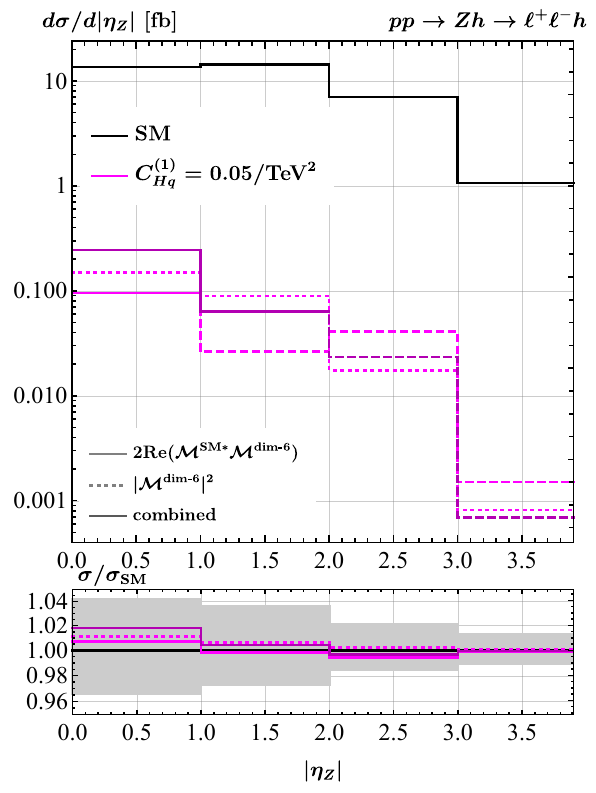} \qquad \quad 
\includegraphics[height=0.575\textwidth]{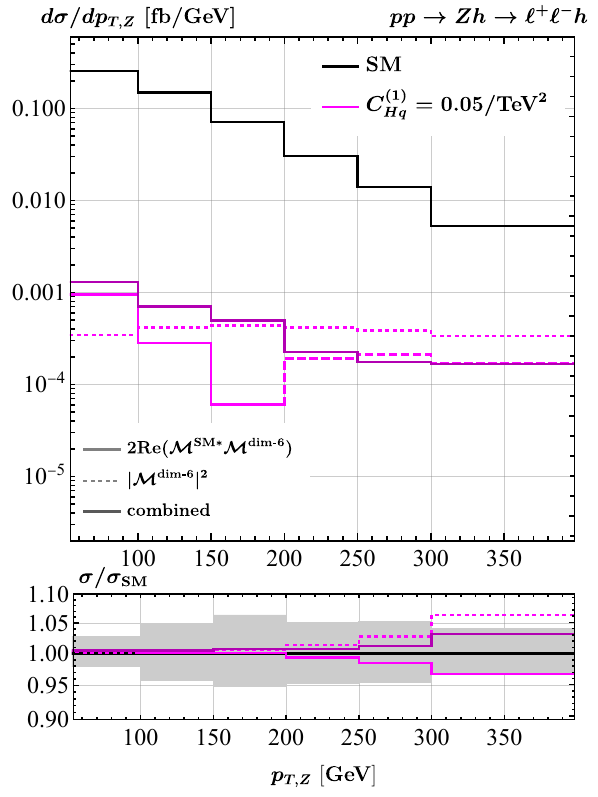}

\vspace{4mm}

\includegraphics[height=0.575\textwidth]{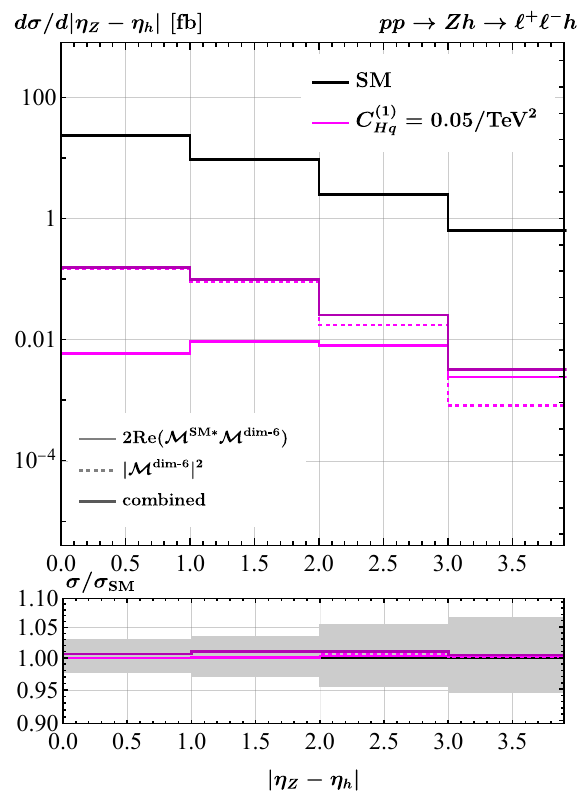} \qquad \quad 
\includegraphics[height=0.575\textwidth]{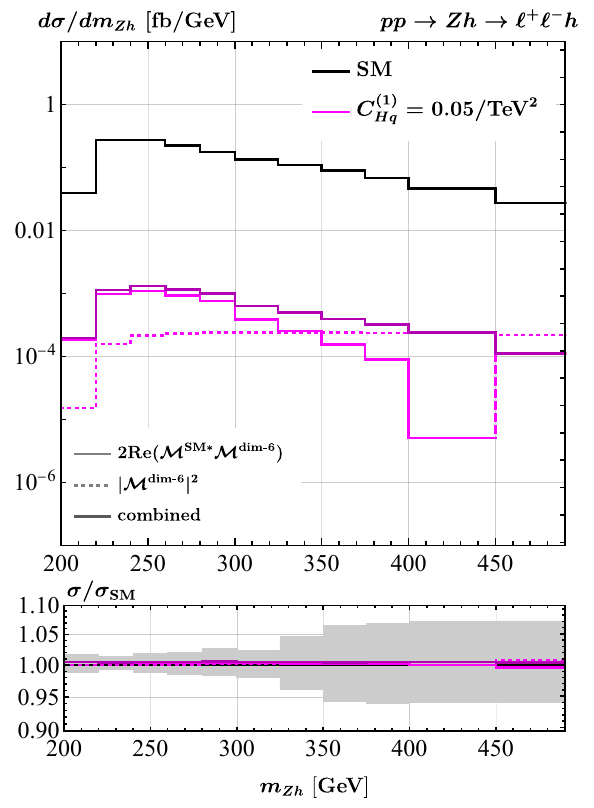}
\end{center}
\vspace{-2mm} 
\caption{\label{fig:bench2} As Figure~\ref{fig:bench1} but for benchmark scenario~(\ref{eq:bench2}) assuming $\Lambda=1\, {\rm TeV}$. The~SMEFT predictions are coloured in magenta. In the case of the linear SMEFT contributions 
the solid (dashed) lines correspond to positive (negative) corrections to the relevant distribution.}
\end{figure}

Figures~\ref{fig:bench3} to~\ref{fig:bench5} contain our \NNLOplusPS~predictions for $pp \to Zh \to \ell^+ \ell^- h$ production in the SMEFT benchmark scenarios~(\ref{eq:bench3}) to~(\ref{eq:bench5}). In all cases we have employed an operator suppression scale of $\Lambda = 1\, {\rm TeV}$. The results depicted in the first three figures show very similar features. In all cases the relative SMEFT corrections are rather flat in the $|\eta_Z|$ and $|\eta_{Z}-\eta_{h}|$ distributions, while in the case of the $p_{T,Z}$ and $m_{Zh}$ spectra they get larger with increasing $p_{T,Z}$ and $m_{Zh}$. The observed high-energy growth is expected from~(\ref{eq:23contribution}) and in all three cases most pronounced in the $p_{T,Z}$ distribution. One also sees that the linear SMEFT effects are largest in the benchmark scenario with $C_{Hq}^{(3)} = 0.05$ where they can exceed $+50\%$ compared to the SM for $p_{T,Z} > 300 \, {\rm GeV}$. The respective effects in the benchmark scenario with $C_{Hd} = -0.1$~($C_{Hu} = 0.1$) just correspond to around $+7\%$~($+20\%$). The observed hierarchy of SMEFT effects can be traced back to the approximate pattern $g_{L}^{d, {\rm SM}} \simeq -g_{L}^{u, {\rm SM}} \simeq -6 \hspace{0.25mm} g_{R}^{d, {\rm SM}} \simeq 3 \hspace{0.25mm} g_{R}^{u, {\rm SM}}$ of left- and right-handed $Z$-boson couplings within the SM and the feature $\delta g_L^d \propto -C_{Hq}^{(3)}$ and $\delta g_L^u \propto C_{Hq}^{(3)}$ --- see~(\ref{eq:DeltagLRdu}). Notice also that the size of the quadratic SMEFT corrections is relatively small in the case of~(\ref{eq:bench3}) while these effects are comparable to or even larger than the linear terms for the benchmark scenarios specified in~(\ref{eq:bench4}) and~(\ref{eq:bench5}). In the case of the SMEFT benchmark scenario~(\ref{eq:bench2}) the pattern of SMEFT deviations turns out to be more complicated. This is illustrated in the four panels of~Figure~\ref{fig:bench2}. One observes that the linear SMEFT effects are typically small and even change sign in some distributions as indicated by the transitions from solid to dashed lines. To~understand these features it is important to realise that $g_{L}^{d, {\rm SM}} \simeq -g_{L}^{u, {\rm SM}}$, $\delta g_L^d \propto -C_{Hq}^{(1)}$ and $\delta g_L^u \propto -C_{Hq}^{(1)}$ and to keep in mind that the down-quark luminosity in a proton is smaller than the up-quark luminosity at large $x$ while the two luminosities are of similar size at small $x$. For the choice $C_{Hq}^{(1)} = 0.05$ the down- and up-quark contributions thus tend to cancel leading to a numerical suppression of the full linear SMEFT effects compared to naive expectation. It is also evident from the shown results that the quadratic SMEFT corrections are as large in magnitude as the linear terms. In fact, in the case of the $p_{T,Z}$ and~$m_{Zh}$ spectra the two types of SMEFT effects have opposite relative signs resulting in very small combined BSM contributions not exceeding the level of $+2\%$ in the case of the $p_{T,Z}$ distribution. Notice finally that due to the energy growth most of the sensitivity to the SMEFT effects considered in~Figures~\ref{fig:bench3}~to~\ref{fig:bench2} comes from the high-energy tails of kinematic distributions such as~the~$p_{T,Z}$ and~$m_{Zh}$ spectra. In such a case the Higgs-boson decay products are significantly boosted, giving rise to very specific kinematic features and providing additional handles to distinguish signal from background events. The articles~\cite{Bishara:2020vix,Bishara:2020pfx,Bishara:2022vsc} have exploited this feature to obtain stringent constraints on the dimension-six operators~(\ref{eq:operators2}) using future hypothetical hadron collider measurements of~$Vh$ production. The \MiNNLOPS~generator~\cite{GitLabPowheg} presented in this work would allow to improve the accuracy of these studies from the \NLOplusPS~to the \NNLOplusPS~level. We leave such an analysis for future study. 

\section{Combining production and decay at \NNLOplusPSbm}
\label{sec:higgsdecay}

Let us finally outline how one can combine the \NNLOplusPS~calculation of $Vh$ production~with the decay of the Higgs boson to bottom quarks at \NNLOplusPS. The starting point is to replace the Higgs boson in each $Vh$ event by the possible decay products (i.e.~$b \bar b g$, $b \bar b q \bar q$ or $b \bar b gg$) taken from a given decay event. Working in the narrow width approximation~(NWA), the full weight $w_{\rm full}$ of each event is then calculated as 
\beq \label{eq:fullweight}
w_{\rm full}^{\rm SMEFT} = \frac{w_{\rm prod}^{\rm SMEFT} \, w_{\rm dec}^{\rm SMEFT}}{\Gamma_h^{\rm SMEFT}} \,, 
\eeq
where $w_{\rm prod}^{\rm SMEFT}$ denotes the weight of the production event obtained at \NNLOplusPS~using the \MiNNLOPS~$pp \to Vh$ generator described earlier in this section, while $w_{\rm dec}^{\rm SMEFT}$ is the weight of the decay event computed at \NNLOplusPS~by means of the \MiNLO~method~\cite{Hamilton:2012np,Hamilton:2012rf} employing the procedure detailed in~Section~3.2 of the paper~\cite{Bizon:2019tfo}. Notice that insertions of the operators~(\ref{eq:operators1}) to~(\ref{eq:operators4}) do not modify the kinematics of any Higgs decay channel involving bottom-quark pairs. In fact, the SMEFT decay weights are simply related to the SM decay weights by 
\beq \label{eq:wdecSMEFT}
w_{\rm dec}^{\rm SMEFT} = \left ( 1 + 2 \hspace{0.25mm} c_{\rm kin} \right ) w_{\rm dec}^{\rm SM} \,, \qquad c_{\rm kin} = \frac{v^2}{\Lambda^2} \left ( C_{H\Box} - \frac{C_{H\hspace{-0.25mm}D}}{4} \right ) \,,
\eeq
where the term $\big ( 1 + 2 \hspace{0.25mm} c_{\rm kin} \big )$ arises from the canonical normalisation of the Higgs kinetic term. The factor $\Gamma_h^{\rm SMEFT}$ in~(\ref{eq:fullweight}) takes into account that the total decay width of the Higgs boson is modified by SMEFT effects. For example, one can employ the result
\beq \label{eq:GammahSMEFT}
\Gamma_h^{\rm SMEFT} = \left ( 1 + 2 \hspace{0.25mm} c_{\rm kin} \right ) \left [ \hspace{0.25mm} 1 - \frac{v^2}{\Lambda^2} \, \Big ( 1.23 \hspace{0.25mm} C_ {H\hspace{-0.3mm}B} + 1.38 \hspace{0.25mm} C_{HW} + 0.12 \hspace{0.25mm} C_{HW\!B} \Big ) \right ] \Gamma_h^{\rm SM} \,, 
\eeq
where we have factorised the contribution $\big ( 1 + 2 \hspace{0.25mm} c_{\rm kin} \big )$ due to the canonical normalisation of the Higgs kinetic term and $\Gamma_h^{\rm SM}$ denotes the total decay width of the $125 \, {\rm GeV}$ Higgs boson within the SM. Notice that the factor $\big ( 1 + 2 \hspace{0.25mm} c_{\rm kin} \big )$ drops out in the ratio of~(\ref{eq:wdecSMEFT}) and~(\ref{eq:GammahSMEFT}) and therefore in~(\ref{eq:fullweight}) when production and decay events are combined. Based on~(\ref{eq:fullweight}) for each event the~Les~Houches event~(LHE) file in addition to the weight also contains the value of the hardest radiation allowed by the shower (i.e.~the scale {\tt scalup}). In the combined~LHE file the value of~{\tt scalup$_{\tt prod}$} for the production process is recorded and, once the event is passed to the~PS, the value of {\tt scalup$_{\tt dec}$} for the specific decay kinematics is recomputed. Emissions of the Higgs decay in all the available phase space is then generated and after the shower is complete it is checked whether the hardness of the splittings is below the veto scale~{\tt scalup$_{\tt dec}$}. If this is not the case the event is again showered until the latter condition~is~met. 

Notice that while above we have considered the SMEFT corrections to $h \to b \bar b$ that arise from~(\ref{eq:operators1}) to~(\ref{eq:operators4}) the modifications required to achieve \NNLOplusPS~accuracy for other operators is relatively straightforward as long as one works in the NWA for the Higgs propagator. A non-trivial example for the combination of Higgsstrahlung and $h \to b \bar b$ at~\NNLOplusPS~has been provided in the recent article~\cite{Haisch:2022nwz}. In fact, this work achieved \NNLOplusPS~precision for the dimension-six operators that contribute to the subprocesses $pp \to Zh$ and $h \to b \bar b$ directly in~QCD. This class of operators includes effective Yukawa- and chromomagnetic dipole-type interactions of the bottom quark that modify the $h \to b \bar b$ decay but do not play a role in $pp \to Zh$ production. Combining the results obtained here and in~\cite{Haisch:2022nwz} via~(\ref{eq:fullweight}) it is now possible to obtain \NNLOplusPS~accurate predictions for the full $pp \to Zh \to \ell^+ \ell^- b \bar b$ and $pp \to Wh \to \ell \nu b \bar b$ processes taking into account 18 different dimension-six SMEFT operators. Since in the NWA the Higgsstrahlungs production processes including~the~leptonic decays of the massive gauge bosons factorises from the subsequent decay of the Higgs boson, extending the general method of combining production and decay~(\ref{eq:fullweight}) to incorporate additional operators modifying $h \to b \bar b$ or considering further Higgs-boson decay modes is relatively simple. 

\section{Conclusions}
\label{sec:conclusions}

In this article, we have presented novel SMEFT predictions for Higgsstrahlung in hadronic collisions. Specifically, we have calculated the NNLO QCD corrections for the complete sets of dimension-six operators that describe the interactions between the Higgs and two vector bosons and the couplings of the Higgs, a $W$ or a $Z$ boson, and light fermions. These~fixed-order predictions have been consistently matched to a PS using the \MiNNLOPS~method and the matching has been implemented into the {\tt POWHEG-BOX}. Our new MC~implementation~\cite{GitLabPowheg} allows for a realistic exclusive description of $Vh$ production at the level of hadronic events including SMEFT effects. This feature makes it an essential tool for future Higgs characterisation studies by the ATLAS and CMS collaborations. Notice that together with the~MC code presented in~\cite{Haisch:2022nwz} one can now simulate the $pp \to Zh \to \ell^+ \ell^- b \bar b$ and $pp \to Wh \to \ell \nu b \bar b$ processes at \NNLOplusPS~including a total number of 18 dimension-six SMEFT operators.

To motivate simple SMEFT benchmark scenarios we have discussed the leading constraints on the Wilson coefficients of the 14~dimension-six operators listed in~(\ref{eq:operators1})~to~(\ref{eq:operators4}). In the case of the effective interactions between the Higgs boson and two vector bosons described by~$Q_{H\hspace{-0.3mm}B}$, $Q_{HW}$ and $Q_{HW\!B}$, we found that the combination of the measurements of the $W$-boson mass and the Higgs signal strength in $h \to \gamma \gamma$ and $h \to \gamma Z$ allow to set stringent constraints on the respective Wilson coefficients. In order to avoid these bounds we have introduced the benchmark scenario~(\ref{eq:bench1})~which as far as the Higgs signal strengths are concerned predicts $h \to WW$, $h \to ZZ$ and $h \to \gamma \gamma$ to be SM-like, while enhancing the $h \to \gamma Z$ rate by a factor of around 2 with respect to the SM. Such a pattern of deviations is presently favoured by LHC data $\big($cf.~(\ref{eq:musexp})$\big)$. In the case of the dimension-six terms that give rise to couplings between the Higgs, a $W$ or a $Z$ boson, and light fermions, we have pointed out that while the leptonic operators~(\ref{eq:operators3}) are in general tightly constrained by SLC and LEP measurements the resulting bounds on the quark operators~(\ref{eq:operators2}) depend sensitively on the flavour assumption in the SMEFT. While in the case of an approximate~$U(3)^5$ flavour symmetry the constraints turn out to be stringent, the limits on the Wilson coefficients of~(\ref{eq:operators2}) are notably relaxed if the assumption of flavour universality only applies to the light but not the heavy down- and up-quark flavours. The remaining operators~(\ref{eq:operators4}) shift the Higgs kinetic term and/or the EW SM input parameters at tree level. The corresponding Wilson coefficients are therefore in general better probed by a global SMEFT fit than by $pp \to Vh$ production alone. 

We have then performed an~\NNLOplusPS~study of the impact of SMEFT contributions on several kinematic distributions in $pp \to Zh \to \ell^+ \ell^- h$ production for a stable Higgs boson considering the simple benchmark scenarios identified earlier. While in our {\tt POWHEG-BOX} implementation~\cite{GitLabPowheg} the user can choose between the $\alpha$, $\alpha_\mu$, and LEP schemes, for concreteness our discussion was based on the LEP scheme which uses~$\{\alpha, G_F, m_Z\}$ as EW input parameters. Another feature of our MC code worth highlighting is that it is able to compute separately both the SMEFT corrections that are linear and quadratic in the Wilson coefficients. Our numerical analysis showed that once the stringent constraints from $m_W$, $h \to \gamma \gamma$ and $h \to \gamma Z$ are imposed the numerical impact of~$C_{H\hspace{-0.3mm}B}$, $C_{HW}$ and $C_{HW\!B}$ on the kinematic distributions in $pp \to Zh \to \ell^+ \ell^- h$ are rather limited, amounting to relative deviations of no more than $5\%$. Future limits on the Wilson coefficients of the effective interaction in~(\ref{eq:operators1}) from $Vh$ production are therefore unlikely to be competitive with the limits that future determinations of the Higgs signal strengths in $h \to \gamma \gamma$ and $h \to \gamma Z$ will allow to set. This will in particular be the case if the latter measurements turn out to be SM-like. The situation turns out to be more promising in the case of the operators~(\ref{eq:operators2}) that induce couplings of the Higgs, a $W$ or a $Z$ boson, and light quarks. The sensitivity of the process $pp \to Zh \to \ell^+ \ell^- h$ to the Wilson coefficients $C_{Hq}^{(1)}$, $C_{Hq}^{(3)}$, $C_{Hd}$ and $C_{Hu}$ arises from the energy growth of the respective amplitudes which results in enhanced high-energy tails of kinematic distributions such as~the~$p_{T,Z}$ and~$m_{Zh}$ spectra. Numerically, we found that these enhancements can reach $50\%$ in the $p_{T,Z}$ spectrum in the region where the Higgs-boson decay products are significantly boosted. As shown in the papers~\cite{Bishara:2020vix,Bishara:2020pfx,Bishara:2022vsc}, future hypothetical HL-LHC measurements of~$Vh$ production can therefore provide constraints on the Wilson coefficients $C_{Hq}^{(1)}$, $C_{Hq}^{(3)}$, $C_{Hd}$ and $C_{Hu}$ that are competitive with the bounds obtained from projected global SMEFT fits. Utilising the \MiNNLOPS~generator presented in this work would allow to improve the accuracy of the studies~\cite{Bishara:2020vix,Bishara:2020pfx,Bishara:2022vsc} from the \NLOplusPS~to the \NNLOplusPS~level. 

Notice that in our phenomenological analysis we have focused on the $0$-jet categories of the stage~1.2 simplified template cross sections~(STXS) framework~\cite{Andersen:2016qtm,Berger:2019wnu,Amoroso:2020lgh} for the $Vh$~production processes. However, it is important to realise that our {\tt POWHEG-BOX} implementation~\cite{GitLabPowheg} of the Higgsstrahlungs processes also allows to simulate the different $1$-jet STXS categories with \NLOplusPS~accuracy. This represents an important improvement compared 
to the MC~code presented in~\cite{Alioli:2018ljm} or the {\tt SMEFT@NLO} package~\cite{Degrande:2020evl} which are only \LOplusPS~accurate for $1$-jet observables in $Vh$ production. Another~novel feature of our implementation of Higgsstrahlung is that it is able to combine production and decay consistently at \NNLOplusPS~including SMEFT effects. To our knowledge, this is presently not possible with any other publicly available tool even if one only aims at~\NLOplusPS~precision. Finally, the presented squared matrix element library~\cite{GitLabAmplitudes} contains all spinor-helicity amplitudes that are needed to obtain~\NNLOplusPS~predictions for Drell-Yan production taking into account the effects of the dimension-six operators~(\ref{eq:operators2}) and~(\ref{eq:operators3}). Modifying the code such that one can calculate the SMEFT effects in diboson production at~\NNLOplusPS~due to operators that induce anomalous triple gauge couplings is also relatively straightforward.

\acknowledgments We thank Fady Bishara, Thomas Gehrmann and Giulia Zanderighi for helpful discussions and communications. Useful feedback from Anke Biek{\"o}tter, Jordy de Vries, Alejo~Rossia and Marion~Thomas concerning the first version of this manuscript are also acknowledged. The~research of~LS is partially supported by the International Max Planck Research School~(IMPRS) on ``Elementary Particle Physics'' as well as the Collaborative Research Center~SFB1258.

\begin{appendix}

\section{Analytic expressions for parameters and couplings}
\label{app:parameters}

In this appendix, we provide the analytic formulae for the parameters and couplings that appear in Section~\ref{sec:calculation}. The presented expressions have been implemented into our MC code~\cite{GitLabPowheg} which allows the user to choose between the $\alpha$, $\alpha_\mu$, and LEP schemes. We refer the interested reader to the articles~\cite{Brivio:2017vri,Brivio:2020onw,Biekotter:2023xle} for additional technical details on EW input schemes in the~SMEFT~context.

In order to write the expression in this appendix as compactly as possible we introduce the following abbreviations
\beq \label{eq:abbreviations}
\begin{split}
& g_\pm = \sqrt{g_1^2 \pm g_2^2} \,, \qquad \Delta m = \sqrt{m_Z^2 - m_W^2} \,, \\[3mm]
& \hspace{-7.5mm} s_w = \sqrt{\frac{1}{2} \left [ 1 - \sqrt{1 - \frac{2 \sqrt{2} \pi \alpha}{G_F \hspace{0.125mm} m_Z^2}} \right ]} \,, \quad c_w = \sqrt{1-s_w^2} \,,
\end{split}
\eeq
where $\alpha$ is the fine-structure constant, $G_F$ is the Fermi constant as extracted from muon decay and $m_Z$~($m_W$) is the mass of the $Z$~($W$) boson in the on-shell scheme. The relevant expressions for 
 the $U(1)_Y$ and $SU(2)_L$ gauge couplings $g_1$ and $g_2$ and the Higgs~VEV~$v$ in terms of the EW input parameters are given in Table~\ref{tab:parameters} for the $\alpha$, the $\alpha_\mu$, and the~LEP~scheme.

In terms of the parameters $g_1$, $g_2$ and $v$ the $Zf\bar f$, $\gamma f\bar f$ and $hZZ$ coupling strengths take the following form in the SM 
\beq \label{eq:gZfgAfghZZ}
g_{Zf}^\pm = \frac{g_1^2 \hspace{0.5mm} Y_f^{\pm} - 2 \hspace{0.25mm} g_2^2 \hspace{0.5mm} T_{f}^{3 \hspace{0.25mm} \pm} }{2 \hspace{0.25mm} g_+} \,, \qquad 
g_{\gamma f}^\pm = - \frac{g_1 g_2 \hspace{0.5mm} Q_f^\pm}{g_+} \,, \qquad 
g_{hZZ} = \frac{v \hspace{0.25mm} g_+^2}{2} \,. 
\eeq 
Notice that these relations are independent of the employed EW input scheme. Here~the symbol $Y_f$ represents the weak hypercharge, $T_f^3$ is the third component of the weak isospin and $Q_f$ denotes the electric charge. The fermions are $f = q, \ell$ with $q = d, u$ and $\ell = e, \nu$, and the helicity states $f_+$ and $f_-$ are identical to the chirality states $f_R$ and $f_L$ in the massless limit. 

\begin{table}
\begin{center}
\begin{tabular}{|c|c|c|c|} 
\hline 
& $g_1$ & $g_2$ & $v$ \\[0.75ex]
\hline \hline
$\begin{matrix} \alpha \hspace{0.125mm}\text{-$\hspace{0.125mm}$scheme} \\ \left \{ \alpha, m_Z, m_W \right \} \end{matrix}$ & \scalebox{1}{$\displaystyle{\sqrt{4 \pi \alpha} \hspace{1mm} \frac{m_Z}{m_W}}$} & \scalebox{1}{$\displaystyle{\sqrt{4 \pi \alpha} \hspace{1mm} \frac{m_Z}{\Delta m}}$} & \scalebox{1}{$\displaystyle{\frac{m_W \Delta m}{\sqrt{\pi \alpha} \hspace{1mm} m_Z}}$} \\[3ex] 
\hline 
$\begin{matrix} \alpha_\mu\hspace{0.125mm}\text{-$\hspace{0.125mm}$scheme} \\ \left \{ G_F, m_Z, m_W \right \} \end{matrix}$ & \rule{0pt}{5ex} \scalebox{1}{$2 \sqrt[4]{2} \hspace{0.25mm} \sqrt{G_F} \hspace{0.25mm} \Delta m$} & \scalebox{1}{$2 \sqrt[4]{2} \hspace{0.25mm} \sqrt{G_F} \hspace{0.5mm} m_W$} & \scalebox{1}{$\displaystyle{\frac{1}{\sqrt[4]{2} \hspace{0.25mm} \sqrt{G_F}}}$} \\[3ex] 
\hline
$\begin{matrix} \text{LEP}\hspace{0.125mm}\text{-$\hspace{0.125mm}$scheme} \\ \left \{ \alpha, G_F, m_Z \right \} \end{matrix}$ & \scalebox{1}{$\displaystyle{\frac{\sqrt{4 \pi \alpha}}{c_w}}$}& \scalebox{1}{$\displaystyle{\frac{\sqrt{4 \pi \alpha}}{s_w}}$} & \scalebox{1}{$\displaystyle{\frac{1}{\sqrt[4]{2} \hspace{0.25mm} \sqrt{G_F}}}$} \\[3ex]
\hline 
\end{tabular}
\end{center}
\vspace{-2mm}
\caption{\label{tab:parameters} The parameters $g_1$, $g_2$ and $v$ expressed in terms of the input parameters for the three EW input schemes implemented in the {\tt POWHEG-BOX} code.}
\end{table}

The relations among the EW input parameters and $g_1$, $g_2$ and $v$ are modified at tree~level by the presence of some of the dimension-six SMEFT~operators listed in~(\ref{eq:operators1}) to~(\ref {eq:operators4}), leading to so-called input scheme corrections. These can be accounted for via the shifts $x \to x + \delta x$ for $x = g_1, g_2, v$. We summarise the relevant shifts in Table~\ref{tab:input_corrections}. The input scheme corrections $\delta g_1$ $\delta g_2$ and $\delta v$ themselves lead to the shifts $\delta g^{(0) \hspace{0.25mm} \pm}_{Zf}$ and $\delta g^{(0)}_{hZZ}$ of the~$Zf\bar f$ and~$hZZ$ couplings, respectively. We find the following scheme-independent results
\bea \label{eq:gZf0ghZZ0}
\begin{split}
\delta g_{Zf}^{(0) \hspace{0.25mm} \pm} & = \frac{g_1^3 \hspace{0.25mm} \delta g_1 Y_{f}^{\pm} - 2 \hspace{0.125mm} g_2^3 \hspace{0.25mm} \delta g_2 \hspace{0.25mm} T_{f}^{3 \hspace{0.25mm} \pm} - g_1^2 \hspace{0.25mm} g_2 \hspace{0.25mm} \delta g_2 \left (Y_f^{\pm} + 4 \hspace{0.25mm} T_{f}^{3 \hspace{0.25mm} \pm} \right ) + 2 g_1 \hspace{0.125mm} g_2^2 \hspace{0.25mm} \delta g_1 \left (Y_f^{\pm} + T_{f}^{3 \hspace{0.25mm} \pm} \right ) }{2 \sqrt[3/2]{g_+}} \,, \hspace{7mm} \\[2mm]
\delta g^{(0)}_{hZZ} & = v \hspace{0.5mm} \big (g_1 \hspace{0.125mm} \delta g_1 + g_2 \hspace{0.125mm} \delta g_2 \big ) \,. 
\end{split}
\eea

At the same time, the SMEFT operators listed in~(\ref{eq:operators1}) to~(\ref {eq:operators4}) give direct contributions to the $Z$-boson couplings to two gauge bosons. We find the following analytic expressions for the non-zero~couplings
\beq \label{eq:ghVV123}
\begin{split}
\delta g_{hZZ}^{(1)} &= \frac{4 \hspace{0,25mm} v}{g_+^2} \, \Big [ \hspace{0.25mm} g_1^2 \hspace{0.5mm} C_ {H\hspace{-0.3mm}B} + g_2^2 \hspace{0.5mm} C_{HW} + g_1 g_2 \hspace{0.5mm} C_{HW\!B} \Big ] \,, \\[2mm]
\delta g_{h\gamma Z}^{(1)} &= \frac{4 \hspace{0.25mm} v}{g_+^2} \left [ \hspace{0.25mm} g_1 g_2 \hspace{0.5mm} C_ {H\hspace{-0.3mm}B} - g_1 g_2 \hspace{0.5mm} C_{HW} - \frac{g_-^2}{2} \hspace{0.5mm} C_{HW\!B} \right ] \,, \\[2mm]
\delta g_{hZZ}^{(3)} & = v^3 \left [ \hspace{0.25mm} g_1 g_2 \hspace{0.25mm} C_{HWB} + \frac{3 \hspace{0.25mm} g_+^2}{8} \hspace{0.5mm} C_{H\Box} + \frac{g_+^2}{2} \hspace{0.5mm} C_{H\hspace{-0.25mm}D} \right ] \,.
\end{split}
\eeq
Furthermore, we obtain $\delta g_{hZZ}^{(2)} = \delta g_{h\gamma Z}^{(2)} = 0$ meaning that the corresponding Dirac structures are not generated at the dimension-six level in the SMEFT. The expressions for the~$hZf\bar{f}$ couplings can finally be written as
\beq \label{eq:ghZf}
\delta g_{hZf}^{(1) \hspace{0.25mm} \pm} = \frac{2 \hspace{0.25mm} \delta g_{Zf}^{(1) \hspace{0.25mm} \pm}}{v} \,,
\eeq
where 
\bea \label{eq:dgZf1}
\begin{split}
\delta g_{Zd}^{(1) \hspace{0.25mm} -} & = \frac{v^2 \hspace{0.25mm} g_+}{2} \left( C_{Hq}^{(1)} + C_{Hq}^{(3)}\right) \,, \qquad 
\delta g_{Zu}^{(1) \hspace{0.25mm} -} = \frac{v^2 \hspace{0.25mm} g_+}{2} \left( C_{Hq}^{(1)} - C_{Hq}^{(3)}\right) \,, \\[2mm]
\delta g_{Ze}^{(1) \hspace{0.25mm} -} & = \frac{v^2 \hspace{0.25mm} g_+}{2} \left( C_{H\ell}^{(1)} + C_{H\ell}^{(3)}\right) \,, \qquad 
\delta g_{Z\nu}^{(1) \hspace{0.25mm} -} = \frac{v^2 \hspace{0.25mm} g_+}{2} \left( C_{H\ell}^{(1)} - C_{H\ell}^{(3)}\right) \,, \\[2mm]
& \hspace{-1.35cm} \delta g_{Zd}^{(1) \hspace{0.25mm} +} = \frac{v^2 g_+}{2} \hspace{0.25mm} C_{Hd} \,, \qquad 
\delta g_{Zu}^{(1) \hspace{0.25mm} +} = \frac{v^2 \hspace{0.25mm} g_+}{2} \hspace{0.25mm} C_{Hu} \,, \qquad 
\delta g_{Ze}^{(1) \hspace{0.25mm} +} = \frac{v^2 \hspace{0.25mm} g_+}{2} \hspace{0.25mm} C_{He} \,, 
\end{split}
\eea
are the relevant direct SMEFT corrections to the $Zf \bar{f}$ couplings. 

\begin{table}
\begin{center}
\begin{tabular}{|c|c|c|c|} 
\hline
 & $\delta g_1/g_1$ & $\delta g_2 / g_2$ & $\delta v / v$ \\[0.75ex]
\hline \hline 
$\begin{matrix} \alpha\hspace{0.125mm}\text{-$\hspace{0.125mm}$scheme} \\ \left \{ \alpha, m_Z, m_W \right \} \end{matrix}$& \scalebox{0.7}{$-\frac{m_W^2 \Delta m^2}{4 \pi \alpha \hspace{0.25mm} m_Z^2} \hspace{0.25mm} C_{H\hspace{-0.25mm}D}$} & \scalebox{0.7}{$\frac{m_W^3 \left( m_W C_{H\hspace{-0.25mm}D} + 4 \hspace{0.25mm} \Delta m \hspace{0.25mm} C_{HW\!B} \right)}{4 \pi \alpha \hspace{0.25mm} m_Z^2}$} & \scalebox{0.7}{$-\frac{m_W^3 \left( m_W C_{H\hspace{-0.25mm}D} + 4 \hspace{0.25mm} \Delta m \hspace{0.25mm} C_{HW\!B} \right)}{4 \pi \alpha \hspace{0.25mm} m_Z^2}$} \\[3ex]
\hline 
$\begin{matrix} \alpha_\mu\hspace{0.125mm}\text{-$\hspace{0.125mm}$scheme} \\ \left \{ G_F, m_Z, m_W \right \} \end{matrix}$ & \rule{0pt}{5ex} \scalebox{0.7}{$-\frac{\frac{m_Z^2 \hspace{0.25mm} C_{H\hspace{-0.25mm}D}}{4 \hspace{0.25mm} \Delta m^2} + C_ {H\ell}^{(3)} - \frac{C_{\ell\ell}}{2} + \frac{m_W \hspace{0.25mm} C_{HW\!B}}{\Delta m}}{\sqrt{2} \hspace{0.25mm} G_F}$} & \scalebox{0.7}{$-\frac{1}{\sqrt{2}\hspace{0.25mm} G_F} \left( C_ {H\ell}^{(3)} - \frac{C_{\ell\ell}}{2} \right)$} & \scalebox{0.7}{$\frac{1}{\sqrt{2} \hspace{0.25mm} G_F} \left( C_ {H\ell}^{(3)} - \frac{C_{\ell\ell}}{2} \right)$} \\[3ex]
\hline
$\begin{matrix} \text{LEP}\hspace{0.125mm}\text{-$\hspace{0.125mm}$scheme} \\ \left \{ \alpha, G_F, m_Z \right \} \end{matrix}$ & \scalebox{0.7}{$\frac{s_w \left[ c_w \hspace{0.25mm} C_{HW\!B} + \frac{s_w \hspace{0.25mm} C_{H\hspace{-0.25mm}D}}{4} + s_w \left ( C_ {H\ell}^{(3)} - \frac{C_{\ell\ell} }{2} \right ) \right]}{\sqrt{2} \hspace{0.25mm} G_F \left (c_w^2-s_w^2 \right )}$}& \scalebox{0.7}{$-\frac{c_w \left[ s_w \hspace{0.25mm} C_{HW\!B} + \frac{c_w \hspace{0,25mm} C_{H\hspace{-0.25mm}D}}{4} + c_w \left ( C_ {H\ell}^{(3)} - \frac{C_{\ell\ell}}{2} \right ) \right ]}{\sqrt{2} \hspace{0.25mm} G_F \left (c_w^2-s_w^2 \right )}$} & \scalebox{0.7}{$ \frac{1}{\sqrt{2} \hspace{0.25mm} G_F} \Big( C_{H\ell}^{(3)} - \frac{C_{\ell\ell}}{2}\Big ) $} \\[3ex]
\hline
\end{tabular}
\end{center}
\vspace{-2mm}
\caption{\label{tab:input_corrections} SMEFT input scheme corrections for the three EW input schemes implemented in our MC code~\cite{GitLabPowheg}.}
\end{table}

\section{SMEFT corrections to $\bm{gg \to Zh}$ process}
\label{app:gginitiated}

Our calculation of $gg \to Zh$ production is based on the spinor-helicity amplitudes for the SM derived in~\cite{Campbell:2016jau} and implemented into~{\tt MCFM}~\cite{Boughezal:2016wmq}. 
In unitary gauge, the expression for the triangle contributions with positive gluon helicities and left-handed fermion chiralities~reads 
\beq \label{eq:A0g2ZHelLplusplustri} 
\begin{split} 
{\cal A}_{\texttt{A0g2Z}\triangle}^q \left( 1_g^+, 2_g^+, 3_\ell^-, 4_{\bar \ell}^+\right) & = -\frac{2 \left [ 2 1 \right ] \big ( \left [ 4 1 \right ] \braket{1 3} + \left [ 4 2 \right ] \braket{2 3} \big)}{\braket{12}} \left(1 - \frac{s_{12}}{m_Z^2} \right) \\[2mm]
& \phantom{xx} \times m_q^2 \hspace{0.5mm} C_0(s_{12},0,0,m_q,m_q,m_q) \,.
\end{split}
\eeq
Notice that we have followed the convention of~\cite{Campbell:2016jau} and written the amplitude for all momenta outgoing. In~(\ref{eq:A0g2ZHelLplusplustri}) the two terms in the last factor in the first line stem from the transversal and longitudinal part of the $Z$-boson propagator in unitary gauge, respectively,~$q$~is the quark running in the loop with mass $m_q$ and $C_0$ is the scalar Passarino-Veltman~(PV) triangle integral defined as in~\cite{Hahn:1998yk,Hahn:2016ebn}. The corresponding~SM~Feynman diagram is displayed on the right-hand side in~Figure~\ref{fig:corrections}. Similarly, we have implemented the box amplitudes
\beq \label{eq:A0g2ZHelLplusplusbox} 
{\cal A}_{\texttt{A0g2Z}\Box}^q\left( 1_g^+, 2_g^+, 3_\ell^-, 4_{\bar \ell}^+\right) \, , \qquad {\cal A}_{\texttt{A0g2Z}\Box}^q\left( 1_g^-, 2_g^+, 3_\ell^-, 4_{\bar \ell}^+\right) \,,
\eeq
which are, however, too lengthy to be reported here but may be inspected in our squared matrix element library~\cite{GitLabAmplitudes}. The remaining non-zero helicity combinations may be obtained via parity and charge conjugation relations. In the case of the triangle contributions, these relations take the form
\beq \label{eq:A0g2ZHel} 
\begin{split}
{\cal A}_{\texttt{A0g2Z}\triangle}^q\left( 1_g^-, 2_g^-,3_\ell^\mp, 4_{\bar \ell}^\pm \right) & = -\overline{{\cal A}_{\texttt{A0g2Z}\triangle}^q\left( 1_g^+, 2_g^+, 4_\ell^\mp, 3_{\bar \ell}^\pm \right)} \,, \\[2mm]
{\cal A}_{\texttt{A0g2Z}\triangle}^q\left( 1_g^\pm, 2_g^\pm,3_\ell^+, 4_{\bar \ell}^- \right) & = {\cal A}_{\texttt{A0g2Z}\triangle}^q\left( 1_g^\pm, 2_g^\pm,4_\ell^-, 3_{\bar \ell}^+ \right) \,,
\end{split}
\eeq
where the overline means that the brackets should be exchanged,~i.e.~$\left [ \ldots \right ] \leftrightarrow \left \langle \ldots \right \rangle$. Analog~relations hold for the box contributions including the cases where the gluons have opposite helicities which are only present for ${\cal A}_{\texttt{A0g2Z}\Box}^q$.

Including the triangle and box contributions the resulting spin-averaged matrix element takes the form 
\beq \label{eq:A0g2Z}
\texttt{A0g2Z} = \frac{\alpha_s^2}{8 \hspace{0.25mm} \pi^2 \hspace{0.5mm} (C_A^2 - 1)^2} \sum_{h_g, h_\ell = \pm} \left | \, \sum_{q=t,b} \hspace{0.5mm} \left( {\cal A}_\triangle^q + \sum_{s = \pm} \hspace{0.25mm} \frac{m_q^2}{m_Z^2} \, {\cal A}_{\Box}^{q,s} \right) \, \right|^2 \,,
\eeq
with
\bea
{\cal A}_\triangle^q & = \displaystyle \frac{ (g_{Zq}^{-} - g_{Zq}^{+}) \hspace{0.5mm} g_{Z\ell}^{h_\ell} \hspace{0.75mm} g_{hZZ}}{D_Z (s_{12}) \hspace{0.5mm} D_Z (s_{34})} \; {\cal A}_{\texttt{A0g2Z}\triangle}^q \left( 1_g^{h_g}, 2_g^{h_g} , 3_\ell^{h_\ell}, 4_{\bar \ell}^{-h_\ell} \right) \hspace{0.25mm} \,, \label{eq:tri} \\[2mm] 
{\cal A}_{\Box}^{q,\pm} & = \displaystyle \frac{(g_{Zq}^{-} - g_{Zq}^{+}) \hspace{0.5mm} g_{Z\ell}^{h_\ell} \hspace{0.75mm} g_{hZZ}}{D_Z (s_{34})} \; {\cal A}_{\texttt{A0g2Z}\Box}^q \left( 1_g^{h_g}, 2_g^{\pm h_g} , 3_\ell^{h_\ell}, 4_{\bar \ell}^{-h_\ell} \right) \hspace{0.25mm} \,. \label{eq:box}
\eea 
Here $D_Z(s)$ has been defined in~(\ref{eq:DZ}) while the expressions for the couplings $g_{Zf}^{\pm}$ and $g_{hZZ}$ can be found in~(\ref{eq:gZfgAfghZZ}). The coupling $g_{hZZ}$ appearing in~(\ref{eq:box}) requires some explanation. In fact, the box contributions do not involve a $hZZ$ vertex but instead the Higgs boson couples directly to the quarks. However, since 
\beq \label{eq:magic}
g_{hZZ} \, \frac{m_q^2}{m_Z^2} = \frac{v \hspace{0.25mm} \big ( g_1^2+g_2^2 \big )}{2} \, \frac{m_q^2}{m_Z^2} = \frac{2 \hspace{0.125mm} m_Z^2}{v} \, \frac{m_q^2}{m_Z^2} = \frac{2 \hspace{0.125mm} m_q^2}{v} \,,
\eeq
with a factor $m_q/v$ coming from the $h q\bar{q}$ vertex and another $m_q$ stemming from the mass insertion in the box diagram the expected mass dependence for ${\cal A}_{\texttt{A0g2Z}\Box}^q$ is recovered.

\begin{figure}[t!]
\begin{center}
\includegraphics[width=\textwidth]{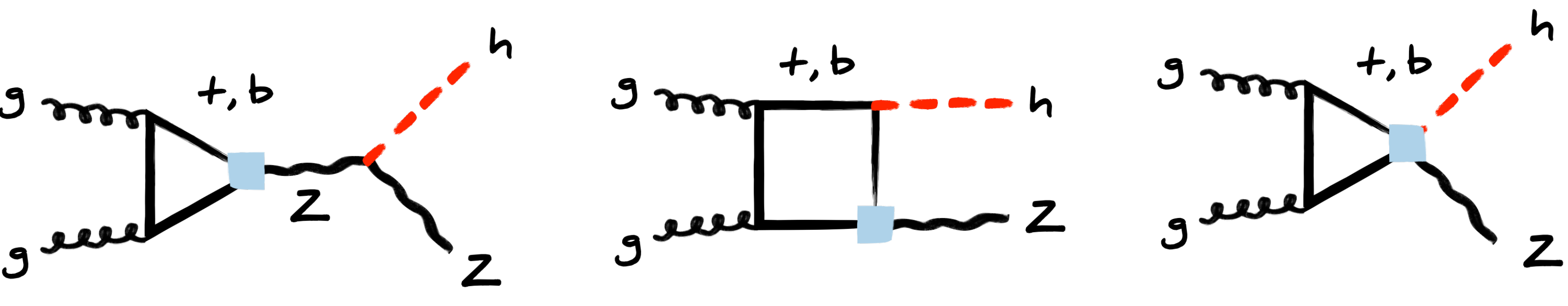}
\end{center}
\vspace{-2mm} 
\caption{\label{fig:anomalous} Examples of contributions to $gg \to Zh$ production within the SMEFT. All~graphs involve an insertion of one of the operators given in~(\ref{eq:operators2}) as indicated by the blue squares. Further details can be found in the main text.}
\end{figure} 

It is important to realise that as a result of the generalised Furry theorem the vector-current coupling of the $Z$ boson, which is proportional to the combination $(g_{Zq}^- + g_{Zq}^+)$ of couplings, does not contribute to the spin-averaged matrix element $\texttt{A0g2Z}$ as given~in~(\ref{eq:A0g2Z}). However, the axial-current part contributes, as~signalled by the factor $(g_{Zq}^- - g_{Zq}^+)$ in both~(\ref{eq:tri}) and~(\ref{eq:box}), and this contribution is directly connected to the $U(1)_A \times SU(3)_c$ gauge anomaly. In fact, a regulator and a loop routing scheme must be introduced to properly define the amplitude ${\cal A}_{\texttt{A0g2Z}\triangle}^q$, rendering its expression scheme-dependent --- for~a~detailed explanation of this point see~for~example~\cite{Fox:2018ldq}. Within the SM, the axial parts of the top- and bottom-quark couplings obey 
\beq \label{eq:SManomalyfree}
(g_{Zt}^{-} - g_{Zt}^{+}) = - (g_{Zb}^{-} - g_{Zb}^{+})\,,
\eeq
and as a result all gauge anomalies are cancelled. It follows that the sum over $q$ that appears in~(\ref{eq:A0g2Z}) evaluates to 
\beq \label{eq:schemeindependenceSM} 
\sum_{q=t,b} \hspace{0.5mm} (g_{Zq}^{-} - g_{Zq}^{+}) \hspace{0.5mm} {\cal A}_{\texttt{A0g2Z}\triangle}^q = (g_{Zt}^{-} - g_{Zt}^{+}) \, \left ({\cal A}_{\texttt{A0g2Z}\triangle}^t - {\cal A}_{\texttt{A0g2Z}\triangle}^b \right ) \,, 
\eeq
and in consequence any scheme-dependent constant shift in the amplitude ${\cal A}_{\texttt{A0g2Z}\triangle}^q$ drops out in the combination $\big({\cal A}_{\texttt{A0g2Z}\triangle}^t - {\cal A}_{\texttt{A0g2Z}\triangle}^b\big)$. Notice that in the degenerate or zero mass case the sum~(\ref{eq:schemeindependenceSM}) vanishes identically. Since we treat the light-quark generations as massless, down-, up-, strange- and charm-quark loops hence do not need to be included in the spin-averaged matrix element~(\ref{eq:A0g2Z}).

The amplitudes including the SMEFT contributions to~(\ref {eq:A0g2Z}) were computed with the procedure outlined in Section \ref{sec:SMEFTfixedorder}. Since the SM amplitudes were derived in unitary gauge, only SMEFT contributions to vertices involving the $Z$ boson have to be considered. We have checked explicitly that in Feynman gauge, the SMEFT effects in the Goldstone diagrams are equivalent to the effects in the longitudinal part of the amplitude in unitary~gauge. In addition to the contributions with an effective $Z q\bar{q}$ or $Z \ell^+ \ell^-$ vertex represented by the diagram on the left in Figure~\ref{fig:anomalous}, there are also contributions with $hZ q\bar{q}$ or $hZ \ell^+ \ell^-$ vertices. A corresponding graph is depicted on the right in Figure~\ref{fig:anomalous}. In these cases only the transversal part of~(\ref {eq:A0g2ZHelLplusplustri}) contributes --- the longitudinal part vanishes because the $Z$ boson couples directly to the leptons that are treated as massless --- and therefore in addition to dropping the factor~$D_Z(s_{12})$ in~(\ref{eq:tri}), one also has to discard the longitudinal part in~(\ref{eq:A0g2ZHelLplusplustri}) by removing the factor $(1-s_{12}/m_Z^2)$. This leads to the following contribution
\bea \label{eq:SMEFTquartic} 
\left [ \frac{\left(\delta g_{hZq}^{(1) \hspace{0.25mm} -}-\delta g_{hZq}^{(1) \hspace{0.25mm} +} \right) g_{Z\ell}^{h_\ell}}{D_Z(s_{34})} +\frac{\left(g_{Zq}^{-} - g_{Zq}^{+} \right) \delta g_{hZ \ell}^{(1) \hspace{0.25mm} h_\ell} }{D_Z(s_{12})} \right ] \, \frac{{\cal A}_{\texttt{A0g2Z}\triangle}^q\left( 1_g^{h_g}, 2_g^{h_g} , 3_\ell^{h_\ell}, 4_{\bar \ell}^{-h_\ell} \right)}{1-\frac{s_{12}}{m_Z^2}} \,, \hspace{8mm}
\eea
from SMEFT diagrams with a $hZ q\bar{q}$ or $hZ \ell^+ \ell^-$ vertex. This contribution can be included by simply adding the expression~(\ref{eq:SMEFTquartic}) to the sum over $q$ in~(\ref{eq:A0g2Z}).

The triangle contributions with a $Z q\bar{q}$ or a $hZ q\bar{q}$ vertex depicted in Figure~\ref{fig:anomalous} deserve further discussion. In fact, in their sum these contributions cancel exactly~\cite{Rossia:2023hen}, which is an interesting feature of the SMEFT. To explicitly see this cancellation we rewrite the SMEFT~$Z q\bar{q}$ contribution to~(\ref{eq:tri}) in the following way 
\beq \label{eq:rewrite}
\begin{split}
\mathcal{A}^q_\triangle &\propto \frac{\left( \delta g_{Zq}^{(1)-} - \delta g_{Zq}^{(1)+}\right) \hspace{0.5mm} g_{Z\ell}^{h_\ell} \hspace{0.75mm} g_{hZZ}}{D_Z(s_{12}) D_Z(s_{34})} = \frac{\frac{v}{2} \left( \delta g_{hZq}^{(1)-} - \delta g_{hZq}^{(1)+}\right) \hspace{0.5mm} g_{Z\ell}^{h_\ell} \hspace{0.75mm} \frac{2 m_Z^2}{v}}{\left (s_{12} - m_Z^2 \right ) D_Z(s_{34})} \\[2mm]
& = -\frac{\left( \delta g_{hZq}^{(1)-} - \delta g_{hZq}^{(1)+}\right) \hspace{0.5mm} g_{Z\ell}^{h_\ell} \hspace{0.75mm}}{D_Z(s_{34})} \, \frac{1}{1-\frac{s_{12}}{m_Z^2}} \,.
\end{split}
\eeq
Here we have used~(\ref{eq:DZ}), (\ref{eq:gZfgAfghZZ}) and~(\ref{eq:ghZf}) in the first step. Notice that the final result in~(\ref{eq:rewrite}) is up to an overall sign and the amplitude ${\cal A}_{\texttt{A0g2Z}\triangle}^q$ equal to the first term in~(\ref{eq:SMEFTquartic}) which proves the cancellation. For simplicity we have treated $m_Z^2$ as real here, however, the discussion does not change if one replaces it by its complex counterpart $m_Z^2 - i \hspace{0.25mm} m_Z \hspace{0.125mm} \Gamma_Z$ in both $D_Z(s_{12})$ and $g_{hZZ}$. The only contributions that remain for the operators in $(\ref{eq:operators2})$ are therefore the box contributions shown in the middle of Figure~\ref{fig:anomalous}. Note that for the operators in $(\ref{eq:operators1})$ both the triangle and box diagrams are non-vanishing.

Notice that the cancellation of the triangle contributions in $pp \to Zh$ production guarantees that both relevant and irrelevant anomalous contributions depending on the Wilson coefficients of the operators~(\ref{eq:operators2}) automatically annul. In fact, it can be shown~\cite{Bonnefoy:2020tyv,Feruglio:2020kfq,Cornella:2022hkc,Cohen:2023gap,Cohen:2023hmq} that the cancellation of relevant anomalous contributions is a general feature of the SMEFT, while the cancellation of irrelevant terms can always be achieved by adding an appropriate local counterterm,~i.e.~a Wess-Zumino term~\cite{Wess:1971yu}, to the SMEFT Lagrangian. As a result, the condition for the cancellation of relevant gauge anomalies in the SMEFT is the same as in the SM and only dependent on the gauge quantum numbers of the fermionic sector,~as~one would naively expect from an effective field theory point of view. The observed cancellation between the triangle contribution with a $Zq \bar q$ and a $hZq \bar q$ vertex hence implies that one does not need to introduce a Wess-Zumino term to obtain a scheme-independent expression for the $gg \to Zh$ amplitudes in the SMEFT.

We finally note that the amplitude for the generalised neutral current proportional to $\delta g_{hZZ}^{(1)}$ as given in~(\ref {eq:new_structures}) vanishes in \texttt{A0g2Z}. Also $\delta g_{h\gamma Z}^{(1)}$ and $\delta g_{h\gamma Z}^{(2)}$ have no effect, since the photon couples vectorially to the quark loop. Only $\delta g_{hZZ}^{(3)}$ as given in~(\ref {eq:ghVV123}) and the corresponding SMEFT operators contribute to $gg \to Zh$ production. This contribution is however not anomalous and hence needs no special treatment. Let us finally mention that we have used {\tt OpenLoops~2}~\cite{Buccioni:2019sur} as well as the implementation {\tt SMEFT@NLO}~\cite{Degrande:2020evl} together with {\tt MadGraph5\_aMC@NLO}~\cite{Alwall:2014hca} to cross check the results presented in this appendix.

\section{SMEFT effects at \NLOplusPSbm~and~\NNLOplusPSbm}
\label{app:NLOPSvsNNLOPS}

NLO QCD correction to $Vh$ production in the SMEFT have been calculated by several groups~\cite{Mimasu:2015nqa,Degrande:2016dqg,Alioli:2018ljm,Bishara:2020vix,Bishara:2020pfx,Bishara:2022vsc}. By now these computations can also be performed automatically by means of the combination of {\tt SMEFT@NLO} and~{\tt MadGraph5\_aMC@NLO}. In what follows, we will use the {\tt POWHEG-BOX} implementation of $pp \to Zh \to \ell^+ \ell^- h$ production presented in~\cite{Alioli:2018ljm} to obtain the relevant \NLOplusPS~predictions. Our physics analysis proceeds as described in the first paragraph of Section~\ref{sec:numerics}. 

\begin{figure}[!t]
\begin{center}
\includegraphics[height=0.575\textwidth]{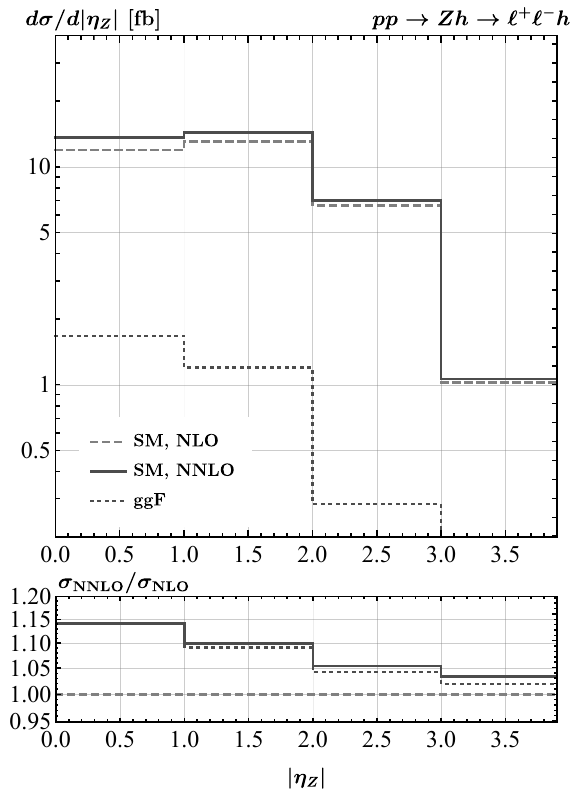} \qquad \quad 
\includegraphics[height=0.575\textwidth]{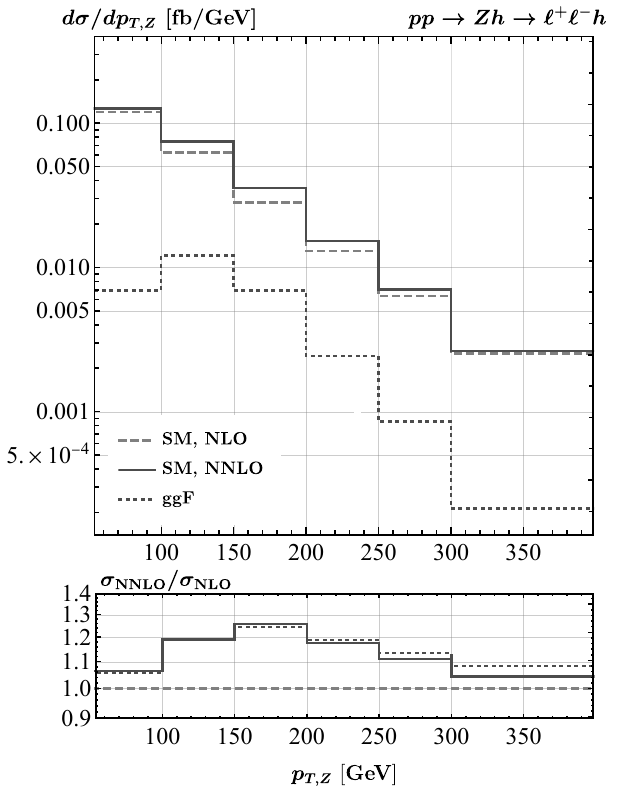}

\vspace{4mm}

\includegraphics[height=0.575\textwidth]{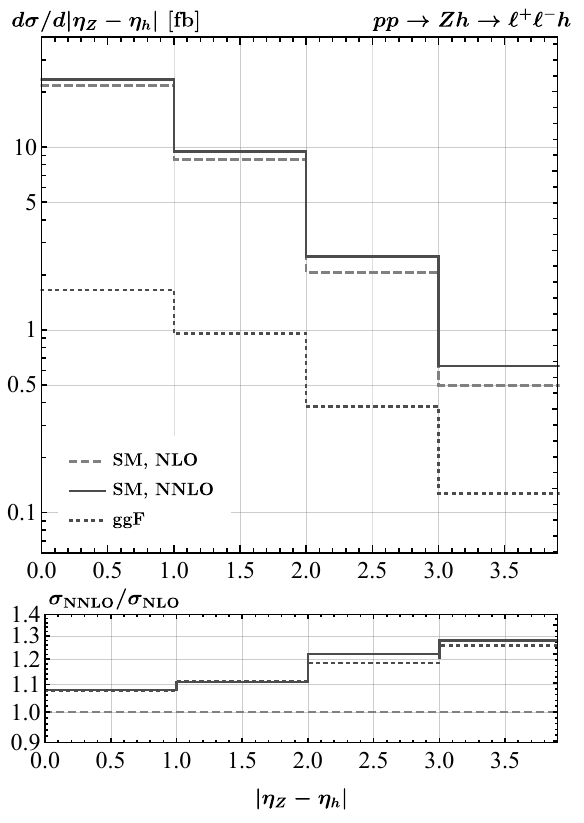} \qquad \quad 
\includegraphics[height=0.575\textwidth]{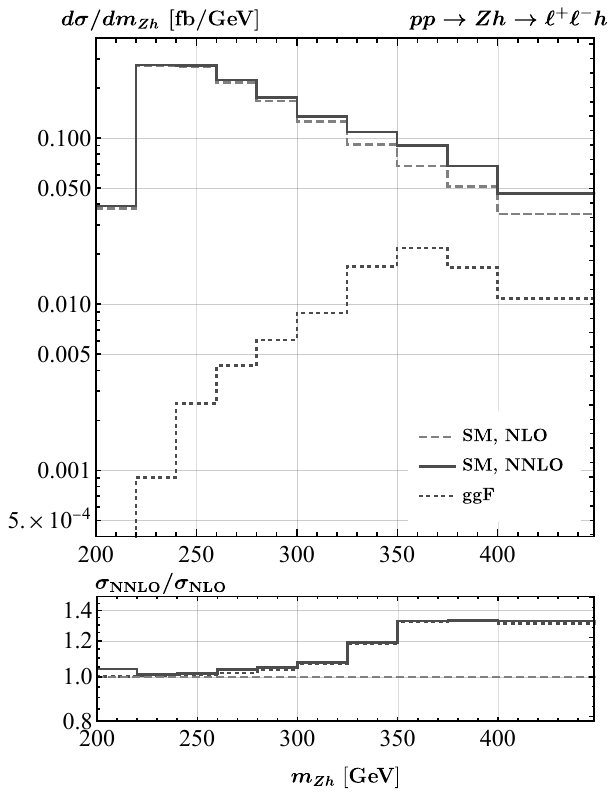}\end{center}
\vspace{-2mm} 
\caption{\label{fig:comp1} SM~\NLOplusPS~and~\NNLOplusPS~results for $pp \to Zh \to \ell^+ \ell^- h$ production. The~$|\eta_Z|$~(upper~left), $p_{T,Z}$~(upper~right), $|\eta_{Z}-\eta_{h}|$~(lower~left) and $m_{Zh}$~(lower~right) spectra are shown. The dashed (solid) lines illustrate the \NLOplusPS~(\NNLOplusPS)~results, while the dotted curves are the \ggF~\NNLOplusPS~corrections. The solid (dotted) lines in the lower panels depict the ratios between the full~\NNLOplusPS~(\NLOplusPS~plus \ggF~\NNLOplusPS) and the \NLOplusPS~results. See main text for further explanations.}
\end{figure}

\begin{figure}[!t]
\begin{center}
\includegraphics[height=0.575\textwidth]{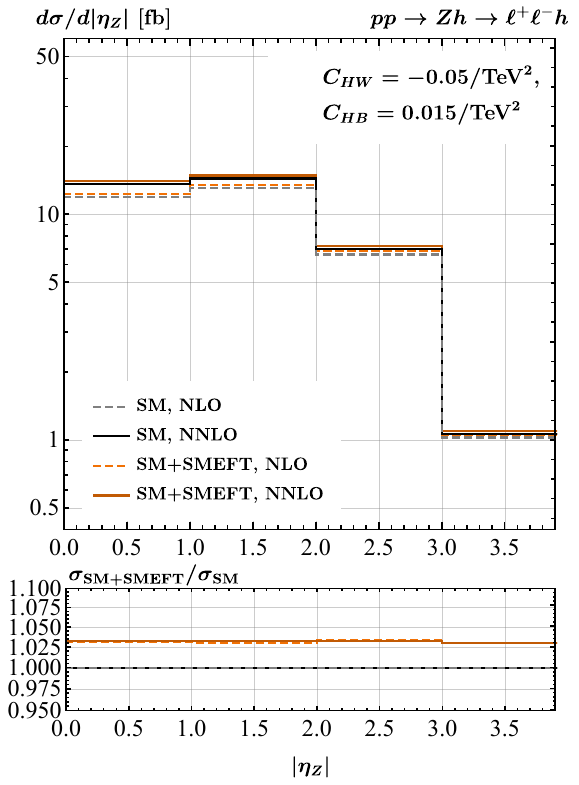} \qquad \quad 
\includegraphics[height=0.575\textwidth]{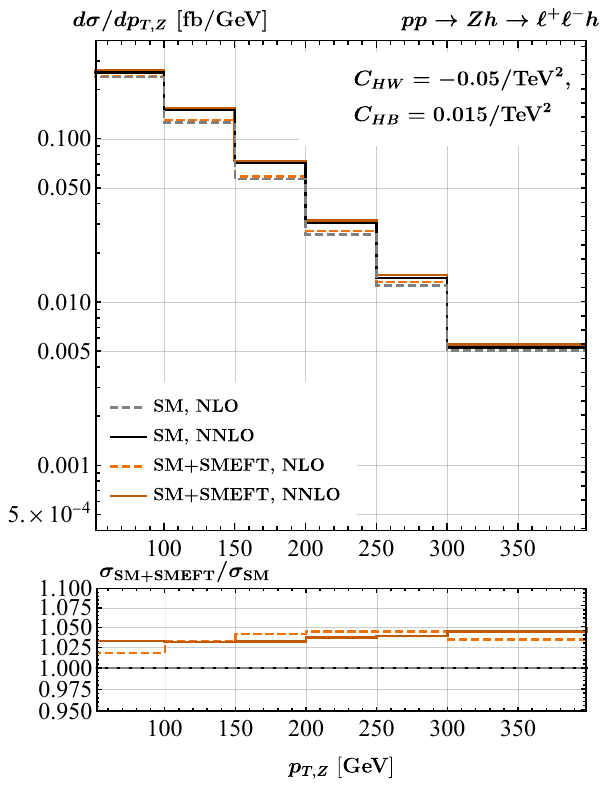}

\vspace{4mm}

\includegraphics[height=0.575\textwidth]{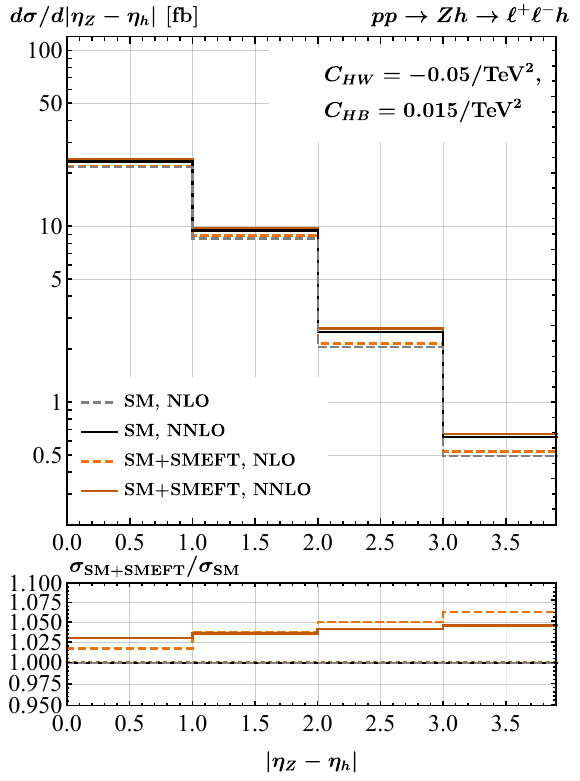} \qquad \quad 
\includegraphics[height=0.575\textwidth]{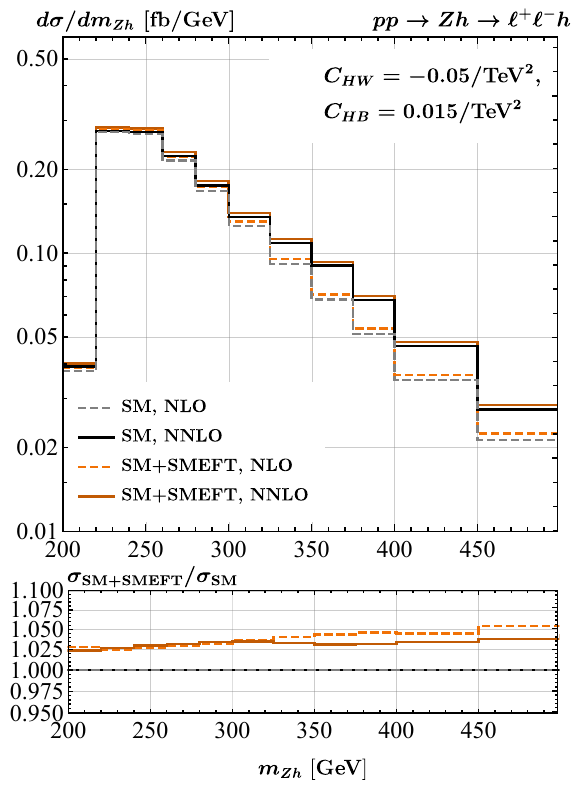}
\end{center}
\vspace{-2mm} 
\caption{\label{fig:comp2} As Figure~\ref{fig:comp1} but for benchmark scenario~(\ref{eq:bench1}). The curves called SM+SMEFT correspond to the full squared matrix elements including the sum of both the SM and SMEFT contributions. The lower panels show the ratios between the SM+SMEFT and the SM predictions at the same order in QCD. For more details consult main text.}
\end{figure}

\begin{figure}[!t]
\begin{center}
\includegraphics[height=0.575\textwidth]{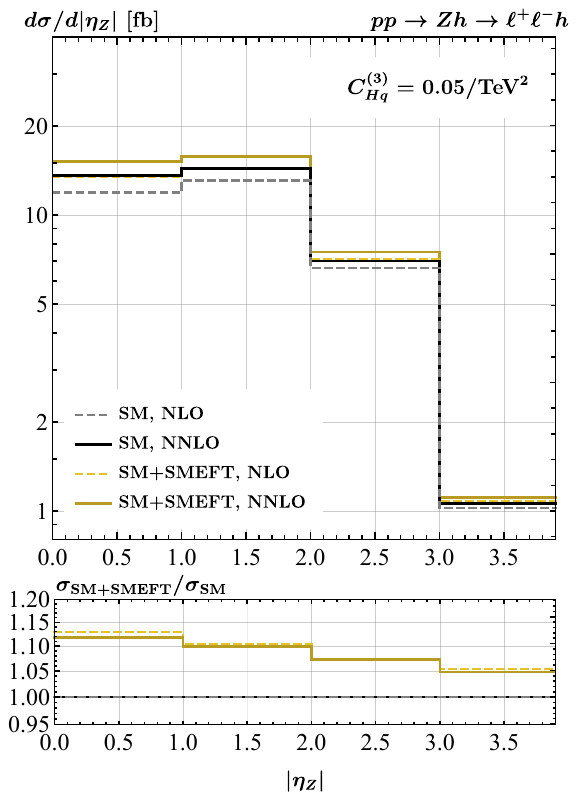} \qquad \quad 
\includegraphics[height=0.575\textwidth]{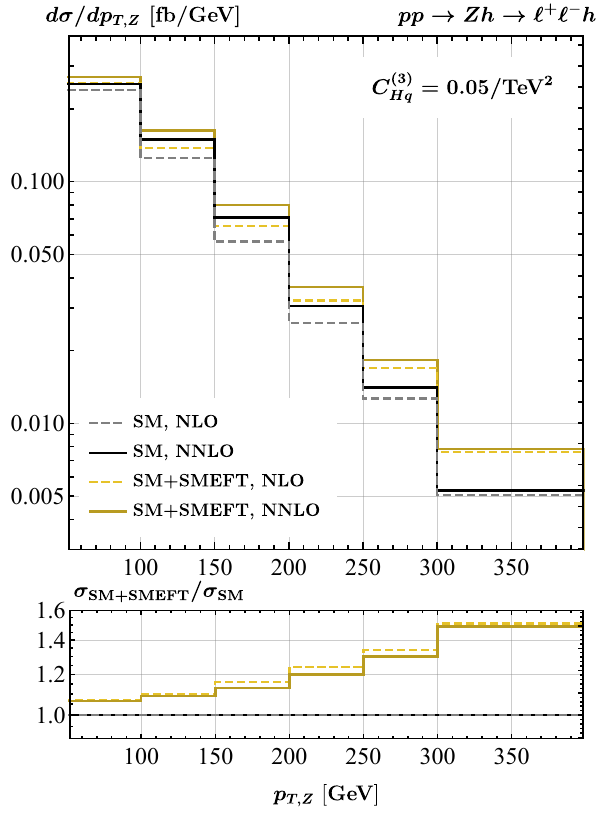}

\vspace{4mm}

\includegraphics[height=0.575\textwidth]{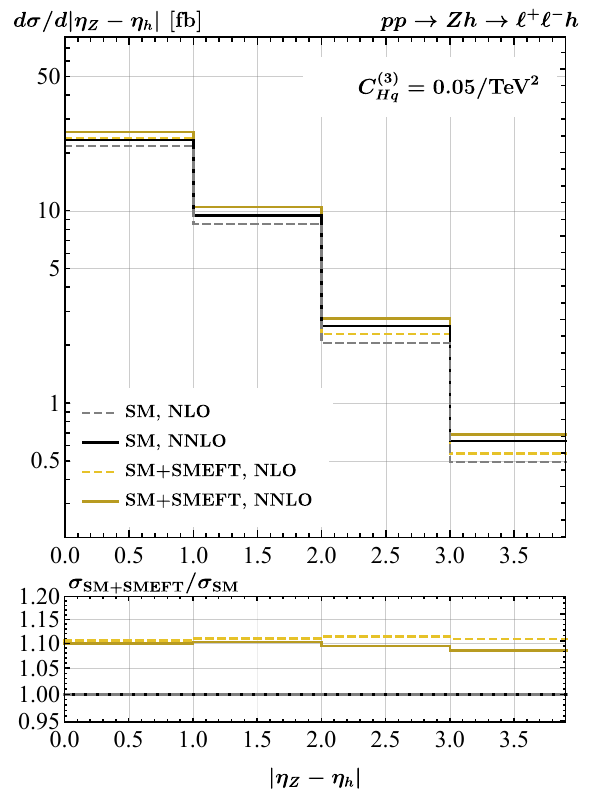} \qquad \quad 
\includegraphics[height=0.575\textwidth]{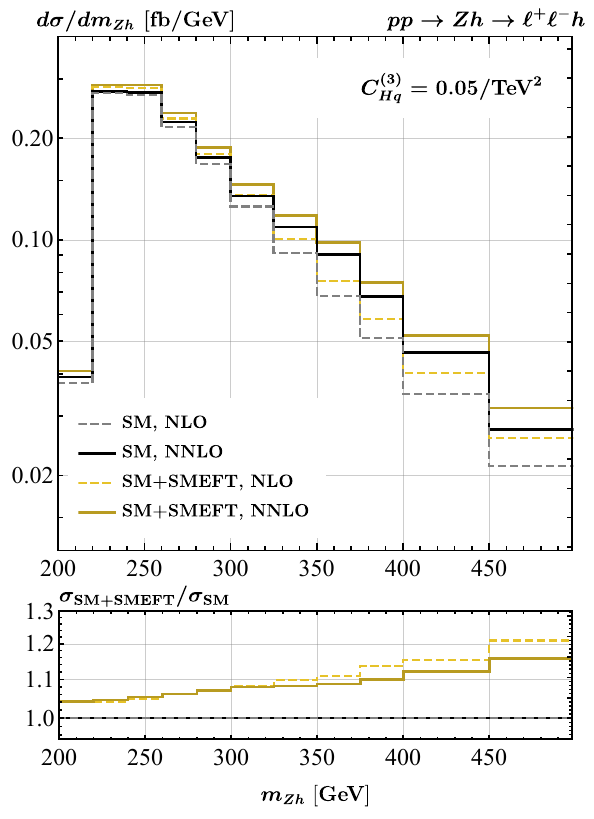}
\end{center}
\vspace{-2mm} 
\caption{\label{fig:comp3} As Figure~\ref{fig:comp2} but for benchmark scenario~(\ref{eq:bench3}).}
\end{figure}

In Figure~\ref{fig:comp1} we compare the SM predictions for the $|\eta_Z|$~(upper~left), $p_{T,Z}$~(upper~right), $|\eta_{Z}-\eta_{h}|$~(lower~left) and $m_{Zh}$~(lower~right) distribution in $pp \to Zh \to \ell^+ \ell^- h$ production obtained at \NLOplusPS~and \NNLOplusPS, respectively. The~dashed (solid) lines correspond to the full~\NLOplusPS~(\NNLOplusPS)~results, while the dotted curves depict the \ggF~contributions that start to contribute at~\NNLOplusPS. From the displayed results it is evident that the NNLO corrections modify the NLO spectra in a non-trivial fashion. The relative size of the NNLO corrections amounts to less than $15\%$ in the $|\eta_Z|$ spectrum, while in the case of the $p_{T,Z}$, $|\eta_Z-\eta_h|$ and $m_{Zh}$ distributions the effects can reach up to around~$30\%$. Notice that for the $p_{T,Z}$ spectrum the NNLO corrections are most pronounced in the vicinity of $p_{T,Z} \simeq m_t$, while in the case of $|\eta_Z-\eta_h|$ and $m_{Zh}$ the largest corrections arise in the tail of the distribution for $|\eta_Z-\eta_h| \gtrsim 3$~and $m_{Zh} \gtrsim 350 \, {\rm GeV}$, respectively. The enhancement of the $p_{T,Z}$ spectrum at $p_{T,Z} \simeq m_t$ is related to the fact that for such transverse momenta the $Z$ boson is able to resolve the top-quark loop in the graph displayed on the right in~Figure~\ref{fig:corrections}. In fact, another feature that is apparent from the solid and dashed lines in the lower panels is that within the SM the \ggF~NNLO effects are in general significantly larger than the \qqF~NNLO counterparts. 

The~\NLOplusPS~and~\NNLOplusPS~predictions corresponding to the SMEFT benchmark scenario~(\ref{eq:bench1}) and~(\ref{eq:bench3}) are given in Figure~\ref{fig:comp2} and Figure~\ref{fig:comp3}, respectively. The dashed grey (solid black) histograms correspond to the~\NLOplusPS~(\NNLOplusPS) results in the SM, while the dashed (solid) coloured results are the corresponding SM+SMEFT predictions. One observes that while the \NLOplusPS~and \NNLOplusPS~results for the full predictions involving the squared matrix elements including the sum of both the SM and SMEFT contributions are notably different, the ratios between the SM+SMEFT and SM results turn out to be essentially independent on whether they are calculated at NLO or NNLO. In~order to~understand this feature one has to recall that in the SM the dominant NNLO corrections to $pp \to Z h \to \ell^+ \ell^- h$ production arise from the $gg \to Zh$ channel, while the NNLO corrections associated to the $q\bar q \to Zh$ and $q g \to Zh$ channels are small. The opposite is the case in the SMEFT, where effects stemming from the $gg \to Zh$ channel are suppressed compared to the SM as a result of the cancellation of triangle contributions discussed in~Appendix~\ref{app:gginitiated}. We add that the comparisons of SM+SMEFT predictions present in this appendix represent a non-trivial validation~of our new \NNLOplusPS~MC~code for Higgsstrahlung~\cite{GitLabPowheg}. 

\end{appendix}


\begin{thebibliography}{10}

\bibitem{ATLAS:2018kot}
{\scshape ATLAS} collaboration, \emph{{Observation of $H \rightarrow b\bar{b}$
 decays and $VH$ production with the ATLAS detector}},
 \href{https://doi.org/10.1016/j.physletb.2018.09.013}{\emph{Phys. Lett. B}
 {\bfseries 786} (2018) 59}
 [\href{https://arxiv.org/abs/1808.08238}{{\ttfamily 1808.08238}}].

\bibitem{CMS:2018nsn}
{\scshape CMS} collaboration, \emph{{Observation of Higgs boson decay to bottom
 quarks}}, \href{https://doi.org/10.1103/PhysRevLett.121.121801}{\emph{Phys.
 Rev. Lett.} {\bfseries 121} (2018) 121801}
 [\href{https://arxiv.org/abs/1808.08242}{{\ttfamily 1808.08242}}].

\bibitem{ATLAS:2018jlh}
{\scshape ATLAS} collaboration, \emph{{Projections for measurements of Higgs
 boson cross sections, branching ratios, coupling parameters and mass with the
 ATLAS detector at the HL-LHC}}, \href{https://cds.cern.ch/record/2652762}{ATL-PHYS-PUB-2018-054}.

\bibitem{CMS:2018qgz}
{\scshape CMS} collaboration, \emph{{Sensitivity projections for Higgs boson
 properties measurements at the HL-LHC}}, \href{https://cds.cern.ch/record/2647699}{CMS-PAS-FTR-18-011}.

\bibitem{Buchmuller:1985jz}
W.~Buchm{\"u}ller and D.~Wyler, \emph{{Effective Lagrangian Analysis of New
 Interactions and Flavor Conservation}},
 \href{https://doi.org/10.1016/0550-3213(86)90262-2}{\emph{Nucl. Phys.}
 {\bfseries B268} (1986) 621}.

\bibitem{Grzadkowski:2010es}
B.~Grzadkowski, M.~Iskrzynski, M.~Misiak and J.~Rosiek, \emph{{Dimension-Six
 Terms in the Standard Model Lagrangian}},
 \href{https://doi.org/10.1007/JHEP10(2010)085}{\emph{JHEP} {\bfseries 10}
 (2010) 085} [\href{https://arxiv.org/abs/1008.4884}{{\ttfamily 1008.4884}}].

\bibitem{Brivio:2017vri}
I.~Brivio and M.~Trott, \emph{{The Standard Model as an Effective Field
 Theory}}, \href{https://doi.org/10.1016/j.physrep.2018.11.002}{\emph{Phys.
 Rept.} {\bfseries 793} (2019) 1}
 [\href{https://arxiv.org/abs/1706.08945}{{\ttfamily 1706.08945}}].

\bibitem{Mimasu:2015nqa}
K.~Mimasu, V.~Sanz and C.~Williams, \emph{{Higher Order QCD predictions for
 Associated Higgs production with anomalous couplings to gauge bosons}},
 \href{https://doi.org/10.1007/JHEP08(2016)039}{\emph{JHEP} {\bfseries 08}
 (2016) 039} [\href{https://arxiv.org/abs/1512.02572}{{\ttfamily
 1512.02572}}].

\bibitem{Degrande:2016dqg}
C.~Degrande, B.~Fuks, K.~Mawatari, K.~Mimasu and V.~Sanz, \emph{{Electroweak
 Higgs boson production in the standard model effective field theory beyond
 leading order in QCD}},
 \href{https://doi.org/10.1140/epjc/s10052-017-4793-x}{\emph{Eur. Phys. J. C}
 {\bfseries 77} (2017) 262}
 [\href{https://arxiv.org/abs/1609.04833}{{\ttfamily 1609.04833}}].

\bibitem{Alioli:2018ljm}
S.~Alioli, W.~Dekens, M.~Girard and E.~Mereghetti, \emph{{NLO QCD corrections
 to SM-EFT dilepton and electroweak Higgs boson production, matched to parton
 shower in POWHEG}},
 \href{https://doi.org/10.1007/JHEP08(2018)205}{\emph{JHEP} {\bfseries 08}
 (2018) 205} [\href{https://arxiv.org/abs/1804.07407}{{\ttfamily
 1804.07407}}].

\bibitem{Bishara:2020vix}
F.~Bishara, P.~Englert, C.~Grojean, M.~Montull, G.~Panico and A.N.~Rossia,
 \emph{{A New Precision Process at FCC-hh: the diphoton leptonic $Wh$ channel}},
 \href{https://doi.org/10.1007/JHEP07(2020)075}{\emph{JHEP} {\bfseries 07}
 (2020) 075} [\href{https://arxiv.org/abs/2004.06122}{{\ttfamily
 2004.06122}}].

\bibitem{Bishara:2020pfx}
F.~Bishara, S.~De~Curtis, L.~Delle~Rose, P.~Englert, C.~Grojean, M.~Montull
 et~al., \emph{{Precision from the diphoton $Zh$ channel at FCC-hh}},
 \href{https://doi.org/10.1007/JHEP04(2021)154}{\emph{JHEP} {\bfseries 04}
 (2021) 154} [\href{https://arxiv.org/abs/2011.13941}{{\ttfamily
 2011.13941}}].

\bibitem{Bishara:2022vsc}
F.~Bishara, P.~Englert, C.~Grojean, G.~Panico and A.N.~Rossia,
 \emph{{Revisiting $Vh(\to b \bar b)$ at the LHC and
 FCC-hh}}, \href{https://doi.org/10.1007/JHEP06(2023)077}{\emph{JHEP}
 {\bfseries 06} (2023) 077}
 [\href{https://arxiv.org/abs/2208.11134}{{\ttfamily 2208.11134}}].

\bibitem{Gauld:2015lmb}
R.~Gauld, B.D.~Pecjak and D.J.~Scott, \emph{{One-loop corrections to $h\to
 b\bar b$ and $h\to \tau\bar \tau$ decays in the Standard Model Dimension-6
 EFT: four-fermion operators and the large-$m_t$ limit}},
 \href{https://doi.org/10.1007/JHEP05(2016)080}{\emph{JHEP} {\bfseries 05}
 (2016) 080} [\href{https://arxiv.org/abs/1512.02508}{{\ttfamily
 1512.02508}}].

\bibitem{Gauld:2016kuu}
R.~Gauld, B.D.~Pecjak and D.J.~Scott, \emph{{QCD radiative corrections for
 $h\to b\bar b$ in the Standard Model Dimension-6 EFT}},
 \href{https://doi.org/10.1103/PhysRevD.94.074045}{\emph{Phys. Rev. D}
 {\bfseries 94} (2016) 074045}
 [\href{https://arxiv.org/abs/1607.06354}{{\ttfamily 1607.06354}}].

\bibitem{Cullen:2019nnr}
J.M.~Cullen, B.D.~Pecjak and D.J.~Scott, \emph{{NLO corrections to $h\to b\bar
 b$ decay in SMEFT}},
 \href{https://doi.org/10.1007/JHEP08(2019)173}{\emph{JHEP} {\bfseries 08}
 (2019) 173} [\href{https://arxiv.org/abs/1904.06358}{{\ttfamily
 1904.06358}}].

\bibitem{Cullen:2020zof}
J.M.~Cullen and B.D.~Pecjak, \emph{{Higgs decay to fermion pairs at NLO in
 SMEFT}}, \href{https://doi.org/10.1007/JHEP11(2020)079}{\emph{JHEP}
 {\bfseries 11} (2020) 079}
 [\href{https://arxiv.org/abs/2007.15238}{{\ttfamily 2007.15238}}].

\bibitem{Haisch:2022nwz}
U.~Haisch, D.J.~Scott, M.~Wiesemann, G.~Zanderighi and S.~Zanoli, \emph{{NNLO
 event generation for $pp \to Zh\to \ell^+ \ell^- b \bar b$ production in the SM
 effective field theory}},
 \href{https://doi.org/10.1007/JHEP07(2022)054}{\emph{JHEP} {\bfseries 07}
 (2022) 054} [\href{https://arxiv.org/abs/2204.00663}{{\ttfamily
 2204.00663}}].

\bibitem{Maltoni:2013sma}
F.~Maltoni, K.~Mawatari and M.~Zaro, \emph{{Higgs characterisation via
 vector-boson fusion and associated production: NLO and parton-shower
 effects}}, \href{https://doi.org/10.1140/epjc/s10052-013-2710-5}{\emph{Eur.
 Phys. J. C} {\bfseries 74} (2014) 2710}
 [\href{https://arxiv.org/abs/1311.1829}{{\ttfamily 1311.1829}}].

\bibitem{Greljo:2017spw}
A.~Greljo, G.~Isidori, J.M.~Lindert, D.~Marzocca and H.~Zhang,
 \emph{{Electroweak Higgs production with HiggsPO at NLO QCD}},
 \href{https://doi.org/10.1140/epjc/s10052-017-5422-4}{\emph{Eur. Phys. J. C}
 {\bfseries 77} (2017) 838}
 [\href{https://arxiv.org/abs/1710.04143}{{\ttfamily 1710.04143}}].

\bibitem{Bizon:2021rww}
W.~Bizo\'n, F.~Caola, K.~Melnikov and R.~R\"ontsch, \emph{{Anomalous couplings
 in associated $VH$ production with Higgs boson decay to massive $b$ quarks at
 NNLO in QCD}}, \href{https://doi.org/10.1103/PhysRevD.105.014023}{\emph{Phys.
 Rev. D} {\bfseries 105} (2022) 014023}
 [\href{https://arxiv.org/abs/2106.06328}{{\ttfamily 2106.06328}}].

\bibitem{Alioli:2010xd}
S.~Alioli, P.~Nason, C.~Oleari and E.~Re, \emph{{A general framework for
 implementing NLO calculations in shower Monte Carlo programs: the POWHEG
 BOX}}, \href{https://doi.org/10.1007/JHEP06(2010)043}{\emph{JHEP} {\bfseries
 06} (2010) 043} [\href{https://arxiv.org/abs/1002.2581}{{\ttfamily
 1002.2581}}].

\bibitem{Monni:2019whf}
P.F.~Monni, P.~Nason, E.~Re, M.~Wiesemann and G.~Zanderighi,
 \emph{{MiNNLO$_{PS}$: a new method to match NNLO QCD to parton showers}},
 \href{https://doi.org/10.1007/JHEP05(2020)143}{\emph{JHEP} {\bfseries 05}
 (2020) 143} [\href{https://arxiv.org/abs/1908.06987}{{\ttfamily
 1908.06987}}].

\bibitem{Monni:2020nks}
P.F.~Monni, E.~Re and M.~Wiesemann, \emph{{MiNNLO$_{\text {PS}}$: optimizing
 $2\rightarrow 1$ hadronic processes}},
 \href{https://doi.org/10.1140/epjc/s10052-020-08658-5}{\emph{Eur. Phys. J. C}
 {\bfseries 80} (2020) 1075}
 [\href{https://arxiv.org/abs/2006.04133}{{\ttfamily 2006.04133}}].
 
 \bibitem{GitLabPowheg} 
R.~Gauld, U.~Haisch, L.~Schnell, \emph{{HZJ SMEFT}}, \href{https://gitlab.com/lucschnell/hzj-smeft}{https://gitlab.com/lucschnell/hzj-smeft} [Accessed: 2024-11-28]. 

\bibitem{Majer:2020kdg}
I.~Majer, \emph{{Associated Higgs Boson Production at NNLO QCD}}, Ph.D. thesis,
 ETH Zurich, 2020,
\href{https://www.research-collection.ethz.ch/handle/20.500.11850/448848}{10.3929/ethz-b-000448848}.
 
 \bibitem{Kramer:1986sg}
G.~Kramer and B.~Lampe, \emph{{Two Jet Cross-Section in $e^+ e^-$ Annihilation}},
 \href{https://doi.org/10.1007/BF01679868}{\emph{Z. Phys. C} {\bfseries 34}
 (1987) 497}.

\bibitem{Hamberg:1990np}
R.~Hamberg, W.L.~van Neerven and T.~Matsuura, \emph{{A complete calculation of
 the order $\alpha_s^{2}$ correction to the Drell-Yan $K$ factor}},
 \href{https://doi.org/10.1016/0550-3213(91)90064-5}{\emph{Nucl. Phys. B}
 {\bfseries 359} (1991) 343}.

\bibitem{Gehrmann:2011ab}
T.~Gehrmann and L.~Tancredi, \emph{{Two-loop QCD helicity amplitudes for $q\bar
 q \to W^\pm \gamma$ and $q\bar q \to Z^0 \gamma$}},
 \href{https://doi.org/10.1007/JHEP02(2012)004}{\emph{JHEP} {\bfseries 02}
 (2012) 004} [\href{https://arxiv.org/abs/1112.1531}{{\ttfamily 1112.1531}}].

\bibitem{Campbell:2016jau}
J.M.~Campbell, R.K.~Ellis and C.~Williams, \emph{{Associated production of a
 Higgs boson at NNLO}},
 \href{https://doi.org/10.1007/JHEP06(2016)179}{\emph{JHEP} {\bfseries 06}
 (2016) 179} [\href{https://arxiv.org/abs/1601.00658}{{\ttfamily
 1601.00658}}].
 
\bibitem{Brein:2011vx}
O.~Brein, R.~Harlander, M.~Wiesemann and T.~Zirke, \emph{{Top-Quark Mediated
 Effects in Hadronic Higgs-Strahlung}},
 \href{https://doi.org/10.1140/epjc/s10052-012-1868-6}{\emph{Eur. Phys. J. C}
 {\bfseries 72} (2012) 1868}
 [\href{https://arxiv.org/abs/1111.0761}{{\ttfamily 1111.0761}}]. 
 
\bibitem{Zanoli:2021iyp}
S.~Zanoli, M.~Chiesa, E.~Re, M.~Wiesemann and G.~Zanderighi,
 \emph{{Next-to-next-to-leading order event generation for $VH$ production with
 $H \to b \bar b$ decay}},
 \href{https://doi.org/10.1007/JHEP07(2022)008}{\emph{JHEP} {\bfseries 07}
 (2022) 008} [\href{https://arxiv.org/abs/2112.04168}{{\ttfamily
 2112.04168}}]. 

\bibitem{Dixon:1996wi}
L.J.~Dixon, \emph{{Calculating scattering amplitudes efficiently}}, in
 \emph{{Theoretical Advanced Study Institute in Elementary Particle Physics
 (TASI 95): QCD and Beyond}}, pp.~539--584, 1, 1996
 [\href{https://arxiv.org/abs/hep-ph/9601359}{{\ttfamily hep-ph/9601359}}].

\bibitem{Maitre:2007jq}
D.~Maitre and P.~Mastrolia, \emph{{S@M, a Mathematica Implementation of the
 Spinor-Helicity Formalism}},
 \href{https://doi.org/10.1016/j.cpc.2008.05.002}{\emph{Comput. Phys. Commun.}
 {\bfseries 179} (2008) 501}
 [\href{https://arxiv.org/abs/0710.5559}{{\ttfamily 0710.5559}}].
 
 \bibitem{GitLabAmplitudes} 
R.~Gauld, U.~Haisch, L.~Schnell, \emph{{Vh Amplitudes}}, \href{https://gitlab.com/lucschnell/vh-amplitudes}{https://gitlab.com/lucschnell/vh-amplitudes} [Accessed: 2024-11-28].

\bibitem{ParticleDataGroup:2022pth}
{\scshape Particle Data Group} collaboration, \emph{{Review of Particle
 Physics}}, \href{https://doi.org/10.1093/ptep/ptac097}{\emph{PTEP} {\bfseries
 2022} (2022) 083C01}.

\bibitem{Gehrmann:2000zt}
T.~Gehrmann and E.~Remiddi,
\emph{{Two loop master integrals for $\gamma^*$ $\to$ 3 jets: The Planar topologies}},
\href{https://doi.org/10.1016/S0550-3213(01)00057-8}{\emph{Nucl. Phys. B} {\bfseries 601} (2001), 248-286}
[\href{https://arxiv.org/abs/hep-ph/0008287}{{\ttfamily hep-ph/0008287}}].

\bibitem{Gehrmann:2001jv}
T.~Gehrmann and E.~Remiddi,
\emph{{Numerical evaluation of two-dimensional harmonic polylogarithms}},
\href{https://doi.org/10.1016/S0010-4655(02)00139-X}{\emph{Comput. Phys. Commun.} {\bfseries 144} (2002), 200-223}
 [\href{https://arxiv.org/abs/hep-ph/0111255}{{\ttfamily hep-ph/0111255}}].
 
 \bibitem{Hahn:1998yk}
T.~Hahn and M.~Perez-Victoria,
\emph{{Automatized one loop calculations in four-dimensions and D-dimensions}},
\href{https://doi.org/10.1016/S0010-4655(98)00173-8}{\emph{Comput. Phys. Commun.} {\bfseries 118} (1999), 153-165}
 [\href{https://arxiv.org/abs/hep-ph/9807565}{{\ttfamily hep-ph/9807565}}].
 
 \bibitem{Hagiwara:1988pp}
K.~Hagiwara and D.~Zeppenfeld,
\emph{{Amplitudes for Multiparton Processes Involving a Current at $e^+ e^-$, $e^\pm p$, and Hadron Colliders}},
\href{https://doi.org/10.1016/0550-3213(89)90397-0}{\emph{Nucl. Phys. B} {\bfseries 313} (1989), 560-594}

\bibitem{Hasegawa:2009tx}
K.~Hasegawa, S.~Moch and P.~Uwer,
\emph{{AutoDipole: Automated generation of dipole subtraction terms}},
\href{https://doi.org/10.1016/j.cpc.2010.06.044}{\emph{Comput. Phys. Commun.} {\bfseries 181} (2010), 1802-1817}
 [\href{https://arxiv.org/abs/0911.4371}{{\ttfamily 0911.4371}}].

\bibitem{Astill:2018ivh}
W.~Astill, W.~Bizo\'n, E.~Re and G.~Zanderighi, \emph{{NNLOPS accurate
 associated $HZ$ production with $H\to b \bar b $ decay at NLO}},
 \href{https://doi.org/10.1007/JHEP11(2018)157}{\emph{JHEP} {\bfseries 11}
 (2018) 157} [\href{https://arxiv.org/abs/1804.08141}{{\ttfamily
 1804.08141}}].

\bibitem{Alioli:2019qzz}
S.~Alioli, A.~Broggio, S.~Kallweit, M.A.~Lim and L.~Rottoli,
 \emph{{Higgsstrahlung at NNLL$^\prime+$NNLO matched to parton showers in GENEVA}},
 \href{https://doi.org/10.1103/PhysRevD.100.096016}{\emph{Phys. Rev. D}
 {\bfseries 100} (2019) 096016}
 [\href{https://arxiv.org/abs/1909.02026}{{\ttfamily 1909.02026}}].

\bibitem{Bizon:2019tfo}
W.~Bizo\'n, E.~Re and G.~Zanderighi, \emph{{NNLOPS description of the $H \to
 b \bar b$ decay with MiNLO}},
 \href{https://doi.org/10.1007/JHEP06(2020)006}{\emph{JHEP} {\bfseries 06}
 (2020) 006} [\href{https://arxiv.org/abs/1912.09982}{{\ttfamily
 1912.09982}}].

\bibitem{Nason:2004rx}
P.~Nason, \emph{{A New method for combining NLO QCD with shower Monte Carlo
 algorithms}},
 \href{https://doi.org/10.1088/1126-6708/2004/11/040}{\emph{JHEP} {\bfseries
 11} (2004) 040} [\href{https://arxiv.org/abs/hep-ph/0409146}{{\ttfamily
 hep-ph/0409146}}].

\bibitem{Frixione:2007vw}
S.~Frixione, P.~Nason and C.~Oleari, \emph{{Matching NLO QCD computations with
 Parton Shower simulations: the POWHEG method}},
 \href{https://doi.org/10.1088/1126-6708/2007/11/070}{\emph{JHEP} {\bfseries
 11} (2007) 070} [\href{https://arxiv.org/abs/0709.2092}{{\ttfamily
 0709.2092}}].

\bibitem{Sjostrand:2014zea}
T.~Sj\"ostrand, S.~Ask, J.R.~Christiansen, R.~Corke, N.~Desai, P.~Ilten et~al.,
 \emph{{An introduction to PYTHIA 8.2}},
 \href{https://doi.org/10.1016/j.cpc.2015.01.024}{\emph{Comput. Phys. Commun.}
 {\bfseries 191} (2015) 159}
 [\href{https://arxiv.org/abs/1410.3012}{{\ttfamily 1410.3012}}].

\bibitem{Boughezal:2016wmq}
R.~Boughezal, J.M.~Campbell, R.K.~Ellis, C.~Focke, W.~Giele, X.~Liu et~al.,
 \emph{{Color singlet production at NNLO in MCFM}},
 \href{https://doi.org/10.1140/epjc/s10052-016-4558-y}{\emph{Eur. Phys. J. C}
 {\bfseries 77} (2017) 7} [\href{https://arxiv.org/abs/1605.08011}{{\ttfamily
 1605.08011}}].
 
 \bibitem{Gauld:2019yng}
R.~Gauld, A.~Gehrmann-De~Ridder, E.W.N.~Glover, A.~Huss and I.~Majer,
 \emph{{Associated production of a Higgs boson decaying into bottom quarks and
 a weak vector boson decaying leptonically at NNLO in QCD}},
 \href{https://doi.org/10.1007/JHEP10(2019)002}{\emph{JHEP} {\bfseries 10}
 (2019) 002} [\href{https://arxiv.org/abs/1907.05836}{{\ttfamily
 1907.05836}}].

\bibitem{Buccioni:2019sur}
F.~Buccioni, J.-N.~Lang, J.M.~Lindert, P.~Maierh\"ofer, S.~Pozzorini, H.~Zhang
 et~al., \emph{{OpenLoops 2}},
 \href{https://doi.org/10.1140/epjc/s10052-019-7306-2}{\emph{Eur. Phys. J. C}
 {\bfseries 79} (2019) 866}
 [\href{https://arxiv.org/abs/1907.13071}{{\ttfamily 1907.13071}}].

\bibitem{Frixione:1995ms}
S.~Frixione, Z.~Kunszt and A.~Signer, \emph{{Three jet cross-sections to
 next-to-leading order}},
 \href{https://doi.org/10.1016/0550-3213(96)00110-1}{\emph{Nucl. Phys. B}
 {\bfseries 467} (1996) 399}
 [\href{https://arxiv.org/abs/hep-ph/9512328}{{\ttfamily hep-ph/9512328}}].

\bibitem{Frixione:1997np}
S.~Frixione, \emph{{A General approach to jet cross-sections in QCD}},
 \href{https://doi.org/10.1016/S0550-3213(97)00574-9}{\emph{Nucl. Phys. B}
 {\bfseries 507} (1997) 295}
 [\href{https://arxiv.org/abs/hep-ph/9706545}{{\ttfamily hep-ph/9706545}}].

\bibitem{Dawson:2019clf}
S.~Dawson and P.P.~Giardino, \emph{{Electroweak and QCD corrections to $Z$ and
 $W$ pole observables in the standard model EFT}},
 \href{https://doi.org/10.1103/PhysRevD.101.013001}{\emph{Phys. Rev. D}
 {\bfseries 101} (2020) 013001}
 [\href{https://arxiv.org/abs/1909.02000}{{\ttfamily 1909.02000}}].

\bibitem{Hamilton:2012np}
K.~Hamilton, P.~Nason and G.~Zanderighi, \emph{{MINLO: Multi-Scale Improved
 NLO}}, \href{https://doi.org/10.1007/JHEP10(2012)155}{\emph{JHEP} {\bfseries
 10} (2012) 155} [\href{https://arxiv.org/abs/1206.3572}{{\ttfamily
 1206.3572}}].

\bibitem{Hamilton:2012rf}
K.~Hamilton, P.~Nason, C.~Oleari and G.~Zanderighi, \emph{{Merging $H/W/Z + 0$
 and $1$~jet at NLO with no merging scale: a path to parton shower + NNLO
 matching}}, \href{https://doi.org/10.1007/JHEP05(2013)082}{\emph{JHEP}
 {\bfseries 05} (2013) 082} [\href{https://arxiv.org/abs/1212.4504}{{\ttfamily
 1212.4504}}].

\bibitem{Chen:2020xot}
L.~Chen and A.~Freitas, \emph{{Mixed EW-QCD leading fermionic three-loop
 corrections at $\mathcal{O}(\alpha_s\alpha^2)$ to electroweak precision
 observables}}, \href{https://doi.org/10.1007/JHEP03(2021)215}{\emph{JHEP}
 {\bfseries 03} (2021) 215}
 [\href{https://arxiv.org/abs/2012.08605}{{\ttfamily 2012.08605}}].

\bibitem{ALEPH:2005ab}
{\scshape ALEPH, DELPHI, L3, OPAL, SLD, LEP Electroweak Working Group, SLD
 Electroweak Group, SLD Heavy Flavour Group} collaboration, \emph{{Precision
 electroweak measurements on the $Z$ resonance}},
 \href{https://doi.org/10.1016/j.physrep.2005.12.006}{\emph{Phys. Rept.}
 {\bfseries 427} (2006) 257}
 [\href{https://arxiv.org/abs/hep-ex/0509008}{{\ttfamily hep-ex/0509008}}].
 
\bibitem{Blennow:2022yfm}
M.~Blennow, P.~Coloma, E.~Fern\'andez-Mart\'\i{}nez and M.~Gonz\'alez-L\'opez,
 \emph{{Right-handed neutrinos and the CDF II anomaly}},
 \href{https://doi.org/10.1103/PhysRevD.106.073005}{\emph{Phys. Rev. D}
 {\bfseries 106} (2022) 073005}
 [\href{https://arxiv.org/abs/2204.04559}{{\ttfamily 2204.04559}}].

\bibitem{Cirigliano:2022qdm}
V.~Cirigliano, W.~Dekens, J.~de~Vries, E.~Mereghetti and T.~Tong,
 \emph{{Beta-decay implications for the W-boson mass anomaly}},
 \href{https://doi.org/10.1103/PhysRevD.106.075001}{\emph{Phys. Rev. D}
 {\bfseries 106} (2022) 075001}
 [\href{https://arxiv.org/abs/2204.08440}{{\ttfamily 2204.08440}}].

\bibitem{Cirigliano:2022yyo}
V.~Cirigliano, A.~Crivellin, M.~Hoferichter and M.~Moulson, \emph{{Scrutinizing
 CKM unitarity with a new measurement of the
 $K_{\mu 3}/K_{\mu 2}$ branching fraction}},
 \href{https://doi.org/10.1016/j.physletb.2023.137748}{\emph{Phys. Lett. B}
 {\bfseries 838} (2023) 137748}
 [\href{https://arxiv.org/abs/2208.11707}{{\ttfamily 2208.11707}}].

\bibitem{Grossman:2019bzp}
Y.~Grossman, E.~Passemar and S.~Schacht, \emph{{On the Statistical Treatment of
 the Cabibbo Angle Anomaly}},
 \href{https://doi.org/10.1007/JHEP07(2020)068}{\emph{JHEP} {\bfseries 07}
 (2020) 068} [\href{https://arxiv.org/abs/1911.07821}{{\ttfamily
 1911.07821}}].

\bibitem{Cirigliano:2023nol}
V.~Cirigliano, W.~Dekens, J.~de~Vries, E.~Mereghetti and T.~Tong,
 \emph{{Anomalies in global SMEFT analyses: a case study of first-row CKM
 unitarity}}, [\href{https://arxiv.org/abs/2311.00021}{{\ttfamily
 2311.00021}}].

\bibitem{DAmbrosio:2002vsn}
G.~D'Ambrosio, G.F.~Giudice, G.~Isidori and A.~Strumia, \emph{{Minimal flavor
 violation: An Effective field theory approach}},
 \href{https://doi.org/10.1016/S0550-3213(02)00836-2}{\emph{Nucl. Phys.}
 {\bfseries B645} (2002) 155}
 [\href{https://arxiv.org/abs/hep-ph/0207036}{{\ttfamily hep-ph/0207036}}]. 

\bibitem{Falkowski:2019hvp}
A.~Falkowski and D.~Straub, \emph{{Flavourful SMEFT likelihood for Higgs and
 electroweak data}},
 \href{https://doi.org/10.1007/JHEP04(2020)066}{\emph{JHEP} {\bfseries 04}
 (2020) 066} [\href{https://arxiv.org/abs/1911.07866}{{\ttfamily
 1911.07866}}].

\bibitem{Brivio:2020onw}
I.~Brivio, \emph{{SMEFTsim 3.0 \textemdash{} a practical guide}},
 \href{https://doi.org/10.1007/JHEP04(2021)073}{\emph{JHEP} {\bfseries 04}
 (2021) 073} [\href{https://arxiv.org/abs/2012.11343}{{\ttfamily
 2012.11343}}].

\bibitem{LHCHiggsWG4}
{\scshape LHC Higgs Cross section Working Group} collaboration, 
 \href{https://twiki.cern.ch/twiki/bin/view/LHCPhysics/CERNYellowReportPageBR}{https://twiki.cern.ch/twiki/bin/view/LHCPhysics/CERNYellowReportPageBR}.

\bibitem{ATLAS:2020qdt}
{\scshape ATLAS} collaboration, \emph{{A combination of measurements of Higgs
 boson production and decay using up to 139~fb$^{-1}$ of proton-proton
 collision data at $\sqrt{s}$~=~13~TeV collected with the ATLAS experiment}}, 
 \href{https://cds.cern.ch/record/2725733}{ATLAS-CONF-2020-027}.

\bibitem{CMS:2020gsy}
{\scshape CMS} collaboration, \emph{{Combined Higgs boson production and decay
 measurements with up to 137~fb$^{-1}$ of proton-proton collision data at
 $\sqrt s$~=~13 TeV}}, \href{http://cds.cern.ch/record/2706103}{CMS-PAS-HIG-19-005}.

\bibitem{CMS:2023mku}
{\scshape ATLAS} and {\scshape CMS} collaborations, \emph{{Evidence for the Higgs boson decay
 to a $Z$ boson and a photon at the LHC}},
 [\href{https://arxiv.org/abs/2309.03501}{{\ttfamily 2309.03501}}].

\bibitem{NNPDF:2017mvq}
{\scshape NNPDF} collaboration, \emph{{Parton distributions from high-precision
 collider data}},
 \href{https://doi.org/10.1140/epjc/s10052-017-5199-5}{\emph{Eur. Phys. J. C}
 {\bfseries 77} (2017) 663}
 [\href{https://arxiv.org/abs/1706.00428}{{\ttfamily 1706.00428}}].

\bibitem{Skands:2014pea}
P.~Skands, S.~Carrazza and J.~Rojo, \emph{{Tuning PYTHIA 8.1: the Monash 2013
 Tune}}, \href{https://doi.org/10.1140/epjc/s10052-014-3024-y}{\emph{Eur.
 Phys. J. C} {\bfseries 74} (2014) 3024}
 [\href{https://arxiv.org/abs/1404.5630}{{\ttfamily 1404.5630}}].

\bibitem{ATLAS:2019yhn}
{\scshape ATLAS} collaboration, \emph{{Measurement of $VH$, $H \to b \bar b$ 
 production as a function of the vector-boson transverse momentum in 13~TeV 
 $pp$ collisions with the ATLAS detector}}, \href{https://doi.org/10.1007/JHEP05(2019)141}{\emph{JHEP}
 {\bfseries 05} (2019) 141}
 [\href{https://arxiv.org/abs/1903.04618}{{\ttfamily 1903.04618}}].

\bibitem{ATLAS:2020fcp}
{\scshape ATLAS} collaboration, \emph{{Measurements of $WH$ and $ZH$ production
 in the $H \to b\bar{b}$ decay channel in $pp$ collisions at 13~TeV
 with the ATLAS detector}},
 \href{https://doi.org/10.1140/epjc/s10052-020-08677-2}{\emph{Eur. Phys. J. C}
 {\bfseries 81} (2021) 178}
 [\href{https://arxiv.org/abs/2007.02873}{{\ttfamily 2007.02873}}].

\bibitem{CMS:2022exo}
{\scshape CMS} collaboration, \emph{{Simplified template cross section
 measurements of Higgs boson produced in association with vector bosons in the
 $H \to b \bar b$ decay channel in proton-proton collisions
 at $\sqrt{s}$~=13~TeV}}, 
 \href{https://cds.cern.ch/record/2827421}{CMS-PAS-HIG-20-001}.


\bibitem{Andersen:2016qtm}
J.R.~Andersen et~al., \emph{{Les Houches 2015: Physics at TeV Colliders
 Standard Model Working Group Report}}, in \emph{{9th Les Houches Workshop on
 Physics at TeV Colliders}}, 5, 2016
 [\href{https://arxiv.org/abs/1605.04692}{{\ttfamily 1605.04692}}].

\bibitem{Berger:2019wnu}
N.~Berger et~al., \emph{{Simplified Template Cross Sections - Stage 1.1}},
 [\href{https://arxiv.org/abs/1906.02754}{{\ttfamily 1906.02754}}].

\bibitem{Amoroso:2020lgh}
S.~Amoroso et~al., \emph{{Les Houches 2019: Physics at TeV Colliders: Standard
 Model Working Group Report}}, in \emph{{11th Les Houches Workshop on Physics
 at TeV Colliders}: {PhysTeV Les Houches}}, 3, 2020
 [\href{https://arxiv.org/abs/2003.01700}{{\ttfamily 2003.01700}}].
 
\bibitem{Degrande:2020evl}
C.~Degrande, G.~Durieux, F.~Maltoni, K.~Mimasu, E.~Vryonidou and C.~Zhang,
 \emph{{Automated one-loop computations in the standard model effective field
 theory}}, \href{https://doi.org/10.1103/PhysRevD.103.096024}{\emph{Phys. Rev.
 D} {\bfseries 103} (2021) 096024}
 [\href{https://arxiv.org/abs/2008.11743}{{\ttfamily 2008.11743}}]. 

\bibitem{Biekotter:2023xle}
A.~Biek\"otter, B.D.~Pecjak, D.J.~Scott and T.~Smith, \emph{{Electroweak input
 schemes and universal corrections in SMEFT}},
 \href{https://doi.org/10.1007/JHEP07(2023)115}{\emph{JHEP} {\bfseries 07}
 (2023) 115} [\href{https://arxiv.org/abs/2305.03763}{{\ttfamily
 2305.03763}}].
 
\bibitem{Hahn:2016ebn}
T.~Hahn, S.~Pa\ss{}ehr and C.~Schappacher, \emph{{FormCalc 9 and Extensions}},
 \href{https://doi.org/10.1088/1742-6596/762/1/012065}{\emph{PoS} {\bfseries
 LL2016} (2016) 068} [\href{https://arxiv.org/abs/1604.04611}{{\ttfamily
 1604.04611}}]. 
 
 \bibitem{Fox:2018ldq}
P.J.~Fox, I.~Low and Y.~Zhang, \emph{{Top-philic $Z'$ forces at the LHC}},
 \href{https://doi.org/10.1007/JHEP03(2018)074}{\emph{JHEP} {\bfseries 03}
 (2018) 074} [\href{https://arxiv.org/abs/1801.03505}{{\ttfamily
 1801.03505}}].
 
 \bibitem{Rossia:2023hen}
A.N.~Rossia, M.O.A.~Thomas and E.~Vryonidou, \emph{{Diboson production in the
 SMEFT from gluon fusion}},
 [\href{https://arxiv.org/abs/2306.09963}{{\ttfamily 2306.09963}}].

\bibitem{Bonnefoy:2020tyv}
Q.~Bonnefoy, L.~Di~Luzio, C.~Grojean, A.~Paul and A.N.~Rossia, \emph{{Comments
 on gauge anomalies at dimension-six in the Standard Model Effective Field
 Theory}}, \href{https://doi.org/10.1007/JHEP05(2021)153}{\emph{JHEP}
 {\bfseries 05} (2021) 153}
 [\href{https://arxiv.org/abs/2012.07740}{{\ttfamily 2012.07740}}].

\bibitem{Feruglio:2020kfq}
F.~Feruglio, \emph{{A Note on Gauge Anomaly Cancellation in Effective Field
 Theories}}, \href{https://doi.org/10.1007/JHEP03(2021)128}{\emph{JHEP}
 {\bfseries 03} (2021) 128}
 [\href{https://arxiv.org/abs/2012.13989}{{\ttfamily 2012.13989}}].

\bibitem{Cornella:2022hkc}
C.~Cornella, F.~Feruglio and L.~Vecchi, \emph{{Gauge invariance and finite
 counterterms in chiral gauge theories}},
 \href{https://doi.org/10.1007/JHEP02(2023)244}{\emph{JHEP} {\bfseries 02}
 (2023) 244} [\href{https://arxiv.org/abs/2205.10381}{{\ttfamily
 2205.10381}}].

\bibitem{Cohen:2023hmq}
T.~Cohen, X.~Lu and Z.~Zhang, \emph{{Anomalies from the covariant derivative
 expansion}}, \href{https://doi.org/10.1103/PhysRevD.107.116015}{\emph{Phys.
 Rev. D} {\bfseries 107} (2023) 116015}
 [\href{https://arxiv.org/abs/2301.00821}{{\ttfamily 2301.00821}}].

\bibitem{Cohen:2023gap}
T.~Cohen, X.~Lu and Z.~Zhang, \emph{{Anomaly cancellation in effective field
 theories from the covariant derivative expansion}},
 \href{https://doi.org/10.1103/PhysRevD.108.056027}{\emph{Phys. Rev. D}
 {\bfseries 108} (2023) 056027}
 [\href{https://arxiv.org/abs/2301.00827}{{\ttfamily 2301.00827}}].
 
 \bibitem{Wess:1971yu}
J.~Wess and B.~Zumino, \emph{{Consequences of anomalous Ward identities}},
 \href{https://doi.org/10.1016/0370-2693(71)90582-X}{\emph{Phys. Lett. B}
 {\bfseries 37} (1971) 95}.
 
 \bibitem{Bardeen:1984pm}
W.A.~Bardeen and B.~Zumino, \emph{{Consistent and Covariant Anomalies in Gauge
 and Gravitational Theories}},
 \href{https://doi.org/10.1016/0550-3213(84)90322-5}{\emph{Nucl. Phys. B}
 {\bfseries 244} (1984) 421}.

\bibitem{Durieux:2018ggn}
G.~Durieux, J.~Gu, E.~Vryonidou and C.~Zhang, \emph{{Probing top-quark
 couplings indirectly at Higgs factories}},
 \href{https://doi.org/10.1088/1674-1137/42/12/123107}{\emph{Chin. Phys. C}
 {\bfseries 42} (2018) 123107}
 [\href{https://arxiv.org/abs/1809.03520}{{\ttfamily 1809.03520}}].

\bibitem{Alwall:2014hca}
J.~Alwall, R.~Frederix, S.~Frixione, V.~Hirschi, F.~Maltoni, O.~Mattelaer
 et~al., \emph{{The automated computation of tree-level and next-to-leading
 order differential cross sections, and their matching to parton shower
 simulations}}, \href{https://doi.org/10.1007/JHEP07(2014)079}{\emph{JHEP}
 {\bfseries 07} (2014) 079} [\href{https://arxiv.org/abs/1405.0301}{{\ttfamily
 1405.0301}}].

\end{thebibliography}


\providecommand{\href}[2]{#2}\begingroup\raggedright\endgroup

\end{document}